\PassOptionsToPackage{hyphens}{url}
\PassOptionsToPackage{colorlinks=true,citecolor=blue,linkcolor=blue,urlcolor=blue}{hyperref}
\documentclass{JFM-FLM_Au}

\usepackage{amsmath,amsfonts,amssymb,bm}
\usepackage[percent]{overpic}
\usepackage{xcolor,xfp}
\usepackage[english]{babel}
\usepackage{csquotes}
\usepackage[final]{microtype}
\usepackage{placeins}
\usepackage{physics}
\microtypesetup{protrusion=true,expansion=true}

\usepackage{xr-hyper}
\usepackage{cleveref}
\usepackage{enumitem}

\renewcommand{\oddfooterflagdefns}[1]{}
\renewcommand{\evenfooterflagdefns}[1]{}

\renewcommand{\oddabsfooterflag}{}
\renewcommand{\evenabsfooterflag}{}

\newcommand{\RN}[1]{%
	\textup{\uppercase\expandafter{\romannumeral#1}}%
}

\providecommand\ie{\emph{i.e.}\ }

\providecommand{\diff}{{\mbox{d}}}
\newcommand{\evec}{\boldsymbol{e}}

\newcommand{\uvec}{\boldsymbol{u}}
\newcommand{\uvecp}{\widetilde{\boldsymbol{u}}}
\newcommand{\pp}{\widetilde{p}}
\newcommand{\uhvec}{\widehat{\boldsymbol{u}}}

\newcommand{\qvecp}{\widetilde{\boldsymbol{b}}}
\newcommand{\qhvec}{\widehat{\boldsymbol{b}}}

\newcommand{\Qvec}{\boldsymbol{b}_0}
\newcommand{\Uvec}{\overline{\boldsymbol{U}}}
\newcommand{\Nvec}{\boldsymbol{N}}
\newcommand{\Tvec}{\boldsymbol{T}}
\newcommand{\Bvec}{\boldsymbol{B}}
\usepackage{caption}
\usepackage{subcaption}
\usepackage{tikz}
\usetikzlibrary{arrows.meta,calc,angles}
\newcommand{\inletplanearrow}[4]{%
	\put(#1,#2){%
		\begin{tikzpicture}[x=1cm,y=1cm]
			\draw[-{Latex[length=3.2mm,width=2.8mm]},line width=2.2pt]
			(0,0) -- (#3,#4);
		\end{tikzpicture}%
	}%
}

\newcommand{\nut}{\nu_t}
\newcommand{\nueff}{\nu_{\mathrm{e}}}
\newcommand{\Smat}{\boldsymbol{S}}
\newcommand{\Sscale}{\mathcal{S}_{\mathrm{sc}}}
\newcommand{\Rmat}{\overline{\boldsymbol{R}}}

\crefname{figure}{Fig.}{Figs.}
\Crefname{figure}{Figure}{Figures}
\crefname{table}{Table}{Tables}
\Crefname{table}{Table}{Tables}
\crefname{equation}{Eq.}{Eqs.}
\Crefname{equation}{Equation}{Equations}
\crefname{section}{Sec.}{Secs.}
\Crefname{section}{Section}{Sections}

\lefttitle{Bagheri, Casali, Becker \& Schlatter}
\righttitle{Disentangling coherent structures and the origin of swirl-switching}

\title{Disentangling coherent structures and the origin of swirl-switching}

\author{
	Eman Bagheri\aff{1},
	Riccardo Casali\aff{1},
	Stefan Becker\aff{1}
	\and
	Philipp Schlatter\aff{1}
}

\affiliation{
	\aff{1}Institute of Fluid Mechanics (LSTM),
	Friedrich--Alexander--Universit\"at (FAU) Erlangen--N\"urnberg,
	91058 Erlangen, Germany
}

\corresau{Philipp Schlatter, \email{philipp.schlatter@fau.de}}

\begin{document}
	\emergencystretch=2em
	
	\maketitle
	
	\begin{abstract}
		The physical origin of swirl-switching in turbulent bent-pipe flow remains the subject of ongoing research. We revisit this phenomenon and investigate whether swirl-switching is an inherent instability of curved-pipe flow. We perform three direct numerical simulations (DNS) of flow through a $180^{\circ}$ bent pipe for curvature $\gamma=0.2$ at $Re_D=5,300$ and $Re_D=10,000$, and for curvature $\gamma=0.4$ at $Re_D=5,300$. Here, $Re_D$ is the Reynolds number based on the bulk velocity $U_b$ and pipe diameter $D$. The DNS data are subsequently used to conduct modal decompositions and local stability analysis (LSA). We discuss the limitations of common modal decomposition methods and introduce a new decomposition method, filtered Hilbert proper orthogonal decomposition (FHPOD), enabling us to isolate different instabilities as distinct FHPOD modes.
		FHPOD reveals seven modes belonging to four distinct families: a low-frequency axial mode, two swirl-switching modes, two swirl-breathing modes and a pair of downstream shear-layer modes. The swirl-switching and swirl-breathing modes are localised to the curved section of the bent-pipe flow and represent the sinuous and varicose instabilities of the Dean vortices, respectively. The DNS data are used to perform local stability analysis of the mean flow using the frozen eddy-viscosity method at the two streamwise wavenumbers associated with the two swirl-switching and swirl-breathing modes. The linear stability analysis reveals six unstable branches, with the swirl-switching mode being the most unstable branch. Another unstable branch represents the swirl-breathing mode. The modes obtained from the local stability analysis are in close agreement with their FHPOD counterparts in both spatial features and the range of Strouhal numbers. The results support the interpretation that swirl-switching arises as an intrinsic instability of the bent-pipe mean flow, which can be excited by incoming turbulent structures.
		
	\end{abstract}
	
	\begin{keywords}
		low-dimensional models, shear-layer instability, pipe flow
	\end{keywords}

	\section{Introduction}
	\label{sec:intro}
	
\begin{figure}[!htbp]
		\centering
		\includegraphics[width=0.80\textwidth]{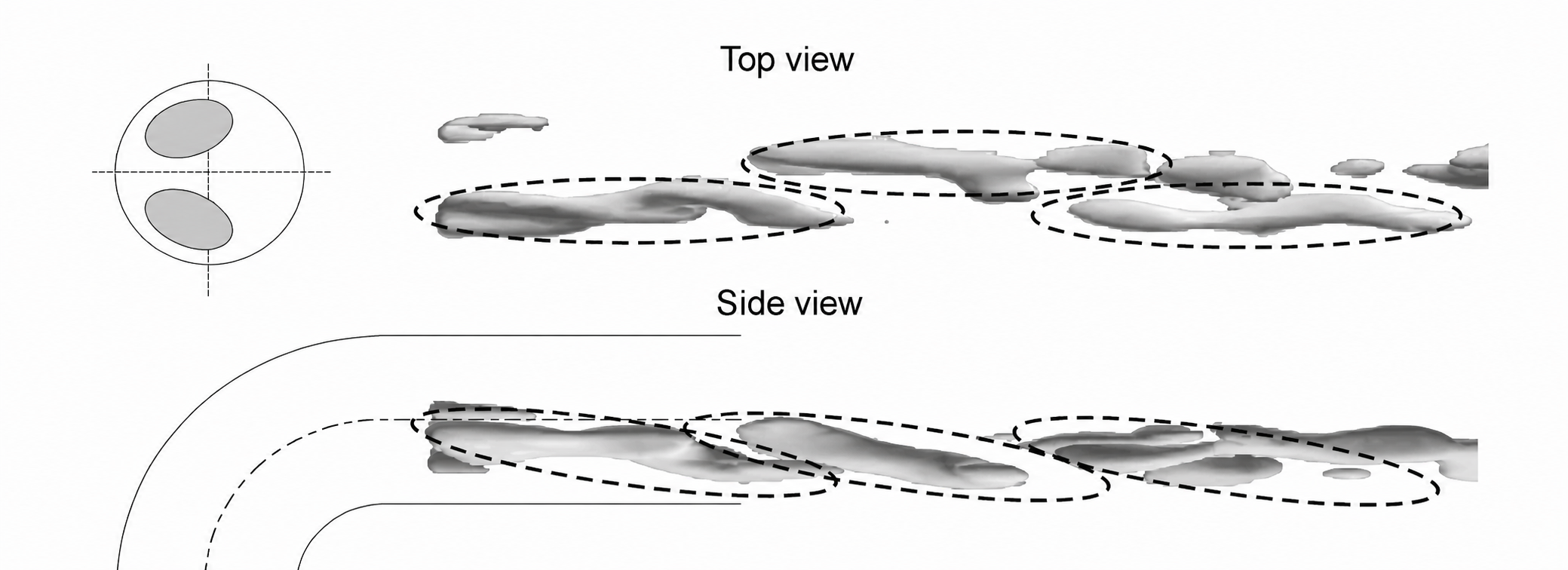}
		\caption{Coherent structures identified by \citet{Brucker1998} as the manifestation of the swirl-switching phenomenon, shown using the spatio-temporal reconstruction of streamwise vorticity isosurfaces at $1.5\,D$ downstream of a $90^\circ$ bend measured over $50$ seconds. Adapted from figure~2(b) of \citet{Brucker1998}, with permission from Christoph Brücker.}
		\label{fig:bruecker_ss}
	\end{figure}

	Flows in curved pipes are found in a wide range of industrial and physiological systems and have been the subject of research for more than a century \citep{Berger1983, Vashisth2008, Vester2016}. The dynamics of the flow and turbulence in curved pipes are fundamentally different from those in straight pipes \citep{NOORANI201316}. Curved pipe flows are characterised by two non-dimensional numbers, namely the Reynolds number,
	\begin{equation}
		Re_D= \frac{U_b D}{\nu}
	\end{equation}
	and the non-dimensional curvature parameter,
	\begin{equation}
		\gamma = \frac{ D}{2R_c} \ .
	\end{equation}
	Here, $D$ denotes the pipe diameter, $U_b$ the bulk velocity, $\nu$ the kinematic viscosity, and $R_c$ the radius of curvature of the bend. As a result of the curvature and the centrifugal force, a Prandtl's secondary flow of the first type forms inside the curved pipe. This cross-stream secondary motion is termed Dean vortices in recognition of the contributions of \citet{Dean1927, Dean1928}, who derived the governing asymptotic solution for weak curvatures. 
	At sufficiently high Reynolds numbers, the Dean vortices become unstable and significantly affect the turbulent flow \citep{bagheri2026}. 
	A key feature of the unstable Dean vortices is the so-called \emph{swirl-switching} phenomenon. \citet{Tunstall1968} were the first to report a low-frequency modulation of the Dean vortices. 
	They studied turbulent flows in mitred $90^{\circ}$ bends and noticed that the Dean pair was replaced by a single vortex that intermittently flipped its axis of rotation. They reported a very low switching frequency, which later reviews commonly express in terms of the Strouhal number.
	\begin{equation}
		St = \frac{fD}{U_b},
		\qquad
		2\times10^{-4} \lesssim St \lesssim 4.5\times10^{-3},
	\end{equation}
	where $f$ is the characteristic modulation frequency. \citet{Brucker1998} utilised time-resolved digital particle image velocimetry (DPIV) in water at $Re_D=2000$ and $5000$ to visualise quasi-periodic oscillations of the secondary flow in a spatially developing $90^{\circ}$ bend with $\gamma=0.5$. From the tangential-velocity signal measured at $1.5D$ downstream of the bend close to the inner wall and in the geometrical plane of symmetry, he found two dominant frequencies at $St\approx0.03$ and $St\approx0.12$, associated with alternate switching of the tangential flow. The corresponding streamwise-vorticity visualisations showed an alternating left and right switching of the vortex pair, which he referred to as ``swirl-switching'' (see the reprinted figure~\ref{fig:bruecker_ss}). A source of confusion, however, emerged in later studies, where the term ``swirl-switching'' was loosely used to refer to similar yet distinct coherent structures with different Strouhal numbers and spatial support from those originally described by \citet{Brucker1998}.
	
	Subsequent work by \citet{Sudo1998, Sudo2000} characterised the streamwise development of the mean flow in $90^{\circ}$ and $180^{\circ}$ bends. \citet{Sudo1998} showed that, in the $90^{\circ}$ bend, the secondary flow weakens downstream, while the influence of the bend remains even at $10D$ downstream of the bend. \citet{Sudo2000} found that, in the $180^{\circ}$ bend, an additional pair of vortices appears just downstream of the bend exit, and that recovery to a straight pipe flow requires a longer downstream distance compared with the $90^{\circ}$ bend. 
	
	Understanding swirl-switching is important both from a flow-physics perspective and from an engineering point of view. The modulation of wall pressure and wall shear stress caused by swirl-switching has practical implications for structural fatigue, flow-induced vibrations and potential resonance in curved-pipe systems \citep{Hufnagel2018, He2021, Zhang2024, Lupi2025}. This practical relevance has helped drive extensive experimental and numerical work in subsequent years. 
	
	\citet{Rutten2001, Rutten2005} performed the first large-eddy simulations (LES) to confirm the existence of this phenomenon. In $90^{\circ}$ bends at $Re_D \approx 2.7\times 10^{4}$, they observed low–frequency oscillations of the Dean-vortex stagnation point and wall forces, with peaks at $St\approx0.0055$ and $0.014$ \citep{Rutten2001}, and later around $St\approx0.01$ \citep{Rutten2005}. They associated the high-frequency oscillations at
	$St \approx 0.2$\,–\,$0.3$ with shear–layer instabilities. They also noted that the low–frequency oscillation can persist even when the curvature is too small to produce a strong separation region.
	
	\citet{Sakakibara2010} applied stereo-PIV and proper orthogonal
	decomposition (POD) in a $90^{\circ}$ bend at high Reynolds number. They identified antisymmetric modes with time-coefficient spectra exhibiting peaks in the range $St \approx 0.02$--$0.07$, depending on the downstream position. In a later study, \citet{SakakibaraMachida2012} performed simultaneous upstream and downstream PIV measurements and found upstream streaks correlated with the downstream displacement of the inner-side stagnation point. They noted that the connection between these streaks and very-large-scale motions (VLSMs) was unclear.
	
	In 2013, several PIV/POD studies focused on the swirl-switching phenomenon in $90^{\circ}$ bends. \citet{Kalpakli2013} used time-resolved stereoscopic PIV (2D3C) at the exit of a $90^{\circ}$ bend
	($\gamma\approx0.31$, $Re_D\approx 3.4\times10^{4}$). Using a snapshot POD of the 2D3C velocity fields, they reconstructed the flow using the first six POD modes and analysed PSDs of the tangential velocity. They found a dominant peak at $St \approx 0.12$, with additional peaks at $St \approx 0.04$ and $St \approx 0.18$. The lower-frequency peak at $St \approx 0.04$ was associated with swirl-switching. \citet{Hellstrom2013} measured the velocity further downstream of a $90^{\circ}$ bend with $\gamma = 0.5$ at $Re_D\approx2.5\times10^{4}$. They stated that the mode at $St \approx 0.16$ is due to the oscillation of the Dean motion, while the higher-frequency mode at $St \approx 0.33$ is the swirl-switching mode. In a follow-up study of mildly and sharply curved $90^{\circ}$ bends, \citet{KalpakliVester2015} reported spectral peaks based on the power spectral density (PSD) of POD time-coefficients at $St \approx 0.04$ (and $0.10$) for the sharp bend case ($\gamma \approx 0.39$). They also showed that conditioning the upstream flow using a honeycomb strongly modified the mean flow and the dominant modes, and the swirl-switching mode did not appear near the bend. However, further downstream, the POD results showed the reappearance of a single-cell structure and a Dean-cell structure, and the authors suggested that swirl-switching might not be eliminated but merely spatially postponed.

	\citet{Carlsson2015} performed LES of flow in configurations similar to that of \citet{Kalpakli2013} with several curvatures using inflow data generated from a separate periodic straight-pipe simulation. Their spectra and POD analysis identified a low-frequency swirl-switching at $St\approx0.003$--$0.01$, with a smaller peak at $St\approx0.13$ for the smaller curvatures, and a high-frequency swirl-switching at $St\approx0.5$--$0.6$ for the sharpest bend. However, their use of an inflow condition from a periodic precursor simulation similar to that of \citet{Rutten2001, Rutten2005} was later found to be the cause of spurious fluctuations due to the imposed inflow periodicity. 
	
	\citet{Noorani2016} performed direct numerical simulation (DNS) in a toroidal pipe and, using POD, showed that coherent structures resembling swirl-switching occur in a geometry with no upstream straight section and no separation bubble. They concluded that swirl-switching may be due to curvature rather than being caused exclusively by upstream large-scale structures or bend separation. They also emphasised that no universal  ``single swirl'' POD mode captures all of the observed dynamics. Nevertheless, the reduced-order reconstructions of the flow using the most energetic modes reproduced a swirl-switching structure.
	
	\citet{Hufnagel2018} addressed two outstanding issues in previous numerical studies and studied spatially developing bent pipe flows. Their DNS employed a synthetic eddy inflow condition to rule out spurious modes arising from the recycling and periodic methods. Furthermore, they applied POD to the full three-dimensional velocity field over the full domain and found a three-dimensional travelling wave with a dominant Strouhal number $St \approx 0.16$ for the bent pipe with $\gamma=0.3$. They also showed that 2D-POD of cross-sections, as typically employed in the previous experimental studies, cannot appropriately separate spatial and temporal modulation of such travelling waves.
	
	\citet{Wang2018} performed DNS for a $90^{\circ}$-bend and compared several methods to detect swirl-switching and found that the reported dominant peak depends on the tracked quantity
	of interest. They found a prominent peak
	around $St \approx 0.5$ when global integral measures such as the lateral
	wall-pressure force and half-sided mass-flow rate were used. They also showed that methods based on local velocity and the stagnation point can be strongly location dependent and may produce multiple competing peaks.

	More recently, using refractive-index-matched PIV (RIM–PIV) in $90^{\circ}$ bends with $\gamma \approx 0.33$,
	\citet{Jain2019} found large-scale oscillations of the secondary flow with
	spectral peaks spanning a broad range from $St \approx 0.03$ up to
	$St \approx 0.6$ including bands around $St \approx 0.17$--$0.33$ and $St \approx 0.4$--$0.55$. In a subsequent study, \citet{Jain2022} addressed a gap in the literature by characterising large-scale oscillations in a moderately curved $180^{\circ}$ bend. They reported low and moderate frequency peaks downstream of the bend, with $St\approx0.02$--$0.05$ and $St\approx0.16$ being dominant, as well as
	other contributions at $St\approx0.33$ and in the higher-frequency range around $St\approx0.4$ and $0.55$.
	
	\citet{He2021} used wall–resolved LES of a $90^{\circ}$ bend for $Re_D = 5.3\times10^{3}$–$4.5\times10^{4}$ and applied POD to velocity fields on cross-sections downstream of the bend. They showed that oscillations associated with swirl-switching and the wall-shear force are dominated by frequencies around $St \approx 0.25$–$0.28$.
	
	Using LES and modal analysis of toroidal pipes, \citet{Zhang2024} further separated swirl-switching frequencies into a low-frequency branch ($St \le 0.05$) and a high-frequency branch ($St > 0.05$). They argued that the toroidal-pipe peaks around $St\approx0.01$, $0.087$, $0.192$ and $0.298$ are linked to the coupling
	between the in-plane secondary flow oscillations and VLSMs, with the low-frequency peak associated with the temporal variation of VLSMs and the higher peaks associated with the spatial scales of VLSMs. 
	
	Most recently, \citet{Lupi2025} investigated swirl-switching in spatially developing $90^{\circ}$ and $180^{\circ}$ bends using DNS and three-dimensional POD. They identified swirl-switching modes for both bend angles, but with different frequencies and mean flow features. They also noted that the swirl-switching persists for both turbulent and laminar inflow conditions, although it was not the dominant mode for the laminar inflow case. In this sense, their results are consistent with earlier toroidal-pipe studies in showing that swirl-switching does not require an inflow boundary and is intrinsic to curvature. They also suggested that swirl-switching could arise from a symmetry-breaking shear-layer instability in the bend, and that the antisymmetric POD modes may be a remnant of that instability.
	
	Throughout this article, the term \emph{mode} refers to a mathematical object. It pertains to either an eigenvector of the covariance matrix of POD or an eigenvector of the linearised operator in the local stability analysis. The terms \emph{coherent structure} and \emph{instability} are reserved for the physical interpretations of those modes.

The paper is organised as follows. \S 2 provides a detailed review of the key questions in bent pipe flow, and provides a list of requirements for a successful classification of swirl-switching in terms of coherent structures. In particular, the necessity of a new decomposition is shown. In \S 3, we provide our working hypothesis of the origins of swirl-switching, and \S 4 summarises our numerical approach for the data generation. Our novel decomposition, FHPOD, is then mathematically described in \S 5, followed by \S 6 which discussed the local stability analysis to be compared against the FHPOD modes. The results are provided in \S 7 and \S 8; starting with the FHPOD (nonlinear) data in \S 7 to extract coherent structures as distinct modes, and the stability analysis in \S 8 to assess whether the swirl-switching may be an unstable mode of the linearised
Reynolds-averaged Navier–Stokes equations. We conclude the paper in \S 9. Four appendices are devoted to certain details of the simulation, algorithms and analysis of the data.

	\begin{table*}
		\centering
		\footnotesize
		\setlength{\tabcolsep}{2.5pt}%
		\resizebox{\textwidth}{!}{%
			\begin{tabular}{@{}llllp{4.8cm}@{}}%
				\hline
				Reference & Geometry & $Re_D$ & $St=fD/U_b$ & Method \\
				\hline
				
				\cite{Tunstall1968} &
				$90^{\circ}$ mitred &
				$(4$--$23)\times10^{4}$ &
				$2\times10^{-4}$--$4.5\times10^{-3}$ &
				Flow visualisation, wall-mounted flags \\[0.3em]
				
				\cite{Brucker1998} &
				$90^{\circ}$, $\gamma \approx 0.5$ &
				$5\times10^{3}$ &
				0.03, 0.12 &
				TR-DPIV, PSD \\[0.3em]
				
				\cite{Rutten2001} &
				$90^{\circ}$, $\gamma \approx 0.17,\;0.5$ &
				$2.7\times10^{4}$ &
				0.0055, 0.014, 0.2, 0.3 &
				LES, PSD \\[0.3em]
				
				\cite{Rutten2005} &
				$90^{\circ}$, $\gamma \approx 0.17,\;0.5$ &
				$2.7\times10^{4}$ &
				0.01, 0.2, 0.3 &
				LES, PSD \\[0.3em]
				
				\cite{Sakakibara2010} &
				$90^{\circ}$, $\gamma \approx 0.75$ &
				$1.2\times10^{5}$ &
				0.02--0.07 &
				TR-SPIV, POD \\[0.3em]
				
				\cite{Kalpakli2013} &
				$90^{\circ}$, $\gamma \approx 0.31$ &
				$(1.4$--$3.4)\times10^{4}$ &
				0.04, 0.12, 0.18 &
				TR-SPIV, POD \\[0.3em]
				
				\cite{Hellstrom2013} &
				$90^{\circ}$, $\gamma \approx 0.5$ &
				$2.5\times10^{4}$ &
				0.16, 0.33 &
				TR-SPIV, POD \\[0.3em]
				
				\cite{KalpakliVester2015} &
				$90^{\circ}$, $\gamma = 0.39$ &
				$2.3\times10^{4}$ &
				0.04, 0.10 &
				TR-SPIV, POD \\[0.3em]
				
				\cite{Carlsson2015} &
				$90^{\circ}$, $\gamma \approx 0.32,\;0.5,\;0.7,\;1$ &
				$3.4\times10^{4}$ &
				0.003--0.01, 0.13, 0.5--0.6 &
				LES, POD \\[0.3em]
				
				\cite{Noorani2016} &
				Toroidal, $\gamma \approx 0.1,\;0.3$ &
				$1.17\times10^{4}$ &
				0.006--0.03, 0.06, 0.087, 0.18, 0.27--0.28, 0.38 &
				DNS, 3D POD \\[0.3em]
				
				\cite{Hufnagel2018} &
				$90^{\circ}$, $\gamma \approx 0.3$ &
				$1.17\times10^{4}$ &
				0.16, 0.32 &
				DNS, 3D POD \\[0.3em]
				
				\cite{Wang2018} &
				$90^{\circ}$, $\gamma = 0.4$ &
				$5.3\times10^{3}$ &
				0.05--0.3 (stagnation-point / streamwise-velocity signals) &
				DNS, PSD/autocorrelation \\[0.15em]
				
				& & &
				$\sim$0.5, $\sim$1.0 (global wall force / half-side mass flow) &
				\\[0.3em]
				
				\cite{Jain2019} &
				$90^{\circ}$, $\gamma \approx 1/3$ &
				$3.48\times10^{4}$ &
				0.033, 0.066, 0.083, 0.10, 0.13, 0.17, 0.22, 0.33, 0.4--0.9 &
				TR RIM-PIV, POD \\[0.3em]
				
				\cite{He2021} &
				$90^{\circ}$, $\gamma = 0.5$ &
				$(2.7,\,4.5)\times10^{4}$ &
				0.01, 0.25, 0.28, &
				LES, POD, wall-shear-force PSD \\[0.3em]\\[0.3em]
				
				\cite{Jain2022} &
				$180^{\circ}$, $\gamma \approx 1/3$ &
				$3.48\times10^{4}$ &
				0.02--0.05, 0.10, 0.16, 0.25, 0.33, 0.40--0.55 &
				TR RIM-PIV, POD \\[0.3em]
				
				\cite{Zhang2024} &
				Toroidal, $\gamma \approx 0.1$ &
				$1.17\times10^{4}$ &
				0.01, 0.087, 0.192, 0.298;
				0.029, 0.096, 0.212, 0.288 &
				LES, POD and wall-shear PSD \\
				
				\cite{Lupi2025} &
				$90^{\circ}/180^{\circ}$, $\gamma \approx 1/3$ &
				$1.17\times10^{4}$ &
				0.12, 0.24 ($90^{\circ}$ with turbulent\ inflow); &
				DNS, 3D POD  (with turbulent \\[0.15em]
				
				& & &
				0.15, 0.30 ($180^{\circ}$ with turbulent\ inflow); &
				and laminar inflow conditions) \\[0.15em]
				
				& & &
				0.23, 0.30, 0.48 ($180^{\circ}$ with laminar\ inflow) &
				\\[0.3em]
				
				\hline
			\end{tabular}%
		}%
		\caption{Summary of selected studies reporting swirl-switching and other
			large-scale coherent structures in curved and bent pipes.}
		\label{tab:swirl_switching_lit}
	\end{table*}
	
	\section{Motivation of the current work}
	\label{sec:motivation}
	
	Despite the extensive body of literature, several key aspects of swirl-switching remain unresolved. The characteristic Strouhal number of the swirl-switching is far from universal. The reported values span more than two orders of magnitude (see table~\ref{tab:swirl_switching_lit}). Excluding the studies that utilise recycling inflow conditions, part of this scatter can be attributed to differences in geometry, Reynolds number, and the location of the instabilities. An equally important reason for this scatter is the method employed to identify the large-scale coherent structure. As \citet{Wang2018} showed, the identified characteristic frequencies depend on the choice of the measured quantity of interest. Furthermore, dimensionality reduction can also lead to a misinterpretation of the characteristic frequency \citep{Hufnagel2018}. In particular, when only a single observable (e.g.\ streamwise vorticity at a given cross-section) is analysed, coherent structures dominated by other velocity components may be overlooked or misidentified.
	
	On the other hand, \citet{Jain2019} showed that the relative contributions of different frequencies depend strongly on the streamwise measurement location. This implies the coexistence of multiple instability mechanisms that dominate at different streamwise positions. Consequently, the phenomenon reported as ``swirl-switching'' in different studies may not correspond to the same universal mechanism, but instead may point to distinct instabilities with similar footprints occurring at different locations within the flow. A part of the confusion in the literature may therefore be due to the use of the same term for different coherent structures with distinct underlying physical mechanisms. To address this issue, we propose the following requirements for the clean separation and the accurate energy quantification of distinct modes:
	\vspace{0.5\baselineskip}
	
	\begin{enumerate}[label=(\arabic*), leftmargin=0pt, labelsep=0.3em, itemindent=*]
		
		\item\label{req:artifactfree}
		
		The data must be free of spurious periodicity or artificial forcing that could contaminate the intrinsic instabilities.
		
		\item\label{req:3Ddomain}
		
		The decomposition and the analysis must be fully three-dimensional and span the entire domain of interest.
		\item\label{req:degeneracy}
		
		Degenerate POD modes represent the same convecting structures, and they must have identical energy and identical spectra of the time-coefficients.
		
		\item\label{req:unimodal}
		
		The temporal spectrum associated with each mode should be unimodal. Multiple peaks in the spectra indicate spectral mixing and inadequate separation.
		
		\item\label{req:interpretation}
		
		The interpretation of mode shapes must consider both rotational (cross-stream) and axial (streamwise) motions to avoid misattribution of mechanisms
		
	\end{enumerate}
	\vspace{0.5\baselineskip}
	
	The use of synthetic-eddy inflow conditions together with 3D-POD in \citet{Hufnagel2018, Lupi2025} addressed the requirements~\ref{req:artifactfree} and~\ref{req:3Ddomain} and clarified some of these earlier discrepancies. However, classical POD does not guarantee a clean separation of distinct instabilities. Thus, requirements~\ref{req:degeneracy} and~\ref{req:unimodal} remain the primary challenges and are crucial to accurately classify different coherent structures. 
	
	The first objective of this work is to satisfy all five requirements outlined above in our modal decompositions, thereby isolating different instabilities of different origins as distinct modes. In this context, a central goal is to isolate the swirl-switching phenomenon from other coexisting structures and determine whether it contains distinct lower- and higher-frequency members. Their separation enables us to determine the characteristic frequency bands, wavelengths, spatial support, and dominant velocity components of the resulting modes. We then investigate whether swirl-switching is an intrinsic instability of the bent-pipe mean flow. In this sense, the present work is not only aimed at improving modal decompositions of turbulent flows, but also at establishing a clearer connection between the observed coherent structures and the mechanisms responsible for them.
	
	In spatially developing flows, travelling structures naturally appear as degenerate POD pairs with a 90-degree phase shift due to their convective nature. In the limit of an infinite number of snapshots, POD satisfies requirement~\ref{req:degeneracy} for statistically stationary flows. In practice, however, particularly in turbulent flows where coherent structures are embedded within a broad range of scales, finite sampling often leads to imperfect pairing of degenerate modes. Consequently, a single convecting structure often appears distributed across multiple modes with unequal energy contents.
	
	Moreover, the classical POD does not satisfy requirement~\ref{req:unimodal} by construction. In this context, the unimodality of a mode refers to a single dominant frequency band in its time coefficient rather than a perfectly harmonic response. In turbulent flows, nonlinear interactions and amplitude modulation can generate spectrally broadened peaks. As a result of violations of requirements~\ref{req:degeneracy} and~\ref{req:unimodal}, the energy-based ranking of distinct instabilities becomes an ill-posed problem. In other words, the energy of a single physical instability may be distributed across several POD modes. This is partially the reason why many studies rely on reduced-order reconstructions of the flow field using multiple modes to recover an instability of interest \citep{Kalpakli2013,Noorani2016}. For instance, \citet{Noorani2016} noted that no single ``swirl mode'' appears among the most energetic POD modes, but the flow reconstruction using multiple modes reproduces a swirl-switching structure. This indicates that spectral mixing distributed the relevant swirl-switching structure across multiple modes in their analysis. This issue is particularly severe in fully turbulent flows, where the modal energy levels are closely spaced. Therefore, truncation to a limited set of leading modes may discard a portion of a physically coherent instability that is distributed across multiple POD modes with lower energy contents. Moreover, the energy of the modes depends strongly on the spatial extent of the computational domain and the duration of the sampled time series. Consequently, a mode whose time coefficient contains several distinct frequency bands may combine contributions from different instabilities, and its energy is ill-defined from a physical point of view. While other decomposition methods such as dynamic mode decomposition (DMD) and spectral proper orthogonal decomposition (SPOD) yield single-frequency modes, such strict monochromaticity is often too restrictive for turbulent flows, given the multiscale and nonlinear nature of turbulence. Here, we instead seek a finite-band representation that captures each instability as a single coherent mode.
	
	To enforce requirement~\ref{req:degeneracy}, we perform POD on the analytic signal of the data, following the Hilbert Proper Orthogonal Decomposition (HPOD) framework introduced by \citet{Kriegseis2021_ISPIV_HPOD} and subsequently applied in \citet{Raiola2024_AIAA_HPODjet,Raiola2025_arXiv_HPOD}. To address requirement~\ref{req:unimodal}, we extract the dominant HPOD modes and analyse their spectral content. A Butterworth filter with sufficiently broad spectral bandwidth is then designed and applied to the snapshots to suppress secondary frequency bands contaminating the dominant mode. The purpose of this step is not to enforce an artificially narrow spectral band but to reduce spectral mixing while preserving the physically relevant frequency content.  As a result, the novel \emph{filtered HPOD (FHPOD) method} can mitigate spectral mixing and limit contamination from structures of different physical origins that are correlated with the mode of interest. The idea of the FHPOD is predicated on the fact that the standard POD and HPOD modes are energy optimal. As a result, to maximise the energy, they often mix different instabilities of different characteristic frequencies into one physically ambiguous mode. In contrast, the \emph{temporal} filtering step of the FHPOD reduces the energy of spectrally adjacent structures in the POD covariance matrix. Due to filtering, most of the energy is concentrated in a finite band. Therefore, this allows the singular value decomposition (SVD) step of FHPOD to isolate a physically meaningful finite-band mode. 
	
	Lastly, requirement~\ref{req:interpretation} is crucial to correctly identify the underlying physical mechanisms. As shown in Section~\ref{sec:results}, some modes carry most of their energy in the streamwise velocity component, whereas others are dominated by cross-stream rotational motion. Thus, we visualise the former using isosurfaces of the streamwise velocity and the latter using isosurfaces of the streamfunction.
	
	The present study also attempts to identify the low-frequency instability at $St \approx 0.03-0.04$ as a distinct mode. Although this low-frequency mode has been observed in multiple experimental studies \citep{Brucker1998, Sakakibara2010, Kalpakli2013, KalpakliVester2015, Jain2019, Jain2022}, the associated instability has not previously been extracted as a separate mode in numerical simulations. \citet{Hufnagel2018} reported that a similar low-frequency peak at $St \approx 0.045$ for $\gamma=0.1$ was present in the PSD of the time-coefficients, but it was not sufficiently resolved to yield a distinct mode. They noted that resolving the large wavelength of this mode would require an extended outlet section that was computationally prohibitive at the time. In the following, we demonstrate that, by extending the straight section downstream of the bend, this low-frequency mode can be adequately isolated and characterised.
	
	Finally, the key objective of this work is to shed light on the physical origin of the swirl-switching phenomenon, as it still remains the subject of ongoing debate. One line of investigation attributes this behaviour to upstream turbulence, suggesting that streaks or VLSMs entering the bend cause or influence the unsteady motion of the Dean vortices \citep{SakakibaraMachida2012,KalpakliVester2015,Carlsson2015}. Similarly, \citet{Zhang2024} linked the swirl-switching phenomenon to VLSM-like streamwise fluctuations generated within the toroidal pipe, arguing that low-frequency switching is associated with the temporal variation of the VLSMs, whereas higher-frequency switching is related to its streamwise spatial scale. In contrast, another body of work interprets swirl-switching as an intrinsic feature of the bent pipe flow itself rather than as a response imposed solely by upstream forcing. Within this perspective, the phenomenon has been associated with bimodal vortex dynamics, inner-wall or shear-layer interactions, and the emergence of a three-dimensional wave-like structure generated within the curved section \citep{Brucker1998, Hellstrom2013, Noorani2016, Hufnagel2018}. Swirl-switching has also been interpreted as a symmetry-breaking process or global instability of the bend shear layer associated with the transition dynamics of the flow, rather than as a response driven solely by upstream forcing \citep{Hufnagel2018, Lupi2025}.
	
	\section{Hypothesis for the origin of swirl-switching}
	\label{sec:swirl_switching_hypothesis}
	
\begin{figure}[!htbp]
		\centering
		\begin{minipage}[b]{0.58\linewidth}
			\centering
			\includegraphics[width=\linewidth]{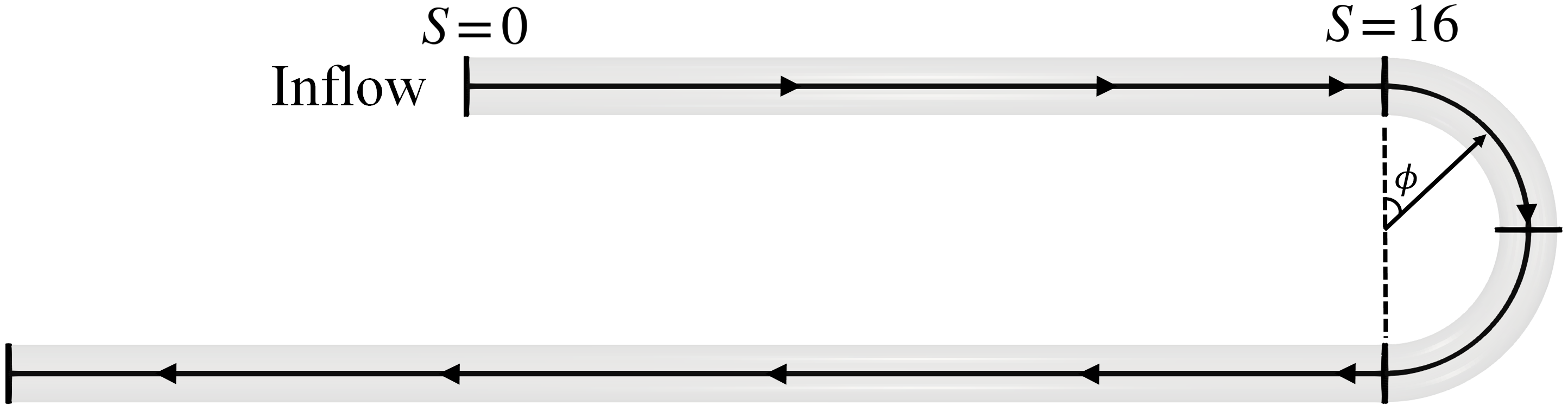}\par
			\makebox[\linewidth][l]{\small\bfseries (a)}
		\end{minipage}\hfill
		\begin{minipage}[b]{0.4\linewidth}
			\centering
			\includegraphics[width=\linewidth]{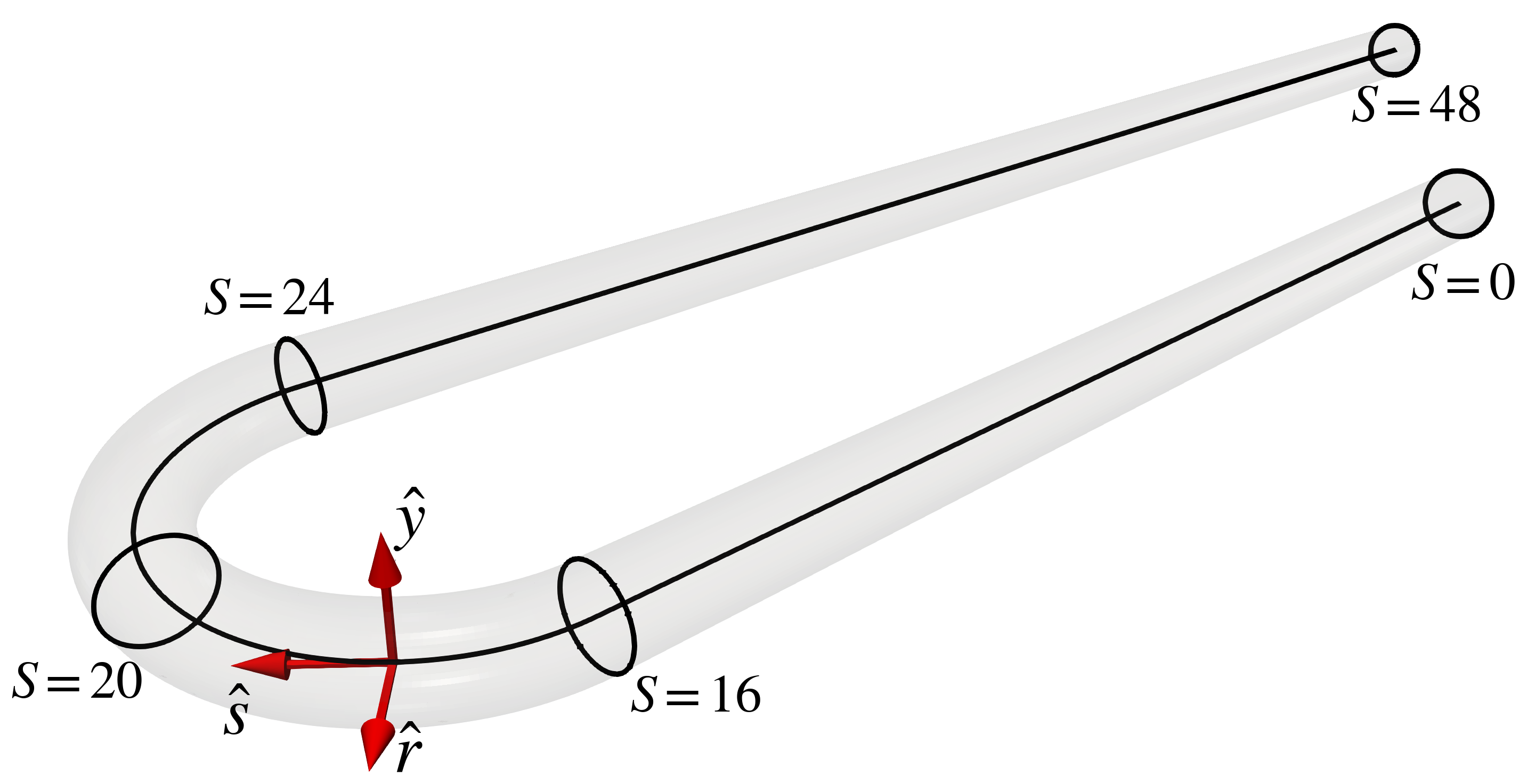}\par
			\makebox[\linewidth][l]{\small\bfseries (b)}
		\end{minipage}
		\caption{\textit{(a)} Geometry of the $180^{\circ}$ bent pipe common to all cases, where $\phi$ denotes the bend angle and the arrows indicate the flow direction. \textit{(b)} Frenet--Serret frame for the $\gamma=0.2$ case, with $\hat{\mathbf{s}}$ tangent to the centreline, $\hat{\mathbf{r}}$ in the radial direction and $\hat{\mathbf{y}}$ in the binormal direction.}
		\label{fig:coordSys}
	\end{figure}

	\citet{bagheri2026} derived the mean vorticity transport equation in the Frenet--Serret coordinate system for toroidal flows. The geometry and coordinate system are shown in figure~\ref{fig:coordSys}. The Frenet--Serret reference frame is spanned by unit vectors $\hat{\mathbf{s}}$ (tangent/streamwise), $\hat{\mathbf{r}}$ (radial/centrifugal), and $\hat{\mathbf{y}}$ (binormal/lateral). The streamwise coordinate $s$ follows the pipe centreline along $\hat{\mathbf{s}}$. Its non-dimensional form is given by $S=s/D$, where $D$ is the pipe diameter. The signed cross-sectional coordinates $r$ and $y$ are measured from the pipe centreline along $\hat{\mathbf{r}}$ and $\hat{\mathbf{y}}$, respectively. They showed that the strength of the Dean vortices is related to the cross-sectional profile of the pipe governed by the centrifugal production term (see \eqref{eq:meanWS}). 
	\FloatBarrier
	
	\begin{align}
		\frac{\mathrm D\overline{\Omega}_s}{\mathrm D t}
		&=
		\underbrace{\partial_y\left(\frac{\kappa}{h_s}\overline{U}_s^2\right)}_{\text{centrifugal production}}
		+\underbrace{\frac{\kappa}{h_s}\overline{U}_r\,\overline{\Omega}_s}_{\text{radial redistribution}}
		+\nu\,(\nabla^2\overline{\boldsymbol{\Omega}})_s
		-\bigl[\nabla\times(\nabla\cdot \Rmat)\bigr]_s 
		\label{eq:meanWS}
	\end{align}
	Here, $\mathrm D/\mathrm D t:=\partial_t+\overline{\mathbf U}\cdot\nabla$ denotes the material derivative following the mean flow, $\kappa=1/R_c$ is the dimensional centreline curvature, $h_s=1+\kappa r$ is the scale factor, and $\Rmat$ denotes the Reynolds-stress tensor. $\overline{\mathbf{U}}$ and $\overline{\boldsymbol{\Omega}}$ represent the mean velocity and mean-vorticity fields, respectively. We use this equation to investigate the effect of streamwise velocity perturbations on the secondary flow. For this purpose, we derive a perturbation equation for the vorticity transport. To describe a short-lived coherent perturbation embedded in turbulence, we employ a triple decomposition as follows:
	\begin{equation}
		\label{eq:triple_decomp}
		\mathbf{u}=\overline{\mathbf{U}}+\widetilde{\mathbf{u}}+\mathbf{u}'',
		\qquad 
		\boldsymbol{\Omega}=\overline{\boldsymbol{\Omega}}+\widetilde{\boldsymbol{\Omega}}+\boldsymbol{\Omega}''.
	\end{equation}
	The total velocity fluctuation and the Reynolds-stress tensor are therefore defined by
	\begin{equation*}
		\mathbf{u}'
		:=\mathbf{u}-\overline{\mathbf{U}}
		=\widetilde{\mathbf{u}}+\mathbf{u}'',
		\qquad
		\Rmat:=\overline{\mathbf{u}'\otimes\mathbf{u}'}.
	\end{equation*}
	Here, $\overline{{(\cdot)}}$ denotes the long-time Reynolds mean, $(\cdot)'$ the total fluctuation about that mean, $\widetilde{(\cdot)}$ the coherent phase-averaged perturbation, and $(\cdot)''$ the remaining incoherent turbulent fluctuation, with $\overline{\widetilde{\mathbf{u}}}=\overline{\mathbf{u}''}=\mathbf{0}$. Assuming the perturbation is sufficiently short-lived that it does not modify the Reynolds-stress tensor $\Rmat$, the perturbation satisfies the following linearised equation for the streamwise $s$-component
	
	\begin{equation}
		\label{eq:vort_tilde_linear_s}
		\left(\partial_t+\overline{\mathbf{U}}\cdot\nabla\right)\widetilde{\Omega}_s
		=
		\big[(\widetilde{\boldsymbol{\Omega}}\cdot\nabla)\overline{\mathbf{U}}\big]_s
		+\big[(\overline{\boldsymbol{\Omega}}\cdot\nabla)\widetilde{\mathbf{u}}\big]_s
		-\big[(\widetilde{\mathbf{u}}\cdot\nabla)\overline{\boldsymbol{\Omega}}\big]_s
		+\nu(\nabla^2\widetilde{\boldsymbol{\Omega}})_s .
	\end{equation}

	\noindent Considering a streamwise-only initial perturbation, a fully developed mean flow, and keeping the leading order terms,  yields
	
	\begin{equation}
		\label{eq:omega_s_tilde_streamwise_only_3D}
		\frac{\mathrm D\widetilde{\Omega}_s}{\mathrm D t}
		=
		\frac{\overline{\Omega}_s}{h_s}\,\partial_s \widetilde{u}_s
		+\partial_y\!\left(\frac{2\kappa}{h_s}\,\overline{U}_s\,\widetilde{u}_s\right)
		+\nu\left(\nabla^2\widetilde{\boldsymbol{\Omega}}\right)_s.
	\end{equation}
	\noindent Note that although $\widetilde{\Omega}_s$ is kinematically zero when velocity perturbations are only streamwise ($\widetilde{u}_y=\widetilde{u}_r=0$), it can be generated dynamically by the source terms on the right-hand side. Finally, we restrict the flow to a 2D cross-section model by neglecting the streamwise variation of the mean flow and the perturbation,
	\begin{equation}
		\label{eq:2D_assumption}
		\partial_s(\cdot)=0 .
	\end{equation}
	Under this assumption, equation~\eqref{eq:omega_s_tilde_streamwise_only_3D} reduces to
	\begin{equation}
		\label{eq:omega_s_tilde_2D}
		\frac{\mathrm D\widetilde{\Omega}_s}{\mathrm D t}
		=
		\partial_y\!\left(\frac{2\kappa}{h_s}\,\overline{U}_s\,\widetilde{u}_s\right)
		+\nu\left(\nabla^2\widetilde{\boldsymbol{\Omega}}\right)_s .
	\end{equation}

	\noindent Equation~\eqref{eq:omega_s_tilde_2D} is a simplified linearised form of the vorticity perturbation equation in which coherent streamwise velocity perturbations appear as a source term on the right-hand side of the equation. Therefore, large-scale and short-lived structures with significant streamwise momentum, such as streaks or VLSMs, can indeed linearly modulate the secondary flow. This is consistent with the observations of \citet{SakakibaraMachida2012}, who stated that upstream low and high-speed streaks are correlated with the downstream displacement of the stagnation point and the switching of the symmetry plane of the Dean vortices. Although equation~\eqref{eq:omega_s_tilde_2D} shows a linear coupling mechanism through which streamwise coherent structures can force the secondary flow modulations, it does not by itself establish whether such structures are the primary cause of swirl-switching or instead excite an existing instability intrinsic to bent pipe flows. In other words, the linear operator in \eqref{eq:omega_s_tilde_2D} admits the possibility that streamwise velocity perturbations can trigger secondary flow modulations. However, determining whether swirl-switching instead originates from an intrinsic instability depends on the spectral properties of the corresponding operator, specifically whether or not it admits unstable or weakly damped eigenmodes of the bent-pipe base flow associated with swirl-switching. Despite variations of Reynolds number, bend angle, curvature, and upstream conditions, the reported peak at $St=0.10$--$0.13$ and qualitative characteristics of this phenomenon remain fairly consistent in both experimental and numerical studies. This suggests that swirl-switching is unlikely to arise solely as a convective response to inflow structures. It instead points to the presence of an intrinsically unstable or weakly damped eigenmode of the bent-pipe mean flow that may be amplified by external streamwise perturbations. To examine this hypothesis, we perform a local linear stability analysis of the mean flow equations in \S\ref{sec:meth_LSA}.

	\section{Direct numerical simulation}
	\label{sec:dns}
	
	We consider three DNS datasets in the $180^{\circ}$ bent-pipe geometry shown in figure~\ref{fig:coordSys}. The primary case, at $Re_D=10,000$ ($Re_\tau=316$) and $\gamma=0.2$, is the main focus of the study and is analysed using both FHPOD and LSA. Here, the wall-friction Reynolds number is defined as
	
	\begin{equation}
		Re_\tau= \frac{u_{\tau} R_p}{\nu},
	\end{equation}
	where $R_p$ is the pipe radius and $u_\tau$ denotes the wall-friction velocity. Two supplementary cases at $Re_D=5300$ ($Re_\tau=181$), with $\gamma=0.2$ and $\gamma=0.4$, are analysed using FHPOD to assess the effects of Reynolds number and curvature on the identified FHPOD modes. For all three cases, the straight section downstream of the bend is extended to be $24D$ long in order to allow the development of the low-frequency instability expected at $St\approx0.03$. The DNS was carried out with $\mathrm{Nek5000}$ \citep{Fischer2008_Nek5000}, which integrates in time the Navier--Stokes equations discretised using the spectral-element method (SEM), with the $\mathrm{P}_N$--$\mathrm{P}_{N-2}$ formulation and seventh-order polynomials. For time integration, we utilised a third-order semi-implicit backward-differentiation scheme ($\mathrm{BDF3}$). Table~\ref{tab:dns_resolution} reports the grid resolution of the three DNS simulations. Here, $\rho$ is the polar radius, $\theta$ is the azimuthal angle and the physical azimuthal spacing is $\rho\,\Delta\theta$. The inflow plane is located $16D$ upstream of the bend at $S=0$. The inflow velocity is prescribed using a divergence-free synthetic eddy method (DFSEM) based on \citet{Poletto2013}. The same inflow condition was used in \citet{Hufnagel2018, Lupi2025}. To remove the transient effects, the first two flow-through times were discarded from the snapshots and the statistics. The statistics and snapshots were obtained over the next $805.95 D/U_b$ convective time units, corresponding to approximately 102 eddy-turnover times based on $R_p/u_{\tau}$ and 24.2 periods of the slowest mode at $St=0.03$.
	More details about this inflow condition and the convergence of the statistics in the DNS simulations are provided in Appendix~\ref{app:inflow_condition}.
	\begingroup
	\makeatletter\def\fps@table{htp}\makeatother
	\begin{table}
		\centering
		\caption{Resolution of the three DNS simulations in wall-units.}
		\label{tab:dns_resolution}
		\begin{tabular*}{\textwidth}{@{\extracolsep{\fill}} c c c c c c @{} }
			$Re_D$ & $\gamma$ & $Re_\tau$ & $\Delta s^+$ & $(\rho\,\Delta\theta)^+$ & $\Delta r^+$ \\
			\hline
			$10,000$ & $0.2$ & $316$ & $2.29$--$11.08$ & $1.01$--$6.11$ & $0.41$--$4.74$ \\
			$5300$ & $0.2$ & $181$ & $2.27$--$12.25$ & $0.98$--$6.06$ & $0.42$--$3.61$ \\
			$5300$ & $0.4$ & $181$ & $0.84$--$12.25$ & $0.83$--$7.72$ & $0.37$--$4.81$ \\
		\end{tabular*}
	\end{table}
	\endgroup
	\section{Filtered Hilbert Proper Orthogonal Decomposition (FHPOD)}
	\label{sec:pod_method}
	
	The FHPOD method is designed to address two limitations of classical POD. The first is the imperfect pairing of degenerate modes associated with a convecting structure. The second is the mixing of multiple frequency bands within a single mode due to energy optimality. For this analysis, we use a sequence of $N_{\mathrm{snap}}=5374$ three-dimensional DNS velocity snapshots $\boldsymbol u(\boldsymbol x,t_n)$ sampled at a constant temporal interval of $\Delta t = 0.15\,D/U_b$. The snapshot sequence therefore has a temporal extent of $805.95 D/U_b$. Modal convergence depends on the sampling duration as well as the number of independent realisations \citep{SchmidtColonius2020,Hekmati2011}. In a DMD sampling study of turbulent flow, \citet{Li2022DMDsampling} observed noticeable modal-frequency variation when the sampling duration covered fewer than ten cycles, while the results stabilised over approximately 15--20 cycles. Similarly, \citet{LiLasagna2026} reported that using approximately 15 oscillation cycles per SPOD block provided a balance between frequency resolution and statistical convergence. In the present study, the sampling duration covers approximately $24.2$ periods of the low-frequency mode at $St=0.03$ and therefore is above the suggested values in the literature ensuring that this instability is sufficiently resolved in the dataset. The sampling interval results in a Nyquist limit $St_{\mathrm{Ny}}=D/(2U_b\Delta t)=3.33$, well above the highest frequency of interest near $St\approx0.5$.
	
	For the FHPOD decomposition, we remove the temporal mean from every snapshot,
	\begin{equation}
		\boldsymbol u'(\boldsymbol x,t_n)
		:=\boldsymbol u(\boldsymbol x,t_n)-\overline{\boldsymbol U}(\boldsymbol x),
		\qquad
		\overline{\boldsymbol U}(\boldsymbol x)
		:=\frac{1}{N_{\mathrm{snap}}}
		\sum_{n=1}^{N_{\mathrm{snap}}}\boldsymbol u(\boldsymbol x,t_n).
		\label{eq:fhpod_mean_subtraction}
	\end{equation}
	All snapshot matrices introduced below are assembled from the fluctuating-velocity snapshots $\boldsymbol u'(\boldsymbol x,t_n)$. Let $\boldsymbol q_n\in\mathbb{R}^{d}$ denote the spectral-element degrees of freedom for the three components of $\boldsymbol u'$, concatenated as
	\begin{equation}
		\boldsymbol q_n=
		\begin{pmatrix}
			\boldsymbol u'_{s,n} \\ \boldsymbol u'_{r,n} \\ \boldsymbol u'_{y,n}
		\end{pmatrix}.
	\end{equation}
	To enforce the symmetry about the equatorial plane of the bent pipe ($y=0$) in the snapshots, each mean-subtracted snapshot is mirrored (reflected) and added to the matrix of snapshots as follows
	\begin{equation}
		\boldsymbol Q
		=
		\big[
		\boldsymbol q_1^{(o)},\dots,\boldsymbol q_{N_{\mathrm{snap}}}^{(o)},
		\boldsymbol q_1^{(m)},\dots,\boldsymbol q_{N_{\mathrm{snap}}}^{(m)}
		\big]
		\in \mathbb{R}^{d\times N_{\mathrm{aug}}}, \qquad N_{\mathrm{aug}}=2N_{\mathrm{snap}}
	\end{equation}
	Here $d=3n_d$, where $n_d$ is the number of scalar velocity degrees of freedom and $\boldsymbol u'_{s,n}$, $\boldsymbol u'_{r,n}$ and $\boldsymbol u'_{y,n}$ are the streamwise, radial and binormal fluctuation velocity components, respectively. $\boldsymbol q_n^{(o)}$ is the original snapshot and $\boldsymbol q_n^{(m)}$ is its reflection about the plane $y=0$. The reflection is performed with respect to the equatorial plane $y=0$ and the reflection operator $\mathcal{R}_y$ is defined as
	\begin{equation}
		\mathcal{R}_y:
		\begin{cases}
			(s,r,y) \mapsto (s,r,-y),\\[2mm]
			(u_s,u_r,u_y)(s,r,y) \mapsto
			(u_s,u_r,-u_y)(s,r,-y).
		\end{cases}
		\label{eq:mirror_operator}
	\end{equation}
	In the next step, we apply a \emph{temporal} band-pass filter to each row of $\boldsymbol Q$. The filtering is performed on the original and the mirrored snapshot sequences separately in order to avoid cross-contamination at the concatenation point. We use a second-order Butterworth section at each active band edge with a forward--backward strategy to achieve zero-phase filtering. The amplitude response produced by the forward--backward application is
	\begin{equation}
		G_{\mathrm{FB}}(f)=
		\left[
		1+\left(\frac{K(f_{\min})}{K(f)}\right)^4
		\right]^{-1}
		\left[
		1+\left(\frac{K(f)}{K(f_{\max})}\right)^4
		\right]^{-1}, \qquad  K(f)=\tan(\pi f/f_{\mathrm{samp}})
		\label{eq:fhpod_filter}
	\end{equation}
	in which $f_{\mathrm{samp}}=1/\Delta t$ is the sampling frequency. This filter is designed to have a soft roll-off so that it attenuates neighbouring spectral content rather than removing it abruptly. Classical POD yields energy-optimal basis vectors, with the consequence that a POD mode often mixes several energetic structures of different frequencies and different origins. By concentrating most of the retained energy within the finite bandwidth of interest, the filter suppresses the tendency of POD to mix multiple energetic structures from other frequency bands into the leading modes. This step results in a filtered snapshot matrix
	\begin{equation}
		\boldsymbol Q_f \in \mathbb{R}^{d\times N_{\mathrm{aug}}}.
	\end{equation}
	The filter bands are selected based on an initial inspection of the POD time-coefficient spectra, with the aim of isolating distinct finite-band structures. The band is placed around the corresponding spectral peak so that it lies in the high-response region. The edges are chosen on the two sides of the peak where the spectral amplitude begins to decay towards neighbouring spectral content. However, the gradual roll-off of the filter makes the resulting modes only weakly sensitive to modest changes in the precise locations of the filter edges. As a sensitivity check, the filter bandwidths selected in this work (see table~\ref{tab:HPOD_modeFamily}) were increased by $20\%$, and the resulting mode shapes remained nearly unchanged. The projected modal energy and peak Strouhal numbers changed by less than $2.47\%$ and $6.41\%$, respectively.

	To pair the degenerate convecting structures into a single complex mode, we first form the analytic signal associated with the filtered snapshots. For a real-valued signal $x(t)$, the analytic signal is defined as
	\begin{equation}
		z(t)=x(t)+ i\,\mathcal{H}_t\{x(t)\},
	\end{equation}
	where $\mathcal{H}_t\{\cdot\}$ is the Hilbert transform in time. Applying the Hilbert transform to $\boldsymbol Q_f$ with the original and mirrored halves treated separately, yields the complex analytic snapshot matrix
	\begin{equation}
		\boldsymbol Q_f^{(\mathcal{H})} \in \mathbb{C}^{d\times N_{\mathrm{aug}}}.
	\end{equation}
	The POD is defined with respect to the inner product.
	\begin{equation}
		\langle \boldsymbol a,\boldsymbol b\rangle_M
		=
		\boldsymbol a^{\dagger}\boldsymbol M \boldsymbol b,
		\label{eq:innerProjct}
	\end{equation}
	in which $\boldsymbol M$ is the diagonal mass matrix induced by Gauss--Lobatto--Legendre quadrature and $(\cdot)^\dagger$ denotes the Hermitian transpose. By introducing the mass-weighted analytic snapshot matrix
	\begin{equation}
		\boldsymbol A_f = \boldsymbol M^{1/2}\boldsymbol Q_f^{(\mathcal{H})}
		\in \mathbb{C}^{d\times N_{\mathrm{aug}}},
		\label{eq:A_def}
	\end{equation}
	the complex singular value decomposition (SVD) problem
	\begin{equation}
		\boldsymbol A_f
		= \boldsymbol W\boldsymbol Z_{\mathrm{POD}}\boldsymbol V^{\dagger},
		\qquad
		\boldsymbol Z_{\mathrm{POD}}
		=\operatorname{diag}(\zeta_1,\zeta_2,\ldots),
		\label{eq:svd}
	\end{equation}
	is solved using a Lanczos-based method in distributed memory \citep{Perez2025}. Here, $\boldsymbol w_j$ and $\boldsymbol v_j$ denote the $j$th columns of $\boldsymbol W$ and $\boldsymbol V$, respectively, and $\zeta_j$ is the corresponding singular value. The FHPOD temporal coefficient is the right singular vector $\boldsymbol v_j$, and the spectra shown in the following section are obtained from its real part. Finally, the corresponding FHPOD mode in physical units can be recovered from the left singular vector $\boldsymbol w_j$ as
	\begin{equation}
		\boldsymbol\xi_j
		= \boldsymbol M^{-1/2}\boldsymbol w_j,
		\qquad
		\boldsymbol\xi_j
		=
		\begin{pmatrix}
			u^{\xi}_{j,s}\\
			u^{\xi}_{j,r}\\
			u^{\xi}_{j,y}
		\end{pmatrix}, 
		\label{eq:unweight}
	\end{equation}
	in which $u^{\xi}_{j,s}$, $u^{\xi}_{j,r}$ and $u^{\xi}_{j,y}$ are the axial, radial and binormal components of the \(j\)th FHPOD mode, respectively. 

	\subsection{Quantifying the energy}
	
	The squared singular value $\zeta_j^2$ represents the energy of the \(j\)th mode within the filtered analytic dataset $\boldsymbol Q_f^{(\mathcal H)}$. To compute the energy of each mode with respect to the unfiltered, mean-subtracted snapshots $\boldsymbol Q$, we first introduce the real, mass-weighted and unfiltered fluctuation-snapshot matrix
	\begin{equation}
		\boldsymbol A_0 := \boldsymbol M^{1/2}\boldsymbol Q \in \mathbb{R}^{d\times N_{\mathrm{aug}}}.
	\end{equation}
	The unfiltered analytic snapshot matrix is defined by
	\begin{equation*}
		\boldsymbol Q^{(\mathcal H)}
		:=\boldsymbol Q+i\,\mathcal H_t\{\boldsymbol Q\}
		\in\mathbb C^{d\times N_{\mathrm{aug}}},
	\end{equation*}
	where $\mathcal H_t$ is applied separately to the original and mirrored halves row-wise in time. The corresponding mass-weighted analytic snapshot matrix is then
	\begin{equation}
		\boldsymbol A
		:=
		\boldsymbol A_0 + i\,\mathcal{H}_t\{\boldsymbol A_0\}
		=
		\boldsymbol M^{1/2}\boldsymbol Q^{(\mathcal{H})}
		\in \mathbb{C}^{d\times N_{\mathrm{aug}}},
		\label{eq:A_unfilt}
	\end{equation}
	Since the mass-weighted left singular vector $\boldsymbol w_j$ corresponding to the physical FHPOD mode $\boldsymbol\xi_j$ was extracted from the filtered matrix $\boldsymbol A_f$ in \eqref{eq:A_def}, it is not in general a left singular vector of the unfiltered matrix $\boldsymbol A$. Therefore, we decompose $\boldsymbol A$ as
	\begin{equation}
		\boldsymbol A = \boldsymbol A_f + \boldsymbol A_{\mathrm{out}},
		\label{eq:decomposition}
	\end{equation}
	where $\boldsymbol A_{\mathrm{out}} = \boldsymbol A - \boldsymbol A_f$ is the remainder matrix containing the residual contributions attenuated by the filter. For the mass-weighted left singular vector $\boldsymbol w_j$, we define the projection coefficients onto the unfiltered analytic snapshots
	\begin{equation}
		\mathcal P_j
		:=
		\boldsymbol w_j^\dagger \boldsymbol A .
		\label{eq:projection}
	\end{equation}
	Substituting the decomposition~\eqref{eq:decomposition} into \eqref{eq:projection} yields
	\begin{equation}
		\mathcal P_j
		=
		\boldsymbol w_j^\dagger \boldsymbol A_f
		+
		\boldsymbol w_j^\dagger \boldsymbol A_{\mathrm{out}}
		=
		\mathcal P_j^{\mathrm{in}}
		+
		\mathcal P_j^{\mathrm{out}} .
		\label{eq:p_decomp}
	\end{equation}
	The first term is the in-band projection,
	\begin{equation}
		\mathcal P_j^{\mathrm{in}}
		=
		\boldsymbol w_j^\dagger \boldsymbol A_f
		=
		\zeta_j \boldsymbol v_j^\dagger ,
	\end{equation}
	where $\zeta_j$ and $\boldsymbol v_j$ are the singular value and right singular vector of the filtered matrix $\boldsymbol A_f$ already obtained by solving the SVD problem \eqref{eq:svd}. The second term is the residual projection,
	\begin{equation}
		\mathcal P_j^{\mathrm{out}}
		=
		\boldsymbol w_j^\dagger \boldsymbol A_{\mathrm{out}} .
	\end{equation}
	The projected energy $E_j$ is then defined by the squared $\ell^2$-norm of the projection coefficient normalised by the number of snapshots,
	\begin{equation}
		E_j
		:=
		\frac{1}{N_{\mathrm{aug}}}\|\mathcal P_j\|_2^2
		=
		\frac{1}{N_{\mathrm{aug}}}
		\mathcal P_j \mathcal P_j^\dagger.
		\label{eq:projected_energy}
	\end{equation}
	Thus, $E_j$ represents the projected energy of the FHPOD spatial mode $\boldsymbol\xi_j$. In the following, the projected energy is expressed as a percentage of the mean-flow energy $E_0$. Let $\overline{\boldsymbol q}$ denote the concatenated mean velocity. The corresponding mean-flow reference energy is
	\begin{equation}
		E_0
		:=\|\overline{\boldsymbol q}\|_M^2
		=\overline{\boldsymbol q}^{\dagger}\boldsymbol M\overline{\boldsymbol q}.
		\label{eq:mean_flow_energy}
	\end{equation}
	It is worth emphasising that the projected energy $E_j$  differs from the modal energy represented by the squared singular values in classical POD. Because FHPOD modes are extracted from different filtered matrices, they are generally not mutually orthogonal and therefore their projected energies are not generally additive.
	In practice, $\mathcal P_j$ is computed without explicitly forming $\boldsymbol Q^{(\mathcal{H})}$ or $\boldsymbol A$. Substituting \eqref{eq:A_unfilt} in \eqref{eq:projection} yields
	\begin{equation}
		\mathcal P_j
		=
		\boldsymbol w_j^\dagger \boldsymbol A
		=
		\boldsymbol w_j^\dagger \boldsymbol A_0
		+
		i\,\boldsymbol w_j^\dagger
		\mathcal{H}_t\{\boldsymbol A_0\}.
		\label{eq:projection_linear}
	\end{equation}
	We first project the real, mean-subtracted and unfiltered snapshots onto the complex mass-weighted left singular vector $\boldsymbol w_j$
	\begin{equation}
		\mathcal P_j^{(0)}
		:=
		\boldsymbol w_j^\dagger \boldsymbol A_0,
		\label{eq:real_val_proj}
	\end{equation}
	which is evaluated in a streaming manner without the need to form the full matrix $\boldsymbol A_0$. As the Hilbert transform is a linear operation performed only in time, and the projection is a spatial linear operation, the two operations commute:
	\begin{equation}
		\boldsymbol w_j^\dagger
		\mathcal{H}_t\{\boldsymbol A_0\}
		=
		\mathcal{H}_t
		\left\{
		\boldsymbol w_j^\dagger\boldsymbol A_0
		\right\}
		=
		\mathcal{H}_t\{\mathcal P_j^{(0)}\}.
		\label{eq:commutative_Hilb_proj}
	\end{equation}
	Substituting \eqref{eq:real_val_proj} and \eqref{eq:commutative_Hilb_proj} into \eqref{eq:projection_linear} yields
	\begin{equation}
		\mathcal P_j
		=
		\mathcal P_j^{(0)}
		+
		i\,\mathcal{H}_t\{\mathcal P_j^{(0)}\},
	\end{equation}
	obtained by applying a one-dimensional Hilbert transform to the projection sequence $\mathcal P_j^{(0)}$ formed from the real snapshot matrix $\boldsymbol A_0$ for the \(j\)th mode.

	\subsection{Cross-sectional streamfunction}
	\label{subsec:streamfunction}
	
	Let $\mathcal{A}(s)$ denote the pipe cross-section at streamwise coordinate $s$ with the plane-normal in $\hat{\boldsymbol s}$ direction. We define the instantaneous streamwise vorticity at time $t$
	\begin{equation}
		\Omega_s(r,y,s,t) := \bigl(\nabla\times\boldsymbol u(r,y,s,t)\bigr)\cdot \hat{\boldsymbol s}(s).
	\end{equation}
	The instantaneous cross-sectional streamfunction $\Psi(r,y,s,t)$ is then obtained by solving the Poisson problem
	\begin{equation}
		\nabla_{\perp}^{2}\Psi(r,y,s,t) = -\Omega_s(r,y,s,t)
	\end{equation}
	with Dirichlet boundary condition at the pipe wall,
	\begin{equation}
		\Psi(r,y,s,t)=0.
	\end{equation}
	Here, $\nabla_{\perp}^{2}$ denotes the in-plane Laplacian operator in the $(r,y)$ plane. In a spatially developing three-dimensional flow, the instantaneous in-plane velocity on $\mathcal{A}(s)$ is not necessarily two-dimensionally divergence-free. Using the Helmholtz decomposition on each cross-section, we obtain
	\begin{equation}
		\boldsymbol u_\perp(r,y,s,t)
		=
		\boldsymbol u_\perp^{\mathrm{sol}}(r,y,s,t) + \nabla_{(r,y)} \Upsilon(r,y,s,t),
		\qquad
		\nabla_{(r,y)}\!\cdot \boldsymbol u_\perp^{\mathrm{sol}}(r,y,s,t)=0,
	\end{equation}
	where $\boldsymbol u_\perp^{\mathrm{sol}}$ is the instantaneous in-plane solenoidal component and $\nabla_{(r,y)}\Upsilon$ contains the non-solenoidal part. Similarly, the streamfunction $\Psi_j$ can be defined for a POD, HPOD or FHPOD mode $\boldsymbol\xi_j$:
	\begin{equation}
		\nabla_{\perp}^{2}\Psi_j
		=-\bigl(\nabla\times\boldsymbol\xi_j\bigr)\cdot\hat{\boldsymbol s},
		\qquad
		\Psi_j\big|_{\partial\mathcal A}=0.
	\end{equation}
	Because the cross-sectional operator and its homogeneous boundary condition are time invariant, the mean-flow streamfunction is defined as
	\begin{equation}
		\overline{\Psi}(r,y,s):=\overline{\Psi(r,y,s,t)},
		\qquad
		\nabla_{\perp}^{2}\overline{\Psi}=-\overline{\Omega}_s,
		\qquad
		\overline{\Psi}\big|_{\partial\mathcal A}=0.
	\end{equation}

	\subsection{Phase of the modes}
    \label{sec:phase_of_modes}

	FHPOD yields complex modes whose global phase is arbitrary. The real and imaginary parts of each complex mode represent the same coherent structure with a $90^\circ$ phase shift. To establish a consistent phase across different cases and remove this arbitrary phase, we can phase-align the modes by applying a phase rotation:
	\begin{equation}
		\boldsymbol\xi_j^{(\vartheta)}
		=\operatorname{Re}\!\left(e^{i\vartheta}\boldsymbol\xi_j\right)
		=\cos\vartheta\,\operatorname{Re}(\boldsymbol\xi_j)
		-\sin\vartheta\,\operatorname{Im}(\boldsymbol\xi_j).
		\label{eq:fhpod_phase_rotation}
	\end{equation}
	Since $|\boldsymbol\xi_j|^2$
	is proportional to the modal kinetic-energy density and thus independent of the global phase, the streamwise coordinate $S_{\mathrm{med}}$ at which the cumulative energy crosses one half of its total value is well defined and is therefore used as a reference. As we visualise the iso-surfaces of the streamfunction of the modes, the reference phase is defined from the area-averaged streamfunction of the mode over the upper half-section of the pipe $y>0$:
	\begin{equation}
		\begin{aligned}
			&\left\langle\Psi_j\right\rangle_{\mathcal A^{+}}
			=
			\frac{1}{\left|\mathcal A^{+}(S_{\mathrm{med}})\right|}
			\iint_{\mathcal A(S_{\mathrm{med}}),\,y>0}
			\Psi_j(r,y,S_{\mathrm{med}})\,\diff r\,\diff y,\\
			&\left|\mathcal A^{+}(S_{\mathrm{med}})\right|
			:=
			\iint_{\mathcal A(S_{\mathrm{med}}),\,y>0}
			\diff r\,\diff y,\\
			&\vartheta_j^{0}
			=
			\arg\!\left(
			\left\langle\Psi_j\right\rangle_{\mathcal A^{+}}
			\right).
		\end{aligned}
		\label{eq:fhpod_phase_zero}
	\end{equation}
	Restricting the integral to $y>0$ prevents cancellations between the upper and lower halves in the antisymmetric modes. Each mode is then rotated by its reference phase, $\boldsymbol\xi_j\mapsto e^{-i\vartheta_j^{0}}\boldsymbol\xi_j$. This removes the initial arbitrary phase and phase-aligns corresponding modes across the three cases. All selected FHPOD mode shapes presented below are displayed after applying this phase rotation.

	\FloatBarrier
	
	\section{Local stability analysis}
	\label{sec:meth_LSA}
	
	\begin{figure}
		\begin{subfigure}[b]{0.4\textwidth}
			\centering
			\begin{tikzpicture}[
				x=1cm,y=1cm,
				line cap=round,
				line join=round,
				>=Latex,
				every node/.style={font=\normalsize},
				]
				\def\R{1.5}
				\def\thet{-30}
				\coordinate (C) at (\R,\R);
				\coordinate (P) at ($(C) + ({\R*sin(\thet)},{\R*cos(\thet)})$);
				\coordinate (T) at ($(C) + ({1.5*\R*sin(130)},{1.5*\R*cos(130)})$ ,0.05);
				
				\draw[very thick,->] ($(C)$) -- ($(C)+(0,1.5*\R)$) node[above] {\small$\textbf{B}$};
				\draw[very thick,->] (C) -- ($(C)+(1.5*\R,0)$) node[right] {\small$\textbf{N}$};
				\draw[very thick,->] (C) -- (T) node[pos=0.76,above right] {\small$\textbf{T}$};
				\draw[thick,->] (C) -- (P) node[pos=0.3,above] {\small$\rho$};
				\draw[thin] ($(C)+(0,{0.75*\R})$) arc[start angle=90,end angle=120,radius=0.75*\R];
				\node at ($(C) + ({0.65*\R*sin(\thet+10)},{0.65*\R*cos(\thet+10)})$) {\small$\theta$};
				\draw[thick] (C) circle [radius=\R];
				\draw[thin,dash dot] ($(C)$) arc[start angle=50,end angle=65,radius=6*\R];
				\draw[thin] ($(C)$) arc[start angle=50,end angle=30,radius=6*\R];
				\draw[thick] ($(C)+({\R*sin(30)},{\R*cos(30)})$) arc[start angle=50,end angle=70,radius=5*\R];
				\draw[thick] ($(C)+({\R*sin(230)},{\R*cos(230)})$) arc[start angle=50,end angle=65,radius=5*\R];
				\node at ($(C)+({1.2*\R},{-1.15*\R})$) {\small$s$};
				\fill (C) circle (1.3pt);
				\fill (P) circle (1.0pt);
			\end{tikzpicture}
			\caption{Toroidal coordinate system.}
		\end{subfigure}
		\hfill
		\begin{subfigure}[b]{0.6\textwidth}
			\centering
			\includegraphics[width=0.7\linewidth]{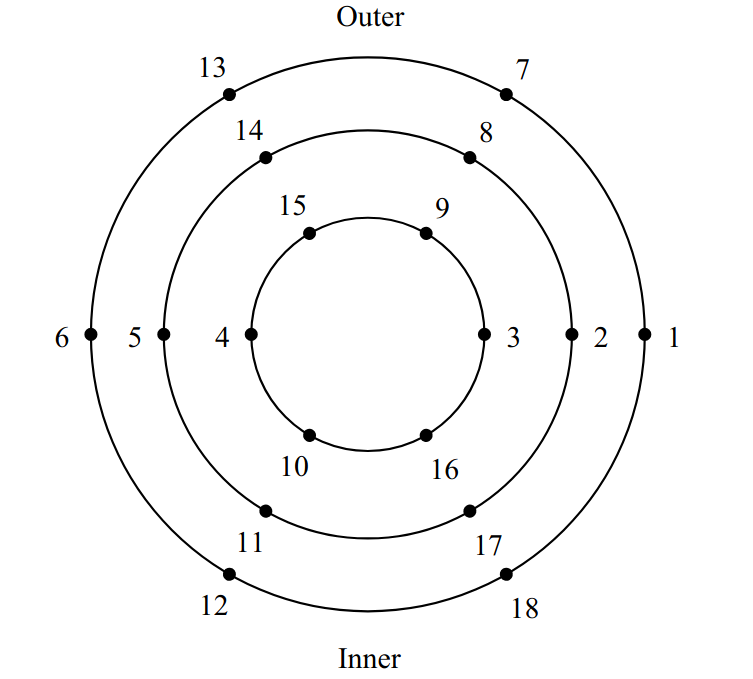}
			\caption{Polar mesh for the Fourier--Chebyshev spatial discretisation on the pipe cross-section. In this example, $n_\rho=3$ and $n_\theta=6$; the node ordering for the radial derivatives is also indicated \citep{Lupi2024_JFM987A40}.}
			\label{fig:grid_Lupi}
		\end{subfigure}
		\caption{Sketches of the coordinate system and mesh for the stability computations.}
		\label{fig:germano_system}
	\end{figure}
	
	The stability of the cross-sectional mean turbulent flow is investigated through temporal modal stability analysis \citep{schmid2002stability, drazin2004hydrodynamic}, \ie by studying the evolution of infinitesimal perturbations $\qvecp = \{\widetilde{\uvec}, \widetilde{p}\}$ applied to the cross-sectional mean turbulent base flow $\Qvec = \{\Uvec, \overline{P}\}$, with $\Uvec$ denoting the mean velocity vector and $\overline{P}$ the mean pressure. Similar mean flow stability approaches have been used to identify instability mechanisms in both unsteady laminar flows \citep{Sipp2007} and turbulent mean flows \citep{Gudmundsson2011,Mettot2014,Rodriguez2015,Tammisola2016,Camarri2017}. For this purpose, the two-dimensional (2D) stability code described in \citet{Casali_thesis}, capable of performing modal stability analysis for toroidal and helical pipe geometries, is used. 
	We perform the local stability analysis of the mean turbulent flow at different pipe cross-sections along $S$, treating each cross-sectional mean flow as the mean flow in a toroidal pipe with the same curvature. Due to this assumption, the streamwise invariance allows a 2D analysis with periodic modes in the streamwise direction, also referred to as BiGlobal \citep{lupi2023thesis}. The code uses the toroidal coordinate system shown in figure~\ref{fig:germano_system}, where $\Tvec$, $\Nvec$ and $\Bvec$ are the tangential, normal and binormal unit vectors. The coordinate system is characterised by the polar coordinate pair $\rho$ and $\theta$ on the pipe cross-section and the curvilinear coordinate $s$. The stability code uses a spectral collocation method, employing Fourier and Chebyshev basis functions in the azimuthal and radial directions, respectively. The Chebyshev nodes are defined for $\left[ -R_p\,,R_p \right]$ and the Fourier set for $ \left[0 \, , 2\pi \right)$. By using an even number of Chebyshev collocation points there is no grid point at the pipe centreline such that any singularity is avoided \cite{Lupi2024_JFM987A40}. The computational domain is sketched in figure \ref{fig:grid_Lupi}. The spatial resolution of the present discretisation is $n_\rho=72$ in the radial and $n_\theta=146$ in the azimuthal direction after performing a grid convergence study.
	
	In general, linear mean-flow analyses can be grouped into no-eddy-viscosity, frozen eddy-viscosity, and perturbed eddy-viscosity approaches \citep{Sarras2024}. In this work, we utilize the frozen eddy-viscosity closure for our mean-flow stability analysis in which the mean turbulent diffusion is retained while the perturbation of the eddy-viscosity is neglected ($\widetilde{\nu}_t=0$). The Reynolds stresses are represented through a spatially varying eddy-viscosity, following the work of \citet{ReynoldsHussain1972} and its later use in turbulent mean-flow stability analysis \citep{Rukes2016,Sarras2024}.
	\begin{equation}
		\label{eq:nueff}
		\nueff = \nu+\nut(\rho,\theta;\phi),
		\qquad
		\Smat(\uvecp)=\tfrac12\left(\grad\uvecp+\grad\uvecp^{\mathsf T}\right),
	\end{equation}
	Here $\nueff$ is the dimensional total effective kinematic viscosity, and $\Smat(\uvecp)$ is the perturbation strain-rate tensor. The eddy-viscosity $\nut$ is estimated at each grid point by fitting the scalar Boussinesq relation to the DNS mean strain and Reynolds stresses using a least-squares method. This tensor fit follows the approach for turbulent mean flow stability analysis used in \citet{Rukes2016,vonSaldern2023}. We first define the deviatoric parts of the Reynolds-stress and mean strain-rate tensors as
	\begin{equation}
		\Rmat_{\mathrm{dev}}
		=\Rmat-\tfrac13\tr(\Rmat)\boldsymbol I_3,
		\qquad
		\Smat_{\mathrm{dev}}
		=\Smat(\Uvec)-\tfrac13\tr\!\left[\Smat(\Uvec)\right]\boldsymbol I_3,
		\label{eq:dev_parts}
	\end{equation}
	where $\boldsymbol I_3$ denotes the three-dimensional identity tensor.
	At each grid-point, the eddy-viscosity is obtained by solving a least-squares problem
	\begin{equation}
		\nut^{\mathrm{raw}}(\rho,\theta;\phi)
		=\underset{\nu_\star\in\mathbb{R}}{\operatorname{arg\,min}}
		\left\|
		\Rmat_{\mathrm{dev}}(\rho,\theta;\phi)
		+2\nu_\star\Smat_{\mathrm{dev}}(\rho,\theta;\phi)
		\right\|_F^2 .
		\label{eq:nut_ls_problem}
	\end{equation}
	Here, $\|\cdot\|_F$ denotes the Frobenius norm. This yields
	\begin{equation}
		\label{eq:nut_raw}
		\nut^{\mathrm{raw}}=
		\begin{cases}
			-\dfrac{\Rmat_{\mathrm{dev}}\!:\!\Smat_{\mathrm{dev}}}
			{2\,\Smat_{\mathrm{dev}}\!:\!\Smat_{\mathrm{dev}}},
			& \Smat_{\mathrm{dev}}\!:\!\Smat_{\mathrm{dev}}>0,\\[1ex]
			0,
			& \Smat_{\mathrm{dev}}\!:\!\Smat_{\mathrm{dev}}=0.
		\end{cases}
	\end{equation}
    However, there are two issues arising from \eqref{eq:nut_raw} in the absence of regularisation. \eqref{eq:nut_raw} becomes numerically ill-conditioned in low-strain regions, and the resulting field can become non-smooth. Furthermore, it can return negative values in regions of local backscatter. To obtain a smooth solution, regularisation \citep{vonSaldern2023} can be applied, while positive clipping has been proposed when the model is intended to retain only dissipative transfer \citep{vonSaldern2024}. In the present work, $\nu_t^{\mathrm{raw}}$ is treated using a smooth positive clipping to obtain the regularised eddy-viscosity field $\nu_t$ used in \eqref{eq:nueff}. Details on the numerical implementation are provided in Appendix~\ref{app:eddy_visc}.
    
    With the eddy-viscosity closure, the evolution of the perturbations $\qvecp$ is governed by the incompressible Reynolds-Averaged Navier--Stokes equations linearised around the turbulent mean flow
	\begin{equation}
		\label{eq:NS_lin}
		\begin{split}
			& \pdv{{\uvecp}}{t}+(\Uvec \vdot \grad)\uvecp + (\uvecp \vdot \grad)\Uvec + \grad \pp - \div\!\left[\,2\,\nueff\,\Smat(\uvecp)\,\right] = 0, \\
			& \div \uvecp = 0 \ ,
		\end{split}
	\end{equation}
	with the no-slip boundary conditions applied at the wall. No boundary condition is required for the pressure, but the system is augmented with equations to cancel out the spurious pressure modes that typically appear in the case of spectral methods \citep{balachandar1992spurious_nasa}, which affect the eigenmodes. The perturbations assume the following form:
	\begin{equation}
		\label{eq:stab_pert}
		\qvecp(s,\rho,\theta,t) = \qhvec_{\alpha}(\rho,\theta) e^{\ \mu t+i\alpha s} \ ,
	\end{equation}
	where $\qhvec_\alpha=\{\uhvec_\alpha,\widehat p_\alpha\}$ comprises the velocity and pressure amplitudes. Here $\alpha=2\pi/l_s$ is the dimensional streamwise wavenumber and $l_s$ is the wavelength. The dimensional complex eigenvalue is $\mu=\sigma+i\omega$, where $\sigma$ is the temporal growth rate, $\omega$ is the angular frequency and the Strouhal number is $St=|\omega|D/(2\pi U_b)$. 
	
	Finally, for each streamwise wavenumber, one obtains the following eigenvalue problem
	\begin{equation}\label{eq:eig}
		-\mu \mathcal{M} \qhvec_\alpha = \mathcal{L}\qhvec_\alpha \ ,
	\end{equation}
	where $\mathcal{M}$ is the mass matrix and $\mathcal{L}$ the linearised Reynolds-Averaged Navier--Stokes operator, defined as
	
	\begin{equation}
		\mathcal{M} =
		\begin{pmatrix}
			\boldsymbol I_u & 0 \\ 0 & 0
		\end{pmatrix},   \ \
		\mathcal{L} =
		\begin{pmatrix}
			\grad_0 \Uvec + \Uvec \vdot \grad_\alpha - \mathcal{D}_\alpha[\nueff] & \ \ \grad_\alpha \\
			\div_\alpha & \ \ 0
		\end{pmatrix} \ ,
	\end{equation}
	\begin{equation}
		\label{eq:Dalpha}
		\mathcal{D}_\alpha[\nueff]\,\uhvec_\alpha := \nu\laplacian_\alpha\uhvec_\alpha + \nabla_\alpha \cdot\!\left[2\nut\Smat_\alpha(\uhvec_\alpha)\right],
	\end{equation}
	where $\boldsymbol I_u$ is the $3 n_{\mathrm{dof}} \cross 3 n_{\mathrm{dof}}$ identity on the discrete velocity space, $n_{\mathrm{dof}} = n_{\theta}n_{\rho}$ is the number of scalar degrees of freedom, and $\nabla_{\alpha}$ denotes that the differential operators depend on the wavenumber $\alpha$, since the term $e^{i\alpha s}$ contributes to $\pdv*{}{s}$, as described in \cite{Casali_thesis}. By solving the eigenvalue problem~\eqref{eq:eig}, we find locally unstable modes when the real part of the eigenvalue is positive ($\sigma>0$). The eigenvalue problem is solved through the shift-and-invert transformation, implemented using the parallel solver \texttt{PETSc}/\texttt{SLEPc} \citep{petsc-user-ref, Hernandez2005SLEPc}.

	\subsection{Tracking the unstable branches along the bend}
	\label{sec:meth_track}
	
	The eigenvalue problem \eqref{eq:eig} is formulated for two-dimensional cross-sections at an individual bend angle $\phi$. Our goal is to identify the unstable modes and track them along the bend angle $\phi$. Throughout this article, the studied interval is $10^{\circ}\le\phi\le160^{\circ}$, sampled at $\Delta\phi=1^{\circ}$. After solving the eigenvalue problem \eqref{eq:eig} for each angle in this interval, the unstable modes are identified, but the connection between them at two different bend angles is still unclear. We track each mode along the bend angle ($\phi$) using Rayleigh quotient iteration (RQI), which provides an automated procedure for following the corresponding eigenpair between successive cross-sections. The details of the RQI method are described in Appendix~\ref{app:RQI}.

	\subsection{Reconstruction of the three-dimensional LSA modes}
	\label{sec:meth_wkb}
	
	The mode-tracking approach introduced in \S\ref{sec:meth_track} provides the continuation of the same eigenpair at successive bend angles $\phi$. Each tracked eigenvector, however, is defined only up to an arbitrary complex phase. Therefore, constructing a three-dimensional field from a sequence of two-dimensional eigenvectors of the same branch requires replacing these arbitrary phases with the physical streamwise phase. In the present reconstruction, the physical phase is computed using a leading-order Wentzel--Kramers--Brillouin (WKB) approximation for slowly varying flows \citep{Huerre1990}. Similar approaches have been used to construct three-dimensional coherent structures in jets and wakes \citep{Oberleithner2011,Oberleithner2014,Siconolfi2017,Li2024}.
	
	At every bend angle $\phi$, the local eigenvector is first normalised to have unit cross-sectional kinetic energy, with $\boldsymbol u$ denoting its three velocity components. In the next step, we phase-lock the individual eigenvectors with respect to a reference plane. We assign the index $0$ to the reference plane and initialise the phase-locking by setting $\qhvec^{\mathrm{lock}}_0=\qhvec_0$ and $\boldsymbol u^{\mathrm{lock}}_0=\boldsymbol u_0$. The choice of the reference plane is arbitrary (in our case $\phi_{\mathrm{lock}}=\pi/2$) and only affects the initial global phase of the reconstructed mode. This initial phase is later removed by applying a phase rotation to each reconstructed mode. Starting from this reference plane, $k=1,2,\ldots$ denote indices of successive neighbouring planes along either the descending or ascending direction ($\phi_k<\phi_{\mathrm{lock}}$ or $\phi_k>\phi_{\mathrm{lock}}$). Then the following recursion is separately applied to both sides of the reference plane:
	\begin{equation}
		\label{eq:phase_lock}
		\begin{aligned}
			\beta_k
			&=\arg\!\left[
			\left(\boldsymbol u_{k-1}^{\mathrm{lock}}\right)^\dagger
			\boldsymbol u_k
			\right],  \qquad
			\qhvec_k^{\mathrm{lock}}
			&=\qhvec_k e^{-i\beta_k},
			\qquad
			\boldsymbol u_k^{\mathrm{lock}}
			=\boldsymbol u_k e^{-i\beta_k}.
		\end{aligned}
	\end{equation}
	Thus, the phase of each eigenvector at each cross-section is recursively connected to the reference plane through its immediate neighbours. By construction, the complex inner product between adjacent phase-locked velocity eigenvectors is real and non-negative, while the eigenvectors themselves remain complex.
	
	After the phase-locking step, the amplitude of the three-dimensional reconstructed mode is determined using the following relation between temporal and spatial growth \citep{Gaster1962}:
	\begin{equation}
		\label{eq:gaster}
		a_0(\phi;\alpha)=\exp\!\left[
		\int_{s_{b,0}}^{s_b(\phi)}
		-\sigma(s_b';\alpha)
		\left[\pdv{\omega(s_b';\alpha)}{\alpha}\right]^{-1}\diff s_b'
		\right],
		\quad
		a(\phi;\alpha)=\frac{a_0(\phi;\alpha)}{\max_{\phi}\left(a_0(\phi;\alpha)\right)}.
	\end{equation}
	In \eqref{eq:gaster}, the dimensional centreline arc length, measured from the bend entrance, is
	\begin{equation}
		s_b(\phi)=\frac{D\phi}{2\gamma},
		\quad
		s_{b,\mathrm{ref}}=s_b\!\left(\frac{\pi}{2}\right)=\frac{\pi D}{4\gamma}.
		\label{eq:centreline_arc}
	\end{equation}
	The lower limit in the integral in \eqref{eq:gaster} is $s_{b,0}=s_b(10^{\circ})$, corresponding to the smallest angle at which the LSA is performed. The quantity $\max_{\phi}\left(a_0(\phi;\alpha)\right)$ is the largest value of $a_0$ over all tracked bend angles at the prescribed $\alpha$, while $\sigma(s_b;\alpha)$ and $\omega(s_b;\alpha)$ represent the growth rates and frequencies along a tracked branch, respectively. The derivative $\partial\omega/\partial\alpha$ is first evaluated using finite differences at discrete angles every 10 degrees. The results are then interpolated along $\phi$. Finally, following the leading-order reconstruction used by \citet{Huerre1990} and \citet{Oberleithner2014}, the three-dimensional reconstruction is obtained from
	\begin{equation}
		\label{eq:wkb}
		\widetilde{\boldsymbol b}_{\mathrm{3D}}(\rho,\theta,\phi,t)
		=a(\phi;\alpha)\,\qhvec^{\mathrm{lock}}(\rho,\theta;\phi)
		\exp\!\left\{i\left[
		\alpha\bigl(s_b(\phi)-s_{b,\mathrm{ref}}\bigr)+\omega_g(\alpha)t
		\right]\right\}.
	\end{equation}
	in which $\omega_g(\alpha)$ is the global frequency and is taken from the local frequency at the bend angle where $a(\phi;\alpha)$ is largest. This is because this location is characterised by the largest accumulated amplification and therefore represents the most strongly amplified part of the tracked instability branch. %
    The global frequency $\omega_g(\alpha)$ in \eqref{eq:wkb} is included for completeness, however, it is not used in the present analysis, which considers only the three-dimensional spatial structure of the reconstructed mode.
	This phase-consistent reconstruction provides an approximation to the global eigenmode by assembling a sequence of local modes of the same branch. As the reconstructed mode also has an arbitrary global phase determined by the choice of $\phi_{\mathrm{lock}}$, we rotate it to match the phase convention of its corresponding FHPOD mode, as outlined in~\eqref{eq:fhpod_phase_rotation}. In the following, the reconstructed mode shapes are shown using isosurfaces of the cross-sectional streamfunction described in \S\ref{subsec:streamfunction} with red and blue indicating positive and negative isovalues. We denote the reconstructed modes from the linear stability analysis with the prefix ``LSA'' to distinguish them from the FHPOD modes. 
	
	\section{Results, analysis and discussion of FHPOD modes}
	\label{sec:results}
	
\begin{figure}[!htbp]
		\centering
		\begin{overpic}[width=1.0\textwidth]{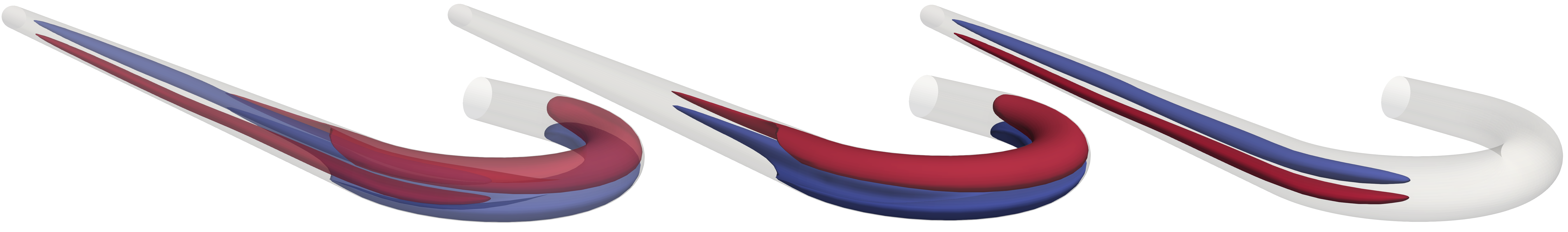}
			\inletplanearrow{20.5}{8.3}{1.25}{-.38}
			\put(30.0,-3){\small\bfseries \textbf{(a) }}
			\put(60,-3){\small\bfseries \textbf{(b) }}
			\put(90,-3){\small\bfseries \textbf{(c) }}
		\end{overpic}
		\vspace{1mm}
		\hfill
		\caption{\textit{(a)} Isosurfaces of the mean-flow streamfunction $\overline{\Psi} = \pm 1.5 \times 10^{-3}\, U_b D$. Panels \textit{(b)} and \textit{(c)} separate the two vortex pairs for clarity. \textit{(b)} The isolated Dean vortices rapidly decay downstream of the bend. \textit{(c)} A new vortex pair with opposite rotation to the Dean vortices emerges downstream of the bend.}
		\vspace{-2mm} 
		\label{fig:meanPsi}
	\end{figure}

	As outlined in Section~\ref{sec:intro}, the literature indicates the presence of multiple instability mechanisms within the bent pipe, whose relative strength depends strongly on the streamwise location \citep{Jain2019, Wang2018, Jain2022}, and each instability is superimposed on distinct underlying mean flow vortices. 
	Therefore, as a first step, we obtain the cross-sectional streamfunction of the mean flow, as described in section~\ref{subsec:streamfunction}. Figure~\ref{fig:meanPsi}\textit{(a)} depicts the resulting mean-flow vortical structure over the entire bend, which comprises two vortex pairs. 
	For a better visualisation, we separate these two vortex pairs, as shown in panels \textit{(b)} and \textit{(c)} of figure~\ref{fig:meanPsi}. 
	As a result of the centrifugal force, a counter-rotating Dean vortex pair forms near the entrance of the bend and dominates within the curved section. 
	As the Dean vortices are driven by the centrifugal force, they rapidly decay downstream of the bend once the curvature becomes zero (figure~\ref{fig:meanPsi}\textit{(b)}). 
	At the bend angle $\phi\approx117^{\circ}$ ($S\approx21.2$), a new counter-rotating vortex pair emerges with the rotation axes opposite to the ones of the Dean vortices and dominates in the downstream straight section near the outer wall (figure~\ref{fig:meanPsi}\textit{(c)}). We refer to this vortex pair as the \emph{downstream vortices} throughout this work. %
	
	\subsection{Identification of modes}
	\label{sec:ident_modes}
	
\begin{figure}[!htbp]
		\centering
		\includegraphics[width=0.80\textwidth]{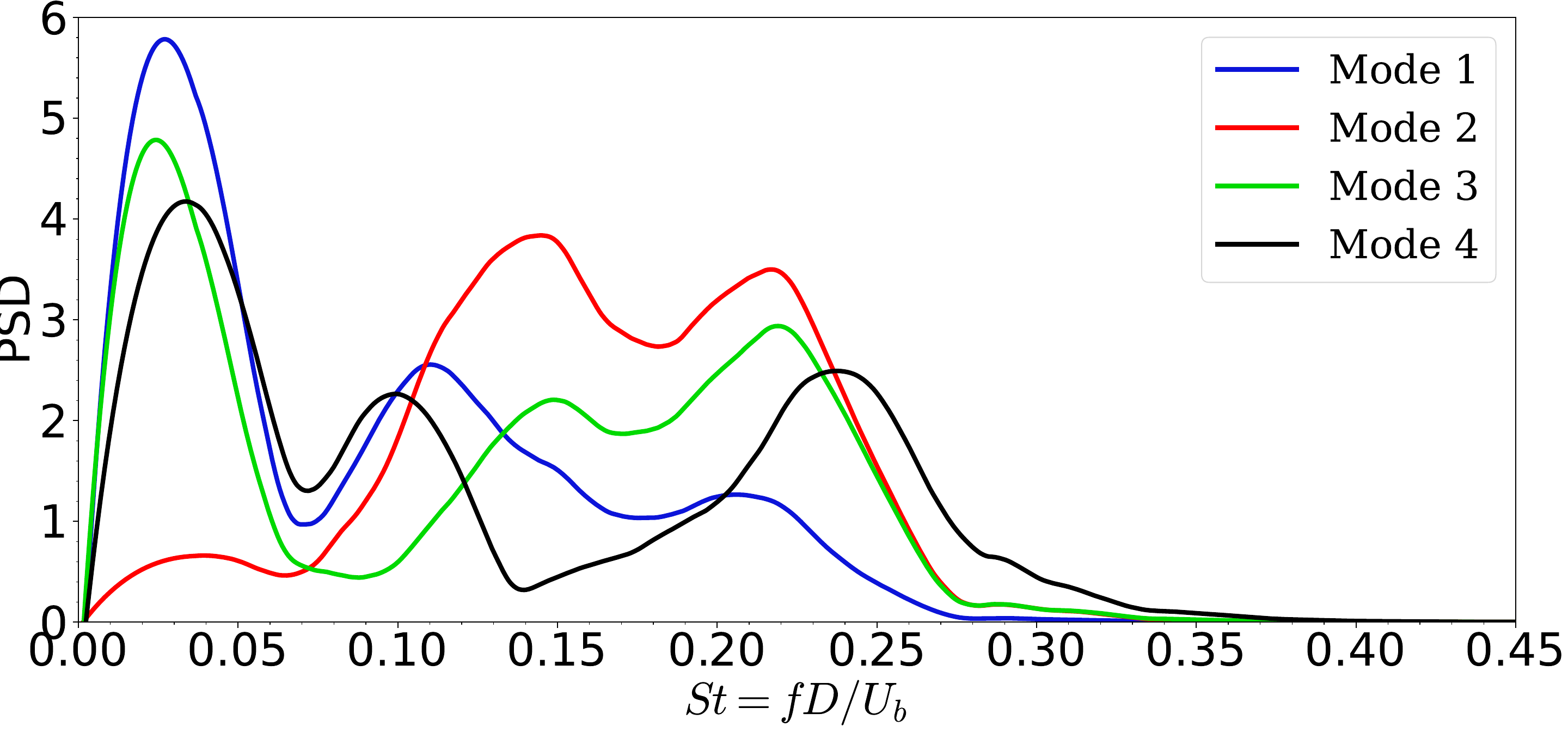}
		\caption{PSD of time-coefficients for the first four POD modes}
		\label{fig:POD_PSD}
	\end{figure}

	Following \citet{Hufnagel2018} and \citet{Lupi2025}, we first perform a three-dimensional POD without applying the Hilbert transform. The PSD of the time-coefficients of the first four modes is shown in figure~\ref{fig:POD_PSD}, where the two major issues associated with classical POD are immediately apparent.
	
    Despite the long sampling duration and large number of snapshots, the modes do not appear as well-defined pairs with equal energy, and requirement~\ref{req:degeneracy} is not satisfied. The breakdown of degenerate pairing is partly due to the energetic low-frequency instability at $St\approx0.03$, whose large streamwise wavelength ($\approx22D$) causes it to contribute to multiple modes. When the POD is repeated on a domain with a shortened downstream extent and fewer snapshots, the influence of this instability is reduced, and the pairing improves, with modes~1 and~2 appearing nearly degenerate. A similar imperfect pairing of degenerate POD modes is also visible in previous POD studies \citep{Hufnagel2018, Lupi2025}.
	
	Modes~2 and~3 exhibit similar spectral content with shared peaks at $St\approx0.15$ and $St\approx0.22$. However, mode~3 also contains a significant contribution from the instability at $St\approx0.03$, showing the contribution of the low-frequency instability to multiple modes. This directly indicates that requirement~\ref{req:unimodal} is not satisfied. When degenerate pairing is imperfect and individual instabilities contribute to several modes, energy-based ranking of the POD modes becomes physically ill-defined.
	
\begin{figure}[!htbp]
		\centering
		\includegraphics[width=0.7\textwidth]{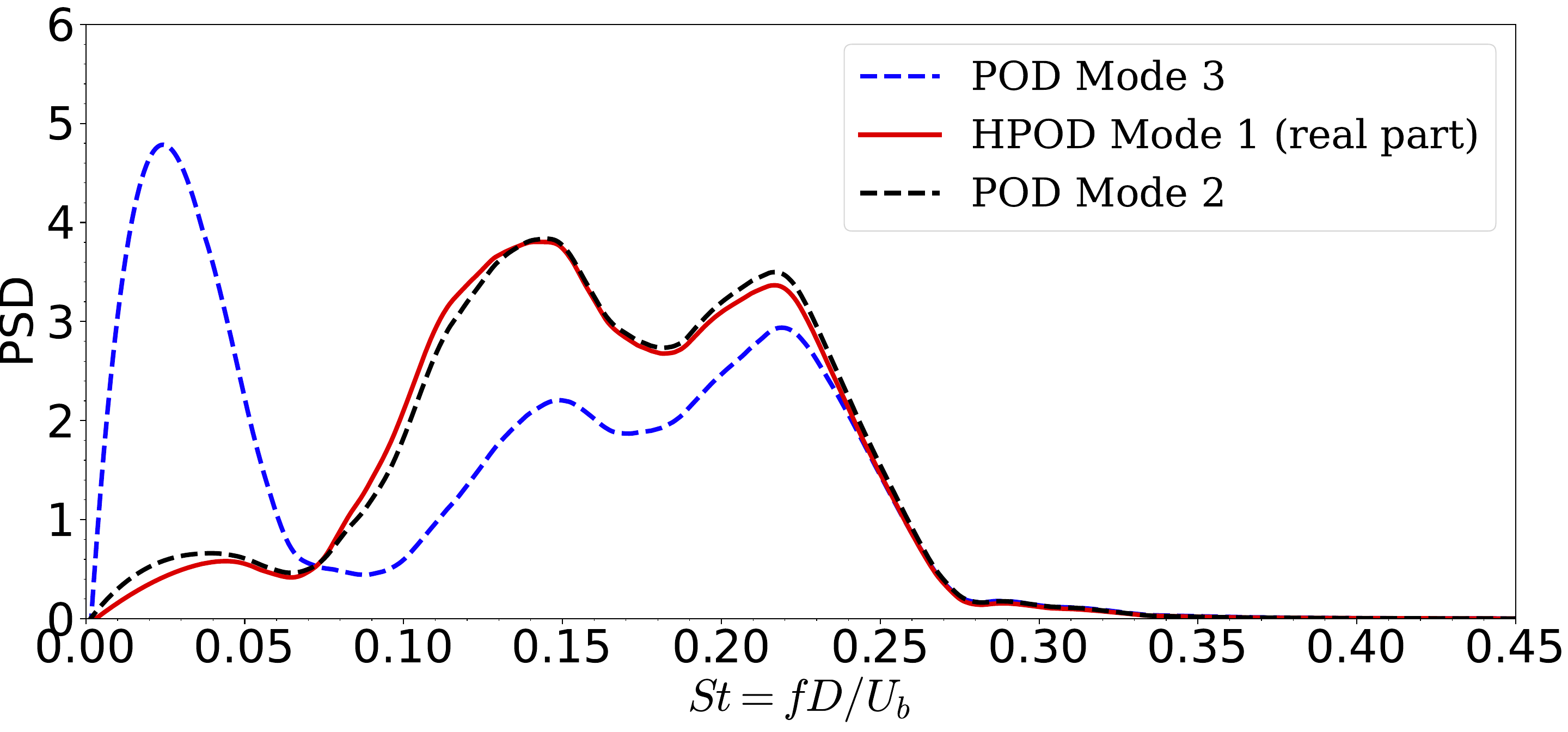}
		\caption{PSD of the time-coefficients of POD modes~2 and~3 and the real part of HPOD mode~1. HPOD collapses the two degenerate POD modes into a single complex mode and reduces the leakage of the low-frequency instability at $St\approx0.03$.}
		\label{fig:hilbertVsNormalPOD}
	\end{figure}

\begin{figure}[!htbp]
		\centering
		
		\begin{overpic}[width=0.8\linewidth]{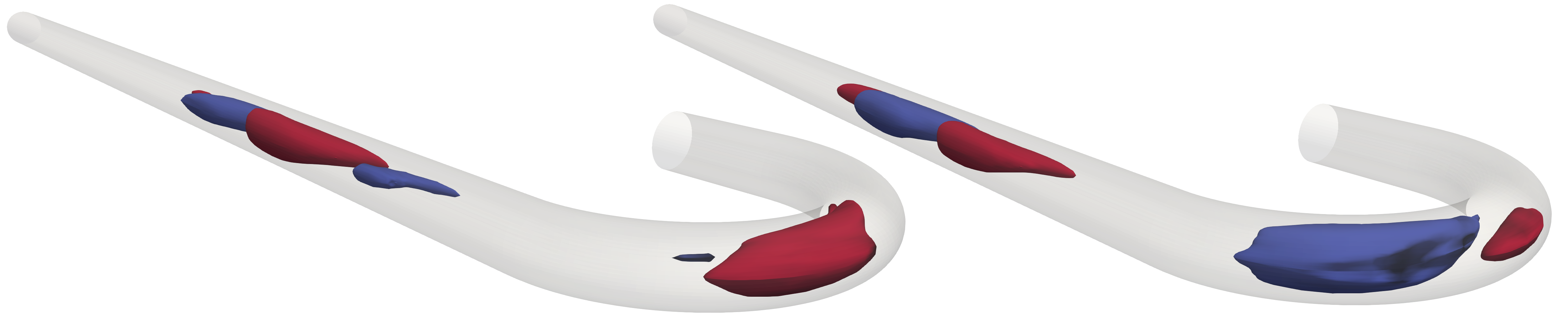}
			\inletplanearrow{29}{11.3}{1.25}{-.38}
			\put(25,1){\small\bfseries (a)}
			\put(68,1){\small\bfseries (b)}  
			\vspace{5mm}
		\end{overpic}
		\vspace{5mm}
		\begin{overpic}[width=0.8\linewidth]{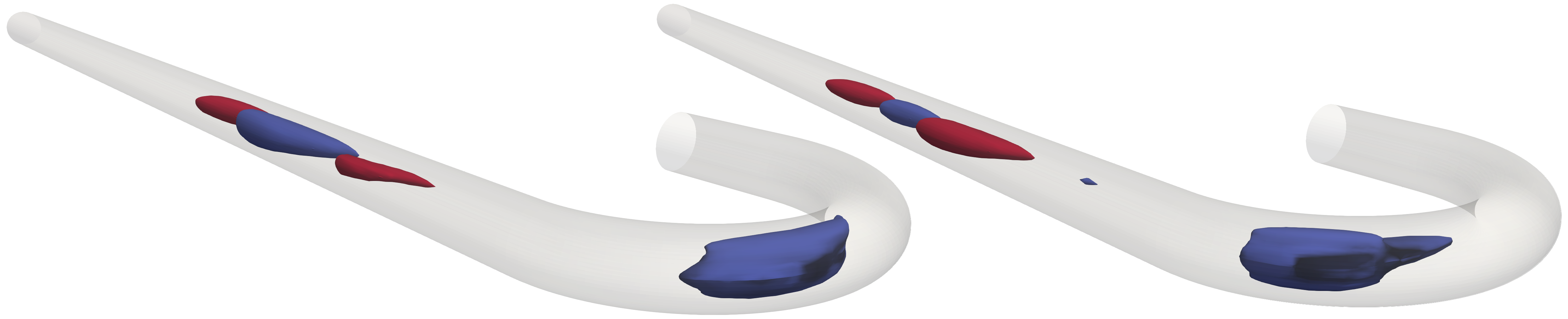}
			\put(25,0){\small\bfseries (c)}
			\put(68,0){\small\bfseries (d)}
		\end{overpic}
		\caption{Isosurfaces of the modal streamfunction $\Psi_j = \pm 0.03 \, U_b D$.
			\textit{(a)} HPOD mode 1 (real part).
			\textit{(b)} HPOD mode 1 (imaginary part).
			\textit{(c)} POD mode 2.
			\textit{(d)} POD mode 3.}
		\label{fig:Psi_POD_HPOD}
	\end{figure}

	To address the issue of degenerate pairing, we apply the Hilbert transform in time. By constructing the analytic signal at each point and assembling the resulting complex fields into the snapshot matrix, we obtain complex POD modes. By construction, the real and imaginary parts of each mode are enforced to be $90^\circ$ phase-shifted and share a single singular value. Consequently, a convecting instability can be represented by a single complex mode with a well-defined energy. This satisfies requirement~\ref{req:degeneracy} as opposed to the classical POD using the same dataset. This is illustrated in figure~\ref{fig:hilbertVsNormalPOD}, where HPOD mode~1 combines POD modes~2 and~3 into a single complex mode while reducing leakage from the low-frequency instability at $St\approx0.03$. Figure~\ref{fig:Psi_POD_HPOD} shows the corresponding mode shapes for POD modes~2 and~3 and HPOD mode~1. Note that the energy-based ordering of the modes also changes, and the same instability that appears in POD modes~2 and~3 emerges as the most energetic complex mode in the HPOD decomposition.
	
\begin{figure}[!htbp]
		\centering
		\begin{overpic}[width=0.45\linewidth]
			{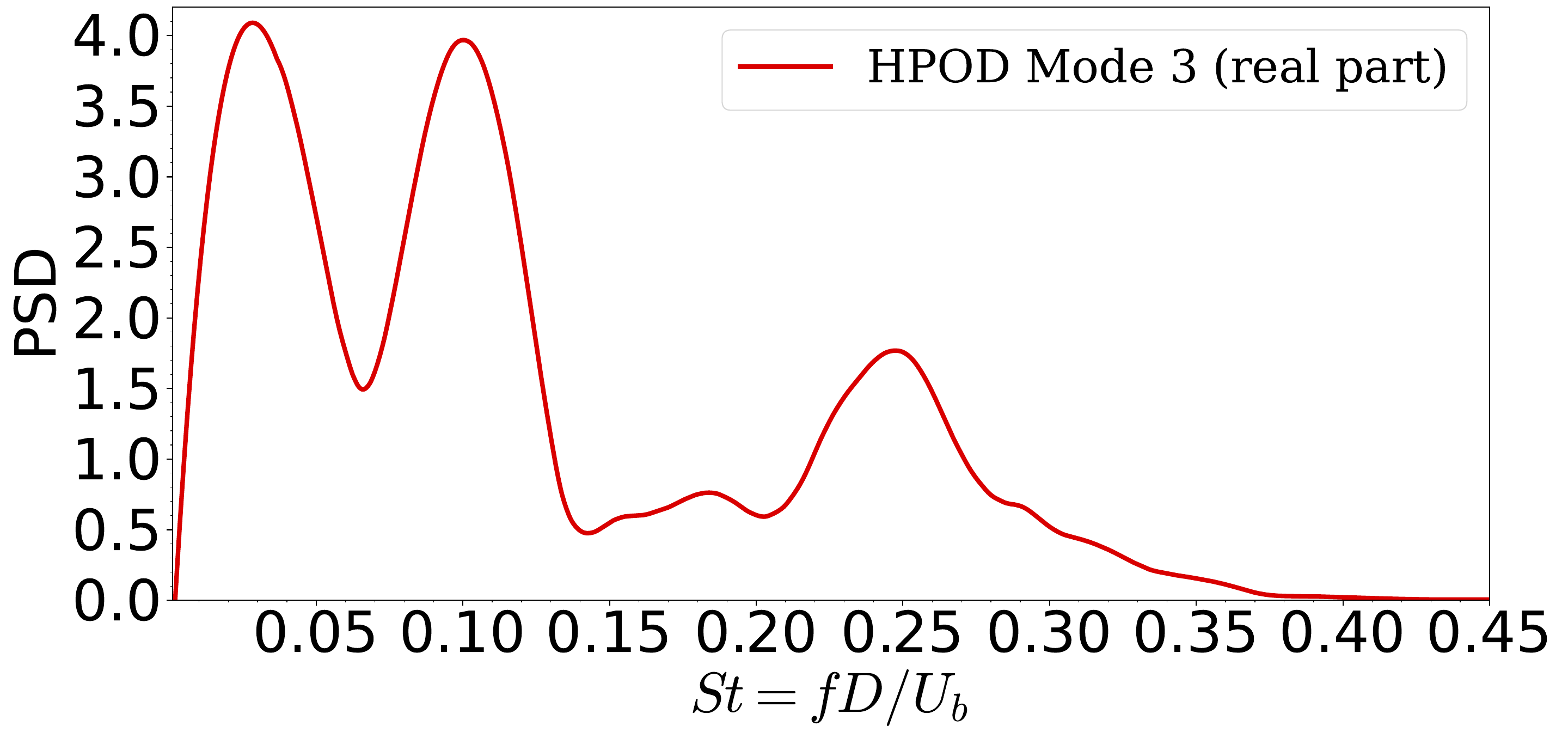}
			\put(2,1){\small\bfseries (a)}
		\end{overpic}
		\raisebox{4mm}{%
			\begin{overpic}[width=0.52\linewidth]
				{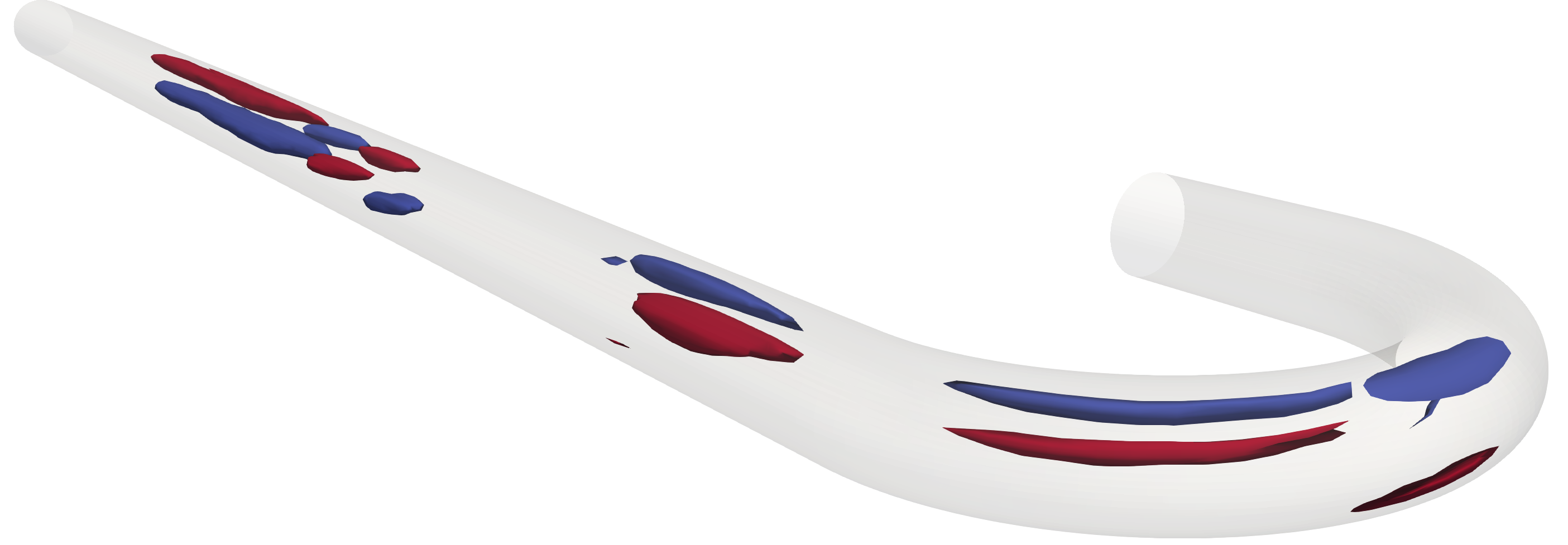}
				\put(45,-4){\small\bfseries (b)}
			\end{overpic}%
		}
		\caption{\textit{(a)} PSD of the time coefficients for the real part of the third HPOD mode.
			\textit{(b)} Isosurface of the streamwise modal component $u^{\xi}_{3,s} = \pm 0.3\, U_b$ for the real part of the third HPOD mode.}
		\label{fig:naiveHilbertMode}
	\end{figure}

	However, HPOD still does not guarantee a clean spectral separation, and requirement~\ref{req:unimodal} remains unresolved. This is clearly visible in figure~\ref{fig:naiveHilbertMode}\textit{(a)}, where the third HPOD mode exhibits nearly equal contributions from $St\approx0.03$ and $St\approx0.1$, along with a smaller peak at $St\approx0.25$. This spectral mixing is also visible in the spatial structure shown in figure~\ref{fig:naiveHilbertMode}\textit{(b)}, where isosurfaces of the streamwise modal velocity component $u^{\xi}_{3,s}$ are shown. The mode shape exhibits structures both within the bend and in the downstream straight section, implying contributions from mechanisms of distinct physical origin associated with different mean-flow vortex pairs (see figure~\ref{fig:meanPsi}).

	\begin{table}
		\centering
		\footnotesize
		\setlength{\tabcolsep}{0pt}
		\renewcommand{\arraystretch}{1.15}
		
		\begin{tabular*}{\textwidth}{@{\extracolsep{\fill}} l c c c l @{}}
			Mode &
			$St$ &
			Filter-band ($St_{\min}$--$St_{\max}$) &
			$100\sqrt{E_j/E_0}$ (\%) &
			Description \\
			\hline
			\textit{Ax} & 0.03 & 0.00--0.08 & 2.14 &
			Axial wave with weak streamwise modal vorticity \\
			\textit{SS1} & 0.13 & 0.07--0.14 & 1.99 &
			Lower-frequency swirl-switching mode \\
			\textit{SS2} & 0.25 & 0.20--0.30 & 1.93 &
			Higher-frequency swirl-switching mode \\
			\textit{DS1} & 0.15 & 0.12--0.18 & 2.23 &
			Lower-frequency downstream shear-layer mode \\
			\textit{DS2} & 0.30 & 0.18--1.00 & 2.04 &
			Higher-frequency downstream shear-layer mode \\
			\textit{SB1} & 0.12 & 0.07--0.14 & 1.93 &
			Lower-frequency swirl-breathing mode \\
			\textit{SB2} & 0.25 & 0.20--0.30 & 1.71 &
			Higher-frequency swirl-breathing mode \\
		\end{tabular*}
		\caption{Summary of FHPOD modes identified in the present study. Here $E_j$ is the snapshot-averaged analytic projected energy of mode $j$ defined in \eqref{eq:projected_energy}, and $E_0$ is the snapshot-averaged mean-flow reference energy defined in \eqref{eq:mean_flow_energy}. Every row reports the same analytic-projection amplitude ratio, $100\sqrt{E_j/E_0}$; the SVD singular values $\zeta_j$ are not reported in this table.}
		\label{tab:HPOD_modeFamily}
	\end{table}

\begin{figure}[!htbp]
		\centering
		\includegraphics[width=0.88\textwidth]{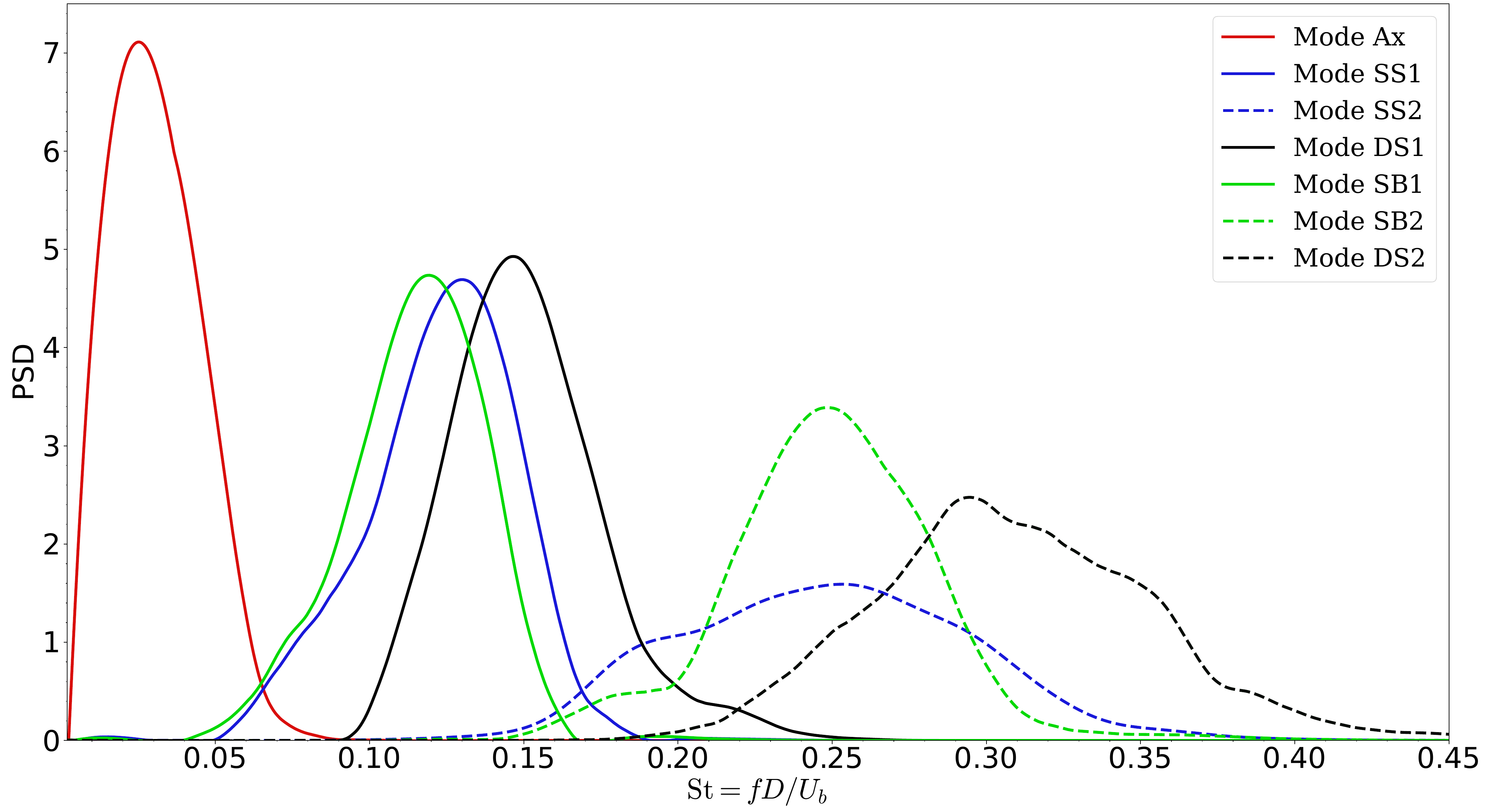}
		\caption{Power spectral density of the real parts of the time coefficients corresponding to the selected FHPOD modes. Within each paired family, the solid curve denotes the lower-frequency member and the dashed curve denotes the higher-frequency member.}
		\label{fig:filteredHPOD_PSD}
	\end{figure}

\begin{figure}[!htbp]
		\centering
		\includegraphics[width=0.88\textwidth]{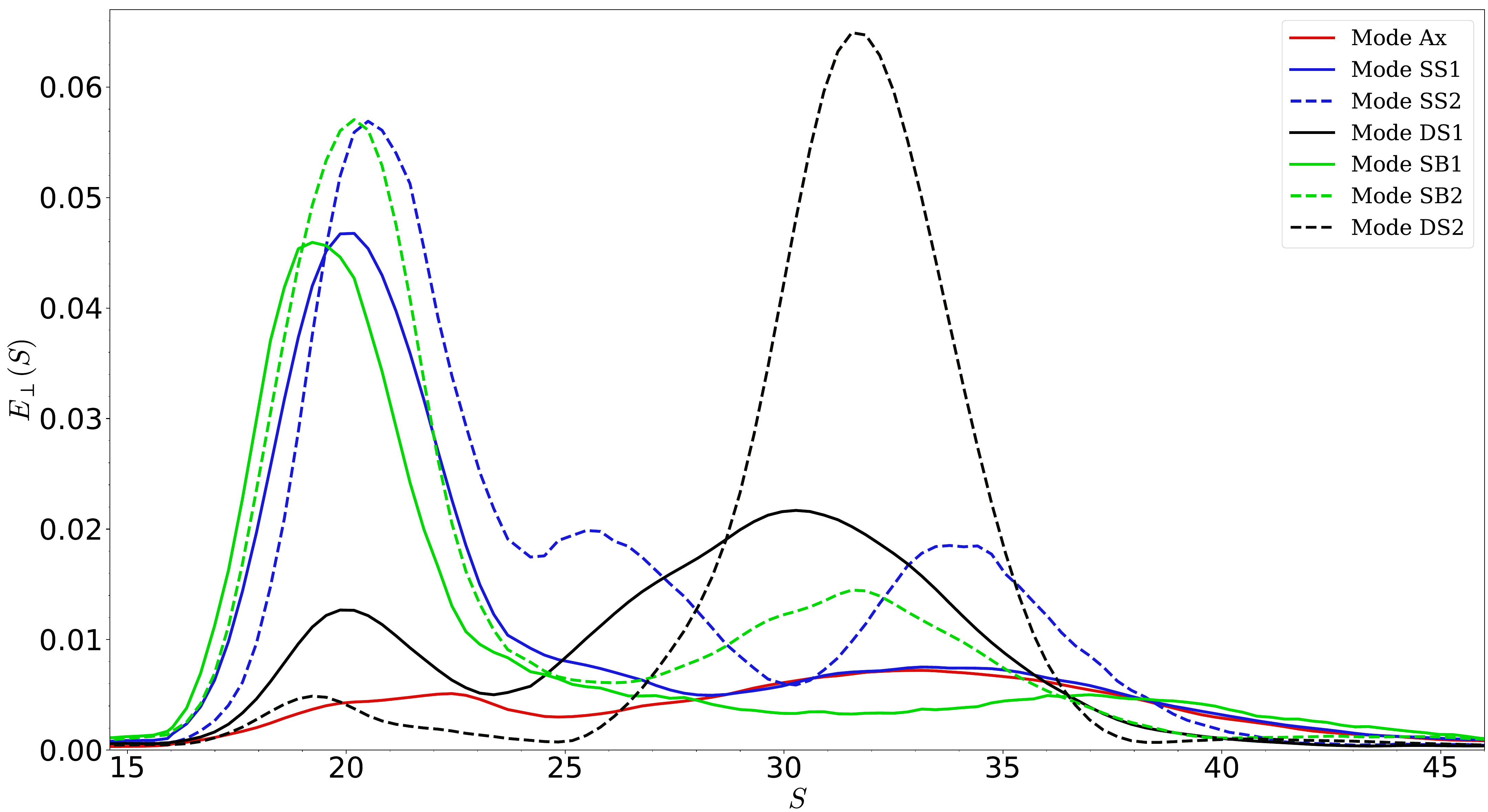}
		\captionsetup{width=0.88\textwidth}
		\caption{Streamwise profile of the cross-stream modal energy $E_{\perp,j}(S)$ of the FHPOD modes.}  \label{fig:filteredHPOD_energy_over_S}
	\end{figure}

	To fulfil requirement~\ref{req:unimodal}, we perform \emph{temporal} band-pass filtering of the snapshots before the Hilbert transformation, as described in section~\ref{sec:pod_method}. The resulting FHPOD approach identifies four physical families, summarised in table~\ref{tab:HPOD_modeFamily}. Mode~\textit{Ax} represents an axial wave. Modes~\textit{SS1} and \textit{SS2} are the lower- and higher-frequency swirl-switching modes within the bend, modes~\textit{SB1} and \textit{SB2} are the corresponding swirl-breathing modes, and modes~\textit{DS1} and \textit{DS2} are the lower- and higher-frequency downstream shear-layer modes. Figure~\ref{fig:filteredHPOD_PSD} shows the PSD of the corresponding time coefficients and demonstrates that unimodality is preserved for all modes. Furthermore, the spatial support of each mode is localised to a specific part of the flow, as shown in figure~\ref{fig:filteredHPOD_energy_over_S}. Here, we quantify the vortical strength of each mode along the pipe axis $S$ by defining the cross-stream modal energy of the $j$th mode as
	
	\begin{equation}
		E_{\perp,j}(S)
		:=
		\iint_{\mathcal{A}(S)}
		\left[
		|u^{\xi}_{j,r}(r,y,S)|^2+
		|u^{\xi}_{j,y}(r,y,S)|^2
		\right]\,\mathrm{d}r\,\mathrm{d}y,
		\label{eq:crossplane_energy}
	\end{equation}
	where $\mathcal A(S)$ denotes the pipe cross-section in the $(r,y)$ plane at the non-dimensional arc length $S=s/D$ and the subscripts $r$ and $y$ denote the radial and binormal components of the complex FHPOD mode, respectively. Figure~\ref{fig:filteredHPOD_energy_over_S} shows that the \textit{SS} and \textit{SB} families are strongest inside the bend, whereas the \textit{DS} family is more prominent in the downstream region and mode~\textit{Ax} has the least cross-stream energy. By enforcing degenerate pairing through the Hilbert transform and ensuring spectral unimodality using filtering, we achieve a clean separation in both space and time, thereby isolating the modes that were previously embedded within others due to spectral mixing. The spectral separation leads to a physically consistent distinction between instabilities occurring inside the bend and those developing downstream. This is illustrated in figures~\ref{fig:SS_modes}, \ref{fig:SB_modes} and~\ref{fig:DownstreamMode}, and by comparison with the classical POD in figure~\ref{fig:Psi_POD_HPOD}. In the classical POD results, downstream structures appear phase-shifted, while the phase relation of the structures within the bend is inconsistent. Furthermore, the streamwise length of these structures inside and downstream of the bend differs, indicating the presence of different physical mechanisms.
	
	\subsection{Discussion of different modes}
	
\begin{figure}[!htbp]
		\centering
		\begin{subfigure}[t]{0.49\textwidth}
			\begin{overpic}[width=\linewidth]{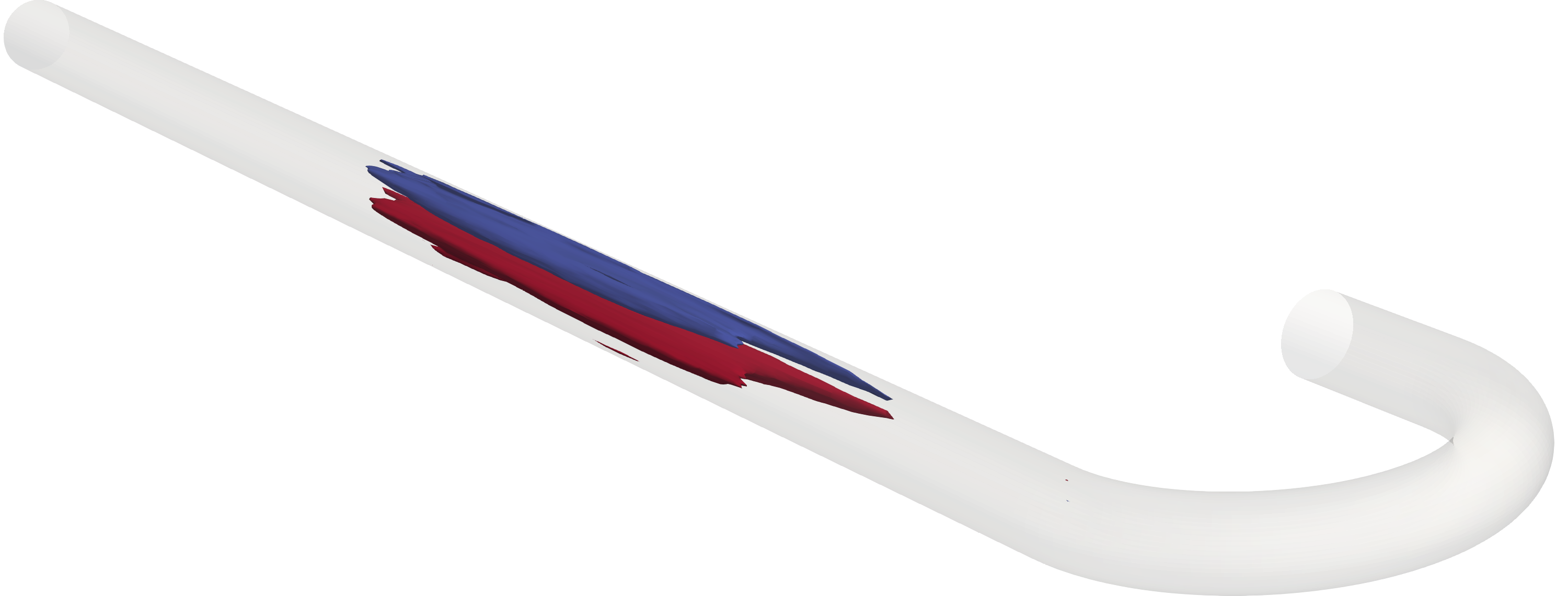}
				\inletplanearrow{62}{17}{1.25}{-.6}
				\put(48,-3){\small\bfseries (a)}
			\end{overpic}
		\end{subfigure}\hfill
		\begin{subfigure}[t]{0.49\textwidth}
			\begin{overpic}[width=\linewidth]{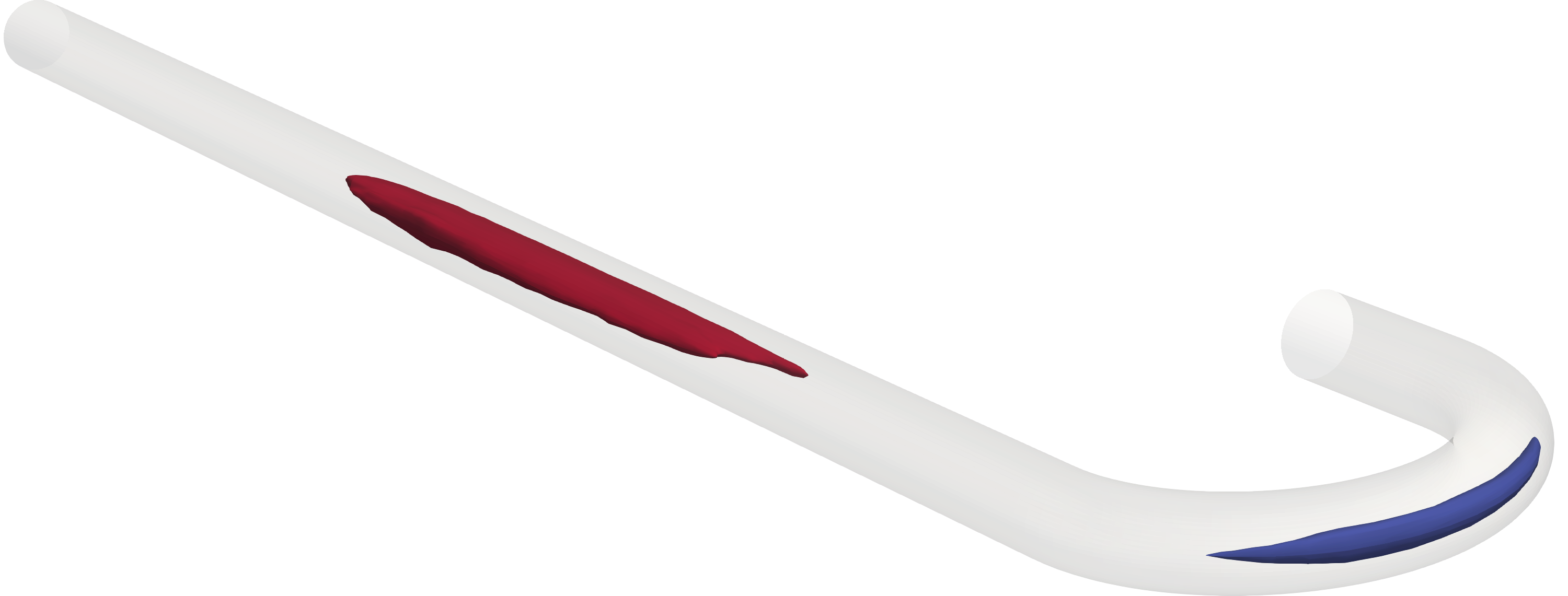}
				\put(48,-3){\small\bfseries (b)}
			\end{overpic}
		\end{subfigure}
		\caption{Phase-aligned signed isosurfaces for mode~\textit{Ax}: \textit{(a)} the streamwise modal velocity $u^{\xi}_{\textit{Ax},s}=\pm0.25\,U_b$ and \textit{(b)} the modal streamfunction $\Psi_j=\pm0.015\,U_bD$.}
		\label{fig:axialMode}
	\end{figure}

	Figure~\ref{fig:axialMode} shows mode~\textit{Ax}, the lowest-frequency mode at $St\approx0.03$, which is the second most energetic mode in the present study (see table~\ref{tab:HPOD_modeFamily}). 
	Low-frequency modes at similar Strouhal numbers have previously been reported in several experimental studies \citep{Brucker1998, Sakakibara2010, KalpakliVester2015, Jain2019, Jain2022}. 
	The in-plane velocity components of this mode, shown in figure~\ref{fig:axialMode}\textit{(b)}, contribute to the swirl-switching of the Dean vortices. This behaviour was identified by \citet{Brucker1998} as a very low-frequency swirl-switching component in the power spectrum of the tangential velocity.
	However, the cross-stream energy of this mode is much smaller than that of other modes reported here (see figure~\ref{fig:filteredHPOD_energy_over_S}). In other words, most of the kinetic energy of this mode is contained in the axial modal component $u^{\xi}_{\textit{Ax},s}$ downstream of the bend, with comparatively smaller contributions from the in-plane modal components $u^{\xi}_{\textit{Ax},r}$ and $u^{\xi}_{\textit{Ax},y}$. For this reason, \textit{Ax} is classified separately from the swirl-switching modes denoted by \textit{SS}, although the weaker cross-stream component of mode~\textit{Ax} slightly modulates the Dean vortices similar to the swirl-switching modes. A very similar structure is observed in the stereo-PIV study of \citet{Sakakibara2010}, where POD was applied to three-dimensional velocity fields on 2D cross-sections. The streamwise component of their first POD mode (see figure~5 in \citet{Sakakibara2010}) closely resembles mode~\textit{Ax} in figure~\ref{fig:axialMode}\textit{(a)}, showing large low-speed and high-speed streaky structures downstream of the bend. Furthermore, the power spectrum of their POD time coefficient (see figure~4 in \citet{Sakakibara2010}) shows a broad low-frequency peak for the first POD time coefficient around $St\approx0.02$--$0.04$, which closely matches the present value of $St\approx0.03$. This agreement is likely due to the fact that \citet{Sakakibara2010} measured all three velocity components using stereo-PIV and considered a sufficiently long downstream section, allowing this low-frequency mode to be captured. In several other experimental studies, however, the mode shapes are primarily represented using cross-stream velocity components, either due to the use of 2D measurements or visualisation choices. 
	In such cases, this mode is often overlooked or underrepresented because most of its energy is contained in the streamwise velocity component downstream of the bend. Additionally, mode~\textit{Ax} is most prominent in the downstream section and exhibits a large streamwise wavelength of approximately $22D$. Therefore, it can only be fully captured if the computational or measurement domain extends sufficiently far downstream of the bend and the simulation time is long enough to resolve such low-frequency content. When the downstream measurement section is too short, this mode may not fully develop and, therefore, remains weak or entirely absent. Especially when classical POD is used, even small spectral mixing from other instabilities that contain much stronger in-plane velocity components can further obscure this mode. To the best of our knowledge, this is the first time this mode has been clearly extracted in a numerical simulation.

	\begin{figure}[!htbp]
		\centering
		\begin{subfigure}[t]{0.49\textwidth}
			\begin{overpic}[width=\linewidth]{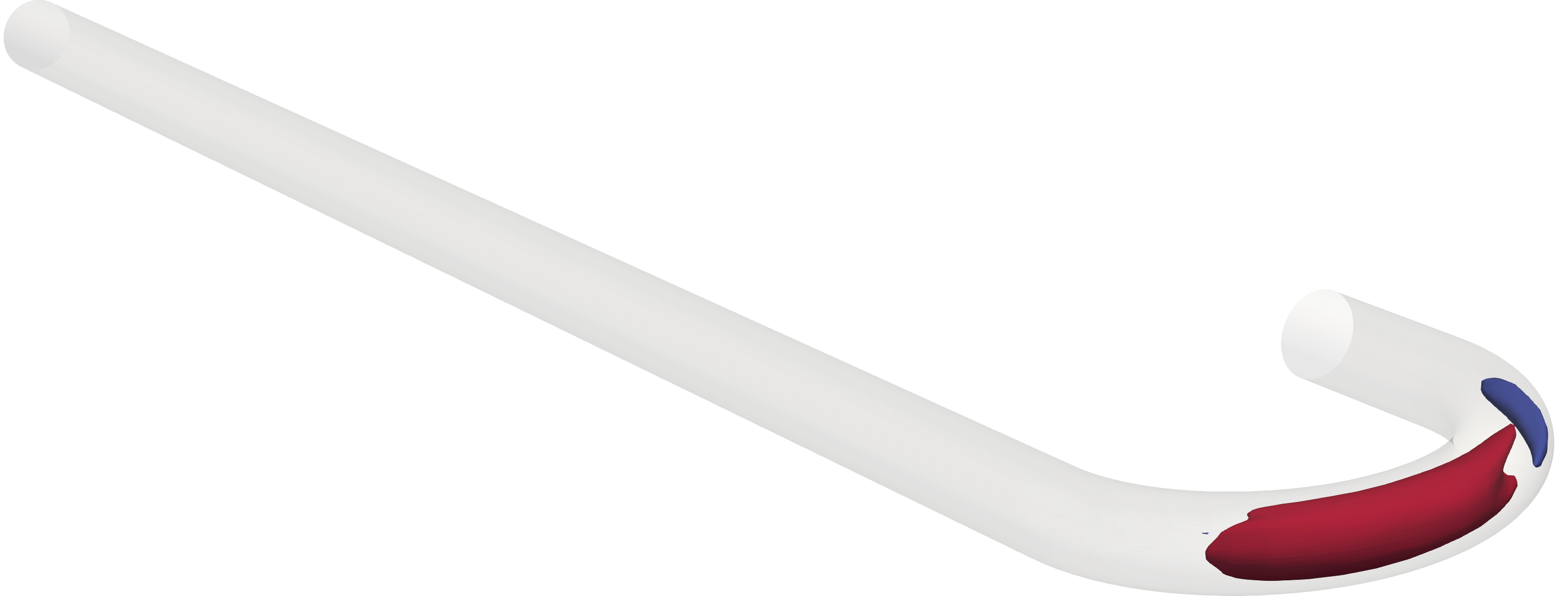}
				\inletplanearrow{62}{17.2}{1.25}{-.6}
				\put(48,-3){\small\bfseries (a)}
			\end{overpic}
		\end{subfigure}\hfill
		\begin{subfigure}[t]{0.49\textwidth}
			\begin{overpic}[width=\linewidth]{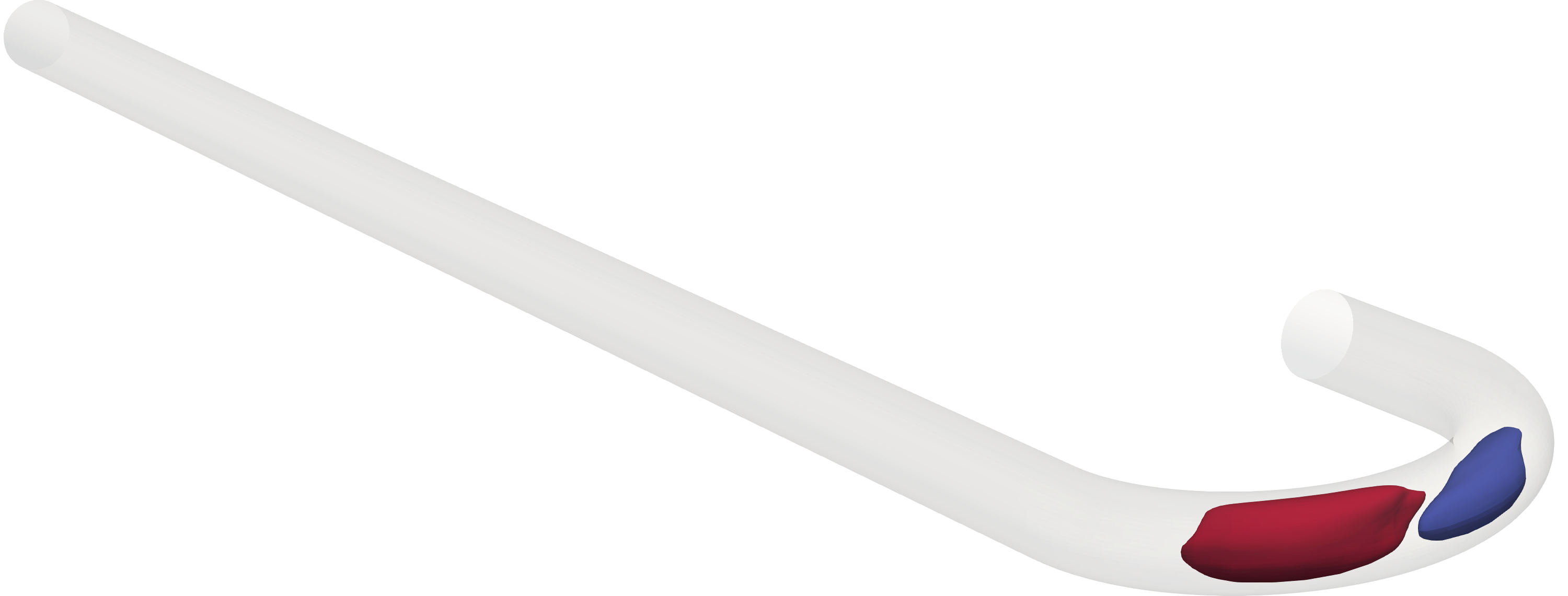}
				\put(48,-3){\small\bfseries (b)}
			\end{overpic}
		\end{subfigure}
		\caption{Phase-aligned signed streamfunction isosurfaces for the swirl-switching modes: \textit{(a)} mode~\textit{SS1} at $\Psi_j=\pm0.025\,U_bD$ and \textit{(b)} mode~\textit{SS2} at $\Psi_j=\pm0.025\,U_bD$.}
		\label{fig:SS_modes}
	\end{figure}
	
	\begin{figure}[!htbp]
		\centering
		\begin{subfigure}[t]{0.49\textwidth}
			\begin{overpic}[width=\linewidth]{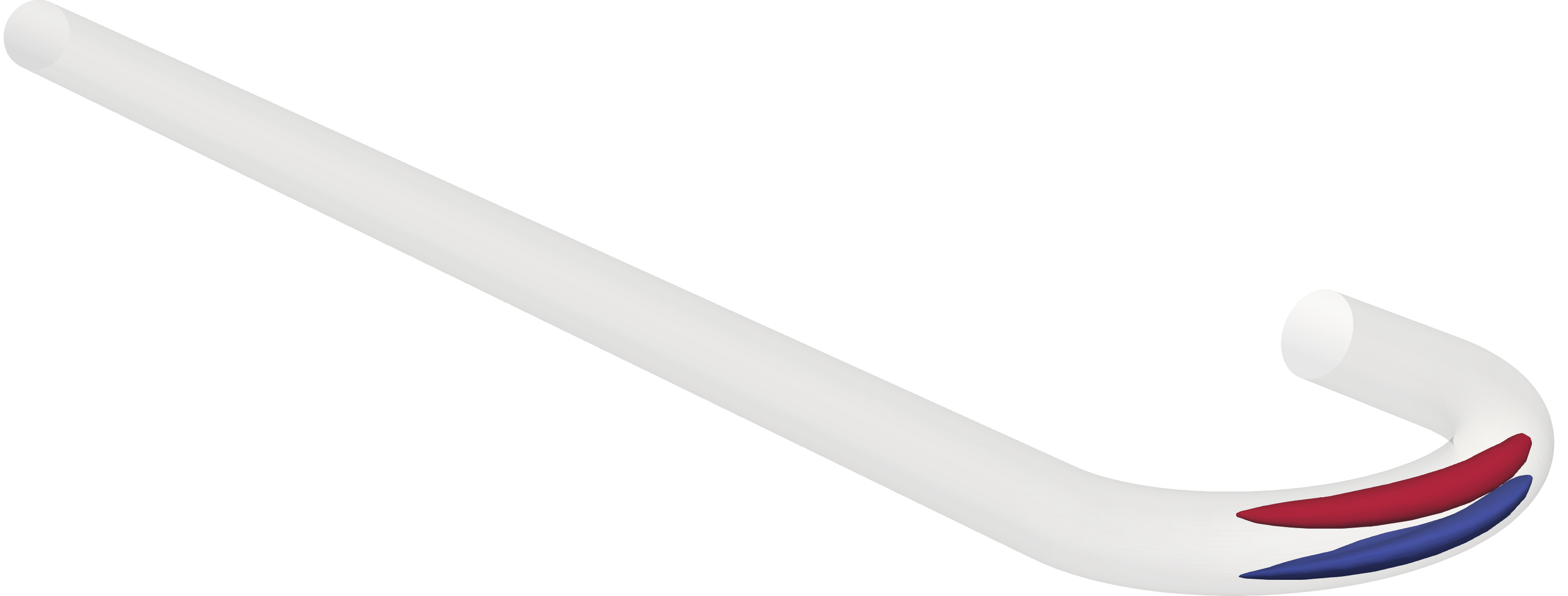}
				\put(48,-3){\small\bfseries (a)}
			\end{overpic}
		\end{subfigure}\hfill
		\begin{subfigure}[t]{0.49\textwidth}
			\begin{overpic}[width=\linewidth]{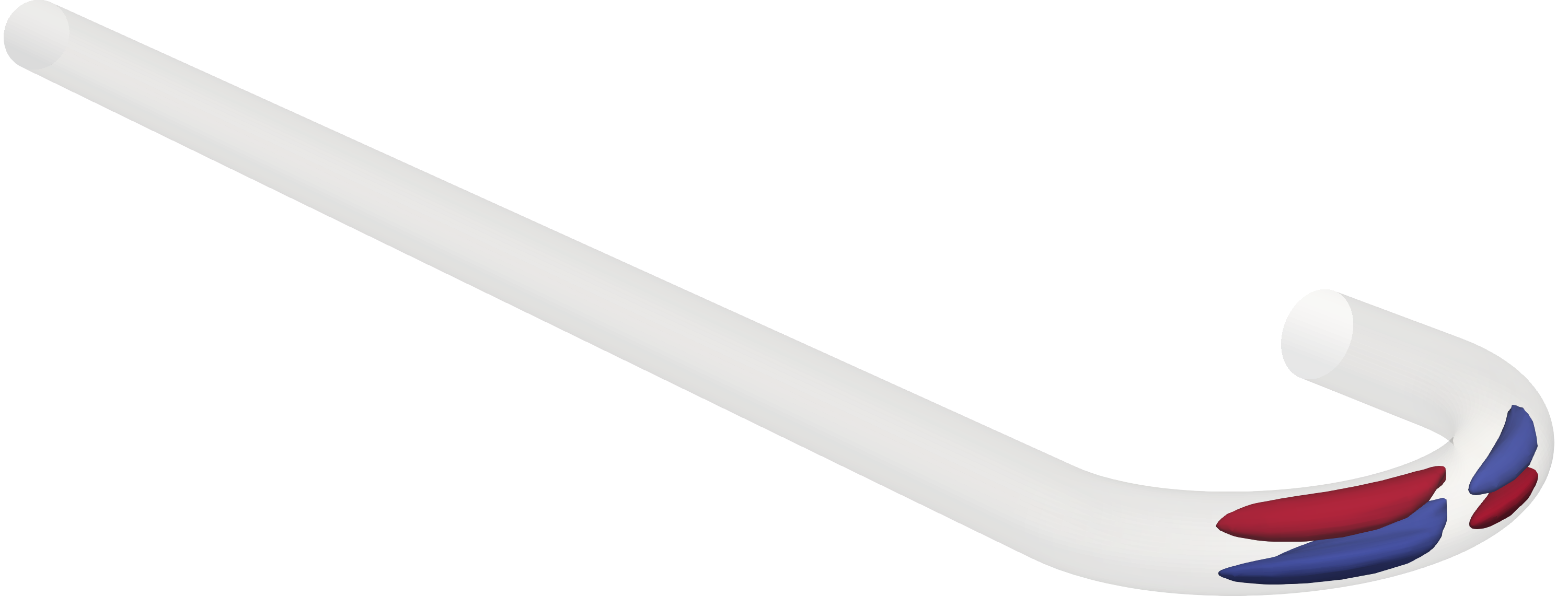}
				\put(48,-3){\small\bfseries (b)}
			\end{overpic}
		\end{subfigure}
		\caption{Phase-aligned signed streamfunction isosurfaces for the swirl-breathing modes: \textit{(a)} mode~\textit{SB1} at $\Psi_j=\pm0.025\,U_bD$ and \textit{(b)} mode~\textit{SB2} at $\Psi_j=\pm0.025\,U_bD$.}
		\label{fig:SB_modes}
	\end{figure}
	
	Figure~\ref{fig:SS_modes} shows the swirl-switching modes \textit{SS1} ($St\approx0.13$) and \textit{SS2} ($St\approx0.25$), whereas figure~\ref{fig:SB_modes} shows the swirl-breathing modes \textit{SB1} ($St\approx0.12$) and \textit{SB2} ($St\approx0.25$). All four modes are confined primarily to the bend and decay rapidly downstream. The \textit{SS} family is sinuous and describes lateral displacement of the Dean vortices. The \textit{SB} family is varicose and describes the simultaneous strengthening and weakening of the two vortices without lateral switching. Their spatial extent is consistent with the Dean-vortex region in figure~\ref{fig:meanPsi}\textit{(b)}, suggesting that both symmetry classes are curvature-driven instabilities of the Dean vortices. Mode~\textit{SS1} closely corresponds to the structure identified by \citet{Brucker1998} at $St\approx0.12$ and resembles the swirl-switching mode reported by \citet{Noorani2016}. As shown in \S\ref{sec:lsa_result}, all four FHPOD modes have corresponding unstable modes in the local stability analysis.
	
	\begin{figure}[!htbp]
		\centering
		\includegraphics[width=0.92\textwidth]{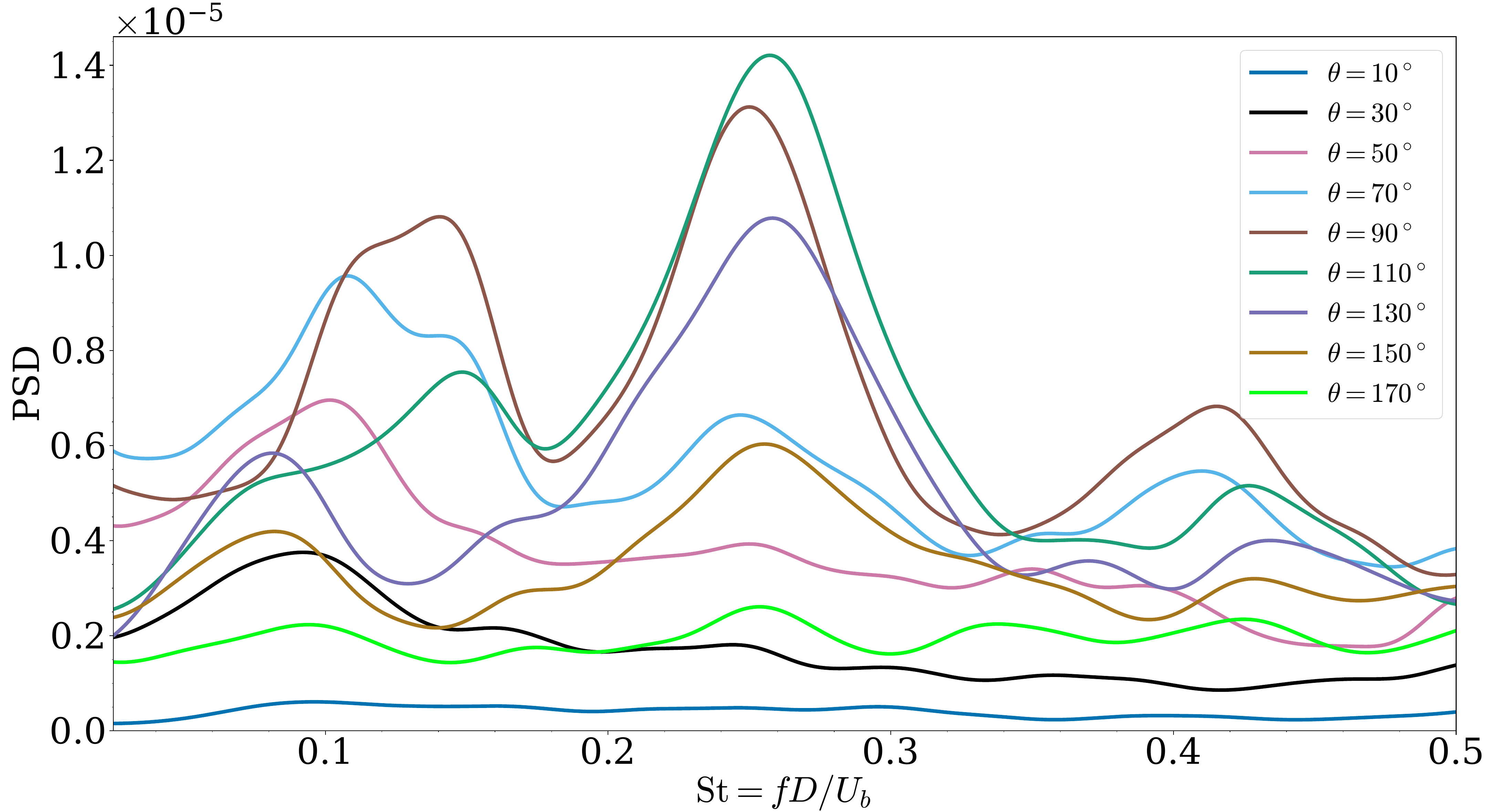}
		\caption{Power spectral density of the cross-sectional average of the streamfunction magnitude $\left\langle |\Psi^*| \right\rangle_{\mathcal A}(\phi,t)$ at selected bend angles.}
		\label{fig:swirl_varicose}
	\end{figure}
	
	The footprint of the swirl-breathing modes is also found in the instantaneous DNS field. Because modes~\textit{SB1} and~\textit{SB2} modulate the strength of the Dean vortices, these modulations are expected to affect the streamfunction at the same frequencies. Let $A_c(\phi)=|\mathcal A(\phi)|$ denote the area of the cross-section. To quantify the strength of the Dean vortices, we define
	
	\begin{equation}
		\left\langle |\Psi^*| \right\rangle_{\mathcal A}(\phi,t)
		=
		\frac{1}{A_c(\phi)}
		\iint_{\mathcal A(\phi)}
		\left|\Psi^*(r,y,\phi,t)\right|
		\mathrm{d}A.
		\label{eq:swirl_breathing_signal}
	\end{equation}%
	$\left\langle |\Psi^*| \right\rangle_{\mathcal A}(\phi,t)$ is the cross-sectional average of the streamfunction magnitude and $\Psi^*:=\Psi/(U_bD)$ is the nondimensional instantaneous streamfunction computed from the DNS. Taking the absolute value of the streamfunction removes the sign of the counter-rotating vortices. As a result, $\left\langle |\Psi^*| \right\rangle_{\mathcal A}(\phi,t)$ reflects their combined instantaneous intensity and can be used as a proxy for the strength of the Dean vortices. In a turbulent field, $\left\langle |\Psi^*| \right\rangle_{\mathcal A}(\phi,t)$ fluctuates over a broad range of frequencies, and its spectrum is expected to be broadband. However, figure~\ref{fig:swirl_varicose} shows two dominant peaks at $St=0.08$--$0.14$ and $St=0.24$--$0.25$ across various bend angles. At lower bend angles ($\phi \leq 70^{\circ}$), the lower-frequency peak is prominent. Near $\phi=90^{\circ}$, the higher-frequency peak prevails, reaches its maximum near $\phi=110^{\circ}$ and then weakens towards the bend exit. The peaks in the spectra closely agree with the dominant frequencies of modes~\textit{SB1} and~\textit{SB2}. The fact that both peaks quickly diminish near the bend exit also supports the conclusion that the swirl-breathing phenomenon is mostly localised to the curved section of the pipe, as shown in figure~\ref{fig:SB_modes}. 
	
	\begin{figure}[!htbp]
		\centering
		\begin{subfigure}[t]{0.49\textwidth}
			\begin{overpic}[width=\linewidth]{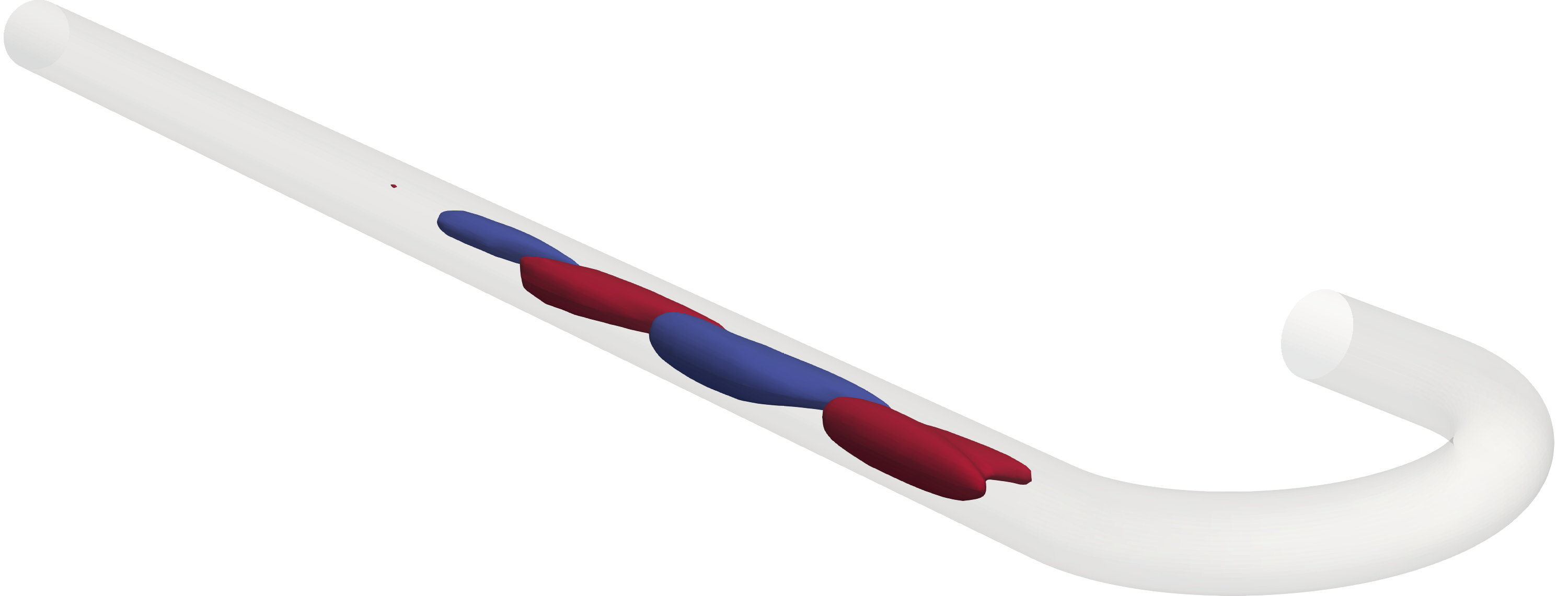}
				\inletplanearrow{62}{17.5}{1.25}{-.6}
				\put(48,-3){\small\bfseries (a)}
			\end{overpic}
		\end{subfigure}\hfill
		\begin{subfigure}[t]{0.49\textwidth}
			\begin{overpic}[width=\linewidth]{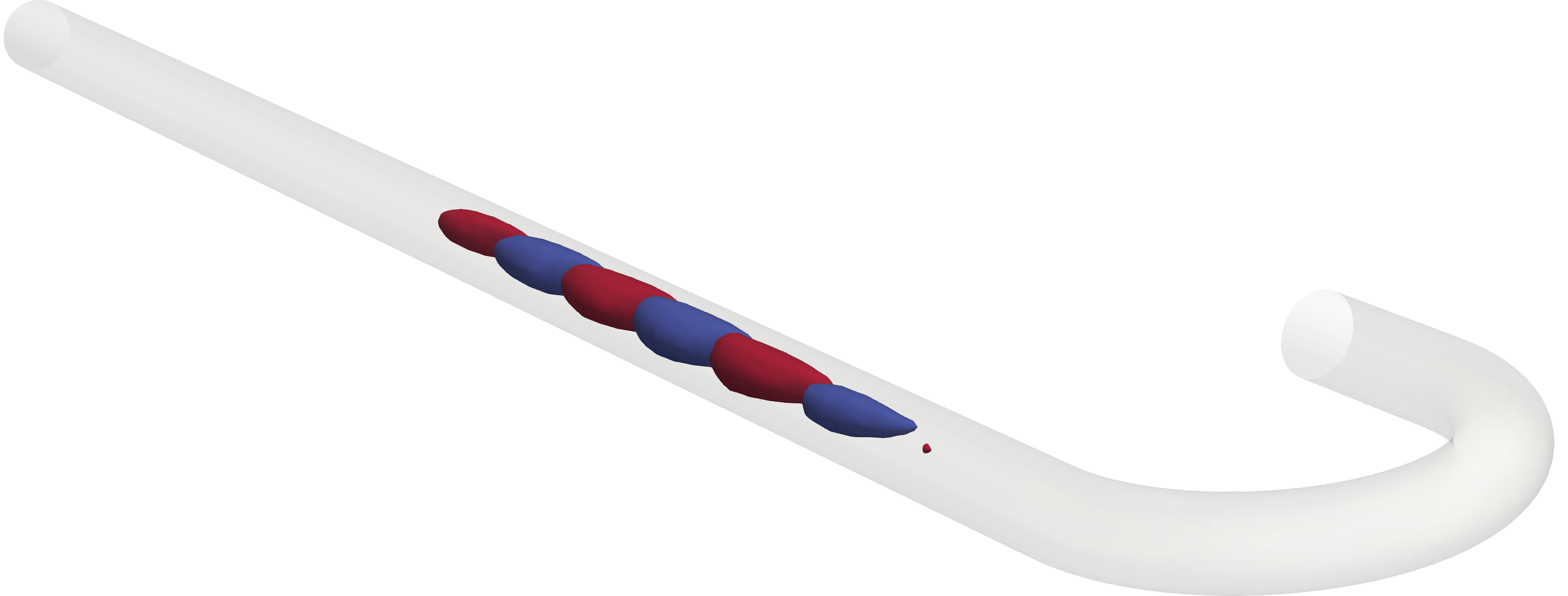}
				\put(48,-3){\small\bfseries (b)}
			\end{overpic}
		\end{subfigure}
		\caption{Phase-aligned signed streamfunction isosurfaces at $\Psi_j=\pm0.012\,U_bD$ for the downstream shear-layer modes: \textit{(a)} mode~\textit{DS1} and \textit{(b)} mode~\textit{DS2}.}
		\label{fig:DownstreamMode}
	\end{figure}
	
\begin{figure}[!htbp]
		\centering
		\begin{overpic}[width=0.7\linewidth]{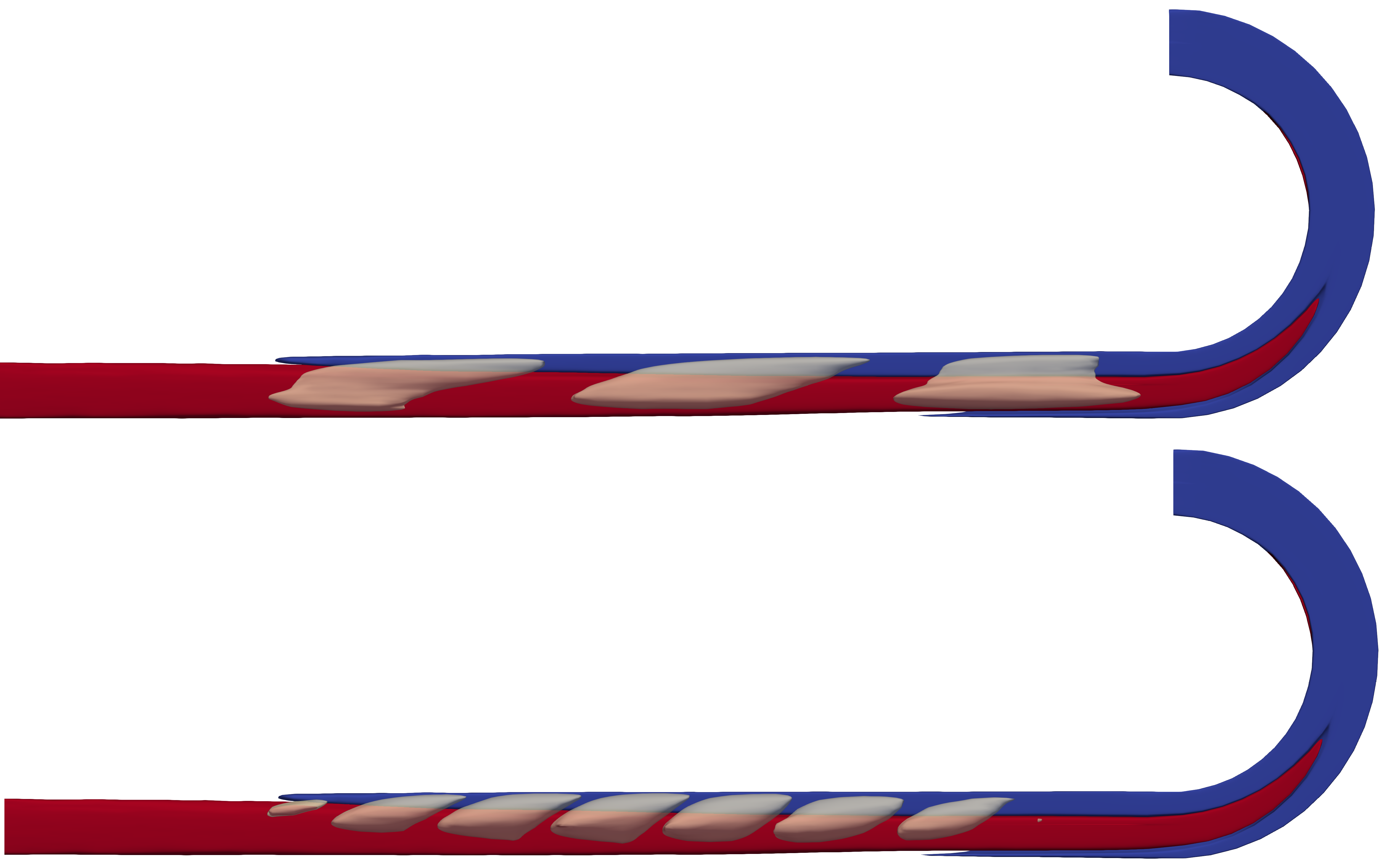}
			\inletplanearrow{62}{58}{2}{0.0}
			\put(45,40){\small\bfseries (a)}
			\put(45,8){\small\bfseries (b)}
		\end{overpic}
		\caption{Isosurfaces of the modal streamfunction $\Psi_j = \pm 0.015\, U_b D$ for \textit{(a)} mode~\textit{DS1} and \textit{(b)} mode~\textit{DS2}, shown in semi-transparent white and overlaid on the mean-flow isosurfaces of $\overline{\Psi} = \pm 5.0 \times 10^{-4}\, U_b D$ in the lower equatorial half of the bend ($y \le 0$). Red indicates a rotation axis aligned with $\hat{\mathbf{s}}$, whereas blue indicates the opposite orientation. Both modes are located near the interface between the counter-rotating four-vortex (two vortex pairs) mean flow structure in the downstream section, where a strong shear layer forms. The modes vanish as the shear layer weakens and the mean flow transitions to a two-vortex configuration.}
		\vspace{2mm}
		\label{fig:DS1_DS2_shearLayer}
	\end{figure}

	Figure~\ref{fig:DownstreamMode} shows the instabilities located in the downstream section denoted by \textit{DS1} and \textit{DS2}, respectively. \citet{Hufnagel2018} reported a variant of \textit{DS1} at $St\approx0.16$ as the dominant mode in a $90^\circ$ bend, while a higher-frequency counterpart (\textit{DS2}) is also visible in the PSD of the time coefficients. Similarly, both \textit{DS1} and \textit{DS2} appear as dominant modes in the $180^\circ$ bend study by \citet{Lupi2025}, with \textit{DS2} being the more energetic of the two. The \textit{DS} family has only been observed in spatially developing $90^{\circ}$ and $180^{\circ}$ bends and has not been reported in toroidal pipes \citep{Noorani2016,Zhang2024}. Our unimodal FHPOD results reveal that \textit{DS1} and \textit{DS2} are confined to the downstream straight region and are distinct from the curvature-driven \textit{SS} family. As shown in figure~\ref{fig:DS1_DS2_shearLayer}, both modes are located in the shear layer formed between the counter-rotating vortex pairs of the mean flow (see the four-vortex configuration in figure~\ref{fig:meanPsi}(a)). Therefore, we refer to these two modes as the \emph{downstream shear-layer} modes. In contrast to \textit{SS1} and \textit{SS2}, which dominate inside the bend and later decay, \textit{DS1} and \textit{DS2} are weak near the bend and prevail farther downstream. This is also visible in \citet{Brucker1998}, in which both peaks at $St\approx0.15$ and $St\approx0.3$ are present in the power spectrum but with much smaller amplitudes than the swirl-switching mode at $1.5D$ downstream of the bend (see figure~3 in \cite{Brucker1998}). A similar streamwise evolution was reported by \citet{Jain2019}, who observed distinct peaks near $St\approx0.07$ on the $70^{\circ}$ and $80^{\circ}$ planes and $0.25D$ downstream, with smaller peaks near $St\approx0.13$ on the latter two planes, while at $1D$ downstream the swirling mode had peaks at $St\approx0.03$ and $St\approx0.17$. Consistently, \citet{Jain2022} reported that the contribution of the swirling mode decays as the flow evolves into the downstream pipe and identified a peak around $St\approx0.16$ in the downstream section. Measurements by \citet{Hellstrom2013} at $5D$, $12D$, and $18D$ downstream of a $90^{\circ}$ bend similarly reported dominant downstream modes at $St\approx0.16$ and $St\approx0.33$. The spatial location of these modes suggests that the \textit{SS} and \textit{DS} families arise from different physical mechanisms. Modes~\textit{SS1} and \textit{SS2} are curvature-driven and dominant inside the bend, whereas \textit{DS1} and \textit{DS2} appear to be related to downstream shear-layer instabilities of the four-vortex mean-flow structure. As the inner-wall vortex pair decays farther downstream and the mean flow transitions to a two-vortex configuration, both \textit{DS1} and \textit{DS2} vanish as the shear layer weakens.

	\FloatBarrier
	\subsection{Effect of Reynolds number and curvature on the FHPOD modes}
	\label{sec:additional_cases}
	
	The reference case in this study has $Re_D=10\,000$ and $\gamma=0.2$. To investigate whether the identified modes are specific to this case, we perform DNS for two additional $180^{\circ}$ cases at $Re_D=5300$. The first case has the same curvature $\gamma=0.2$ as the investigated reference case and is used to examine the effect of Reynolds number on the identified modes. The second case has a curvature $\gamma=0.4$ and is used to examine the effect of curvature at fixed Reynolds number. All three cases use the same number of snapshots ($N_{\mathrm{snap}}=5374$), sampling interval ($\Delta t=0.15D/U_b$) and temporal extent ($805.95D/U_b$). The additional cases are decomposed using FHPOD, and the corresponding modes are shown below. All seven modes found in the reference case are also identified in these two additional cases. 
	
	\begin{figure}[!htbp]
		\centering
		\begin{subfigure}[t]{0.49\textwidth}
			\begin{overpic}[width=\linewidth]{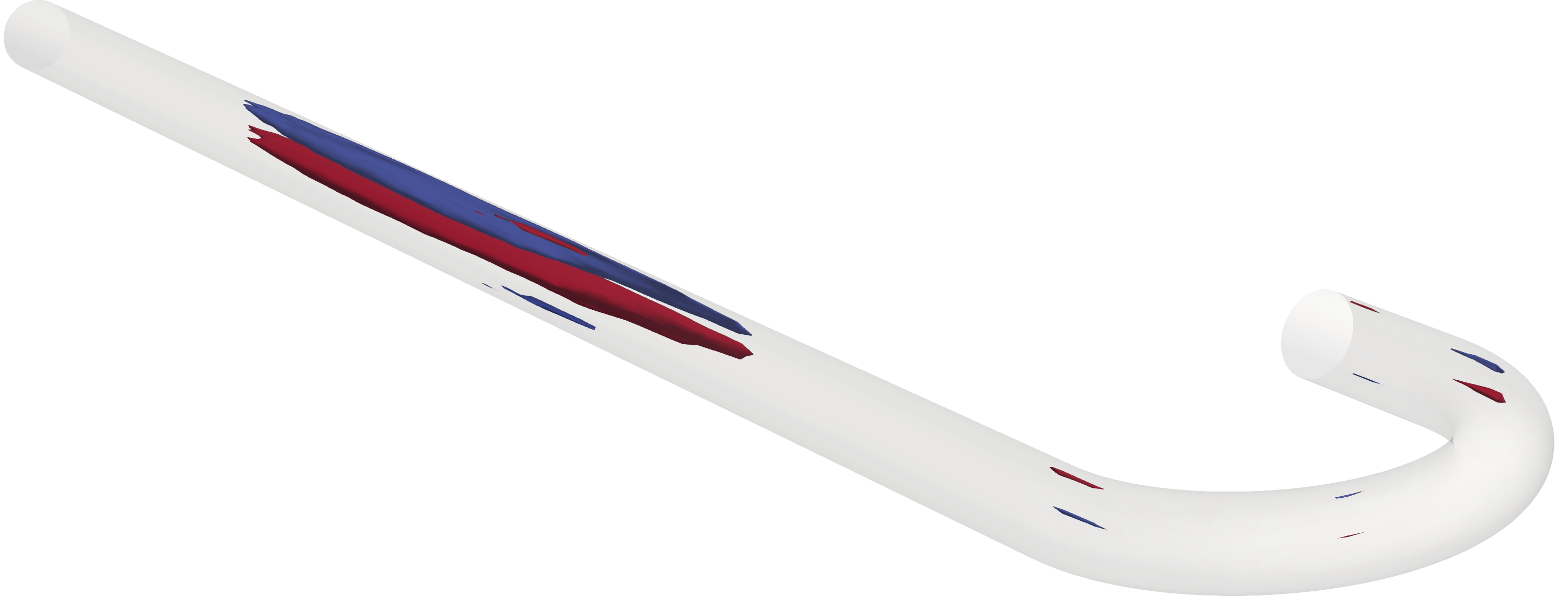}
				\inletplanearrow{62}{17.2}{1.25}{-.6}
				\put(48,-3){\small\bfseries (a)}
			\end{overpic}
		\end{subfigure}\hfill
		\begin{subfigure}[t]{0.49\textwidth}
			\begin{overpic}[width=\linewidth]{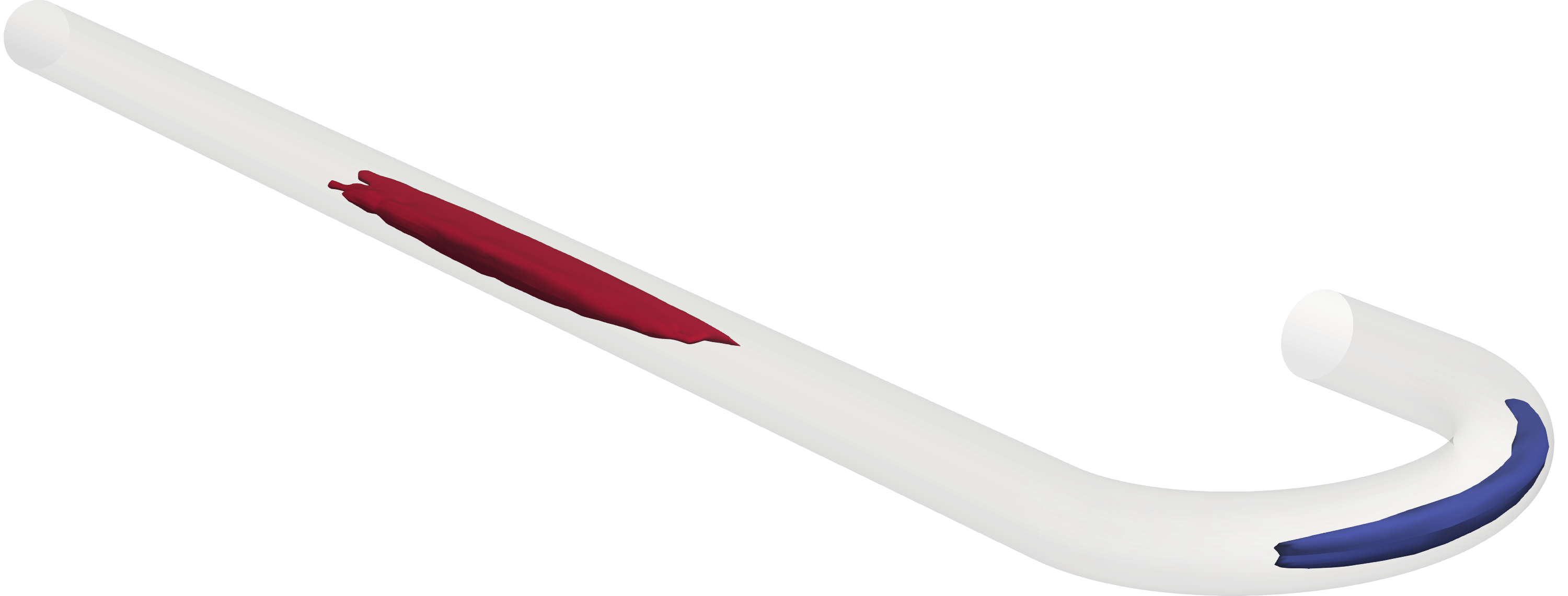}
				\put(48,-3){\small\bfseries (b)}
			\end{overpic}
		\end{subfigure}
		\caption{Phase-aligned signed isosurfaces for mode~\textit{Ax} ($St=0.03$) at $Re_D=5300$ and $\gamma=0.2$: \textit{(a)} the streamwise modal velocity $u^{\xi}_{\textit{Ax},s}=\pm0.25\,U_b$ and \textit{(b)} the modal streamfunction $\Psi_j=\pm0.015\,U_bD$.}
		\label{fig:additional_cases_gamma0p2_axial_mode}
	\end{figure}
	
	\begin{figure}[!htbp]
		\centering
		\begin{subfigure}[t]{0.49\textwidth}
			\begin{overpic}[width=\linewidth]{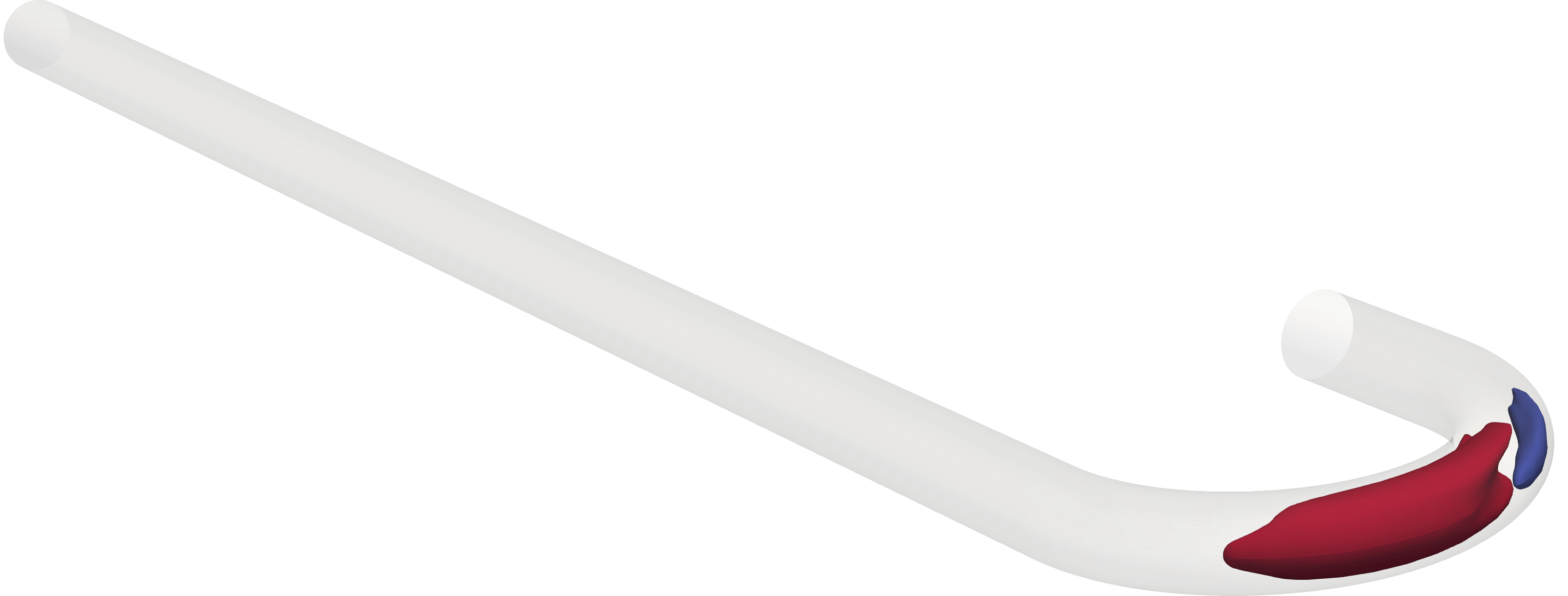}
				\put(48,-3){\small\bfseries (a)}
			\end{overpic}
		\end{subfigure}\hfill
		\begin{subfigure}[t]{0.49\textwidth}
			\begin{overpic}[width=\linewidth]{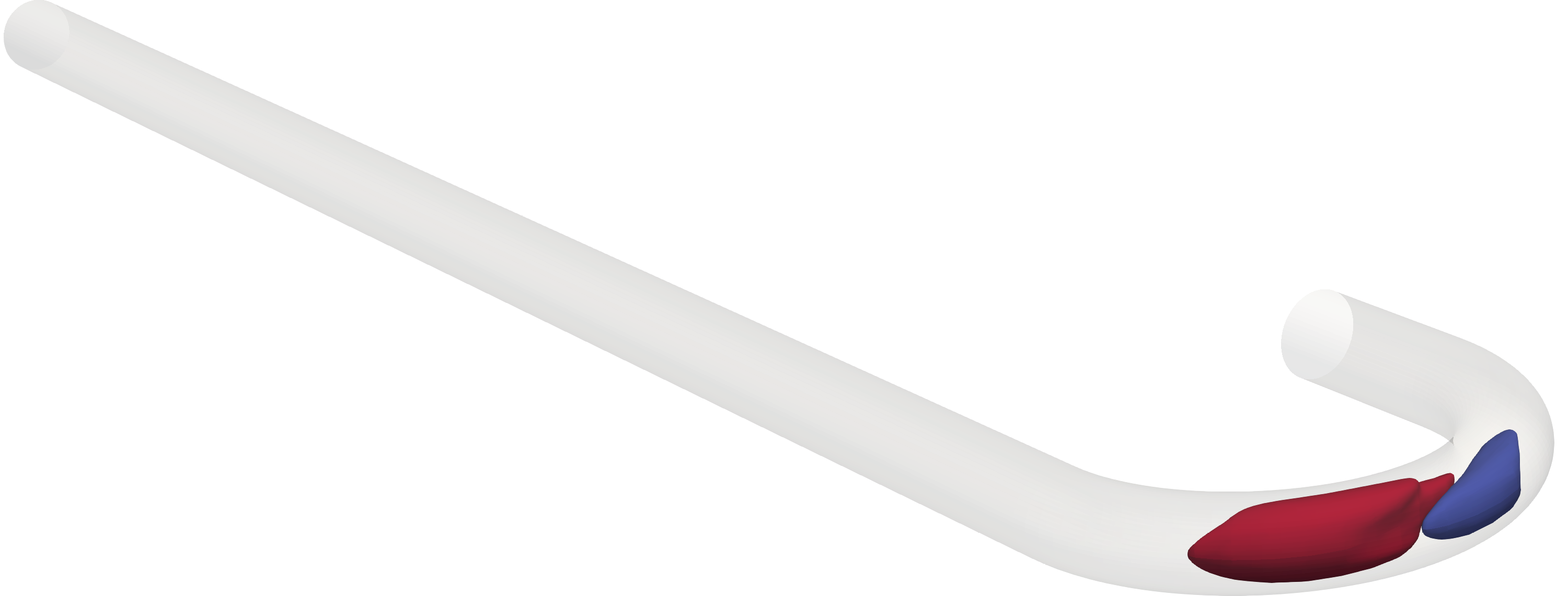}
				\put(48,-3){\small\bfseries (b)}
			\end{overpic}
		\end{subfigure}
		\caption{Phase-aligned signed streamfunction isosurfaces for the swirl-switching modes at $Re_D=5300$ and $\gamma=0.2$: \textit{(a)} mode~\textit{SS1} ($St=0.13$) at $\Psi_j=\pm0.025\,U_bD$ and \textit{(b)} mode~\textit{SS2} ($St=0.25$) at $\Psi_j=\pm0.025\,U_bD$.}
		\label{fig:additional_cases_gamma0p2_ss_modes}
	\end{figure}
	
	\begin{figure}[!htbp]
		\centering
		\begin{subfigure}[t]{0.49\textwidth}
			\begin{overpic}[width=\linewidth]{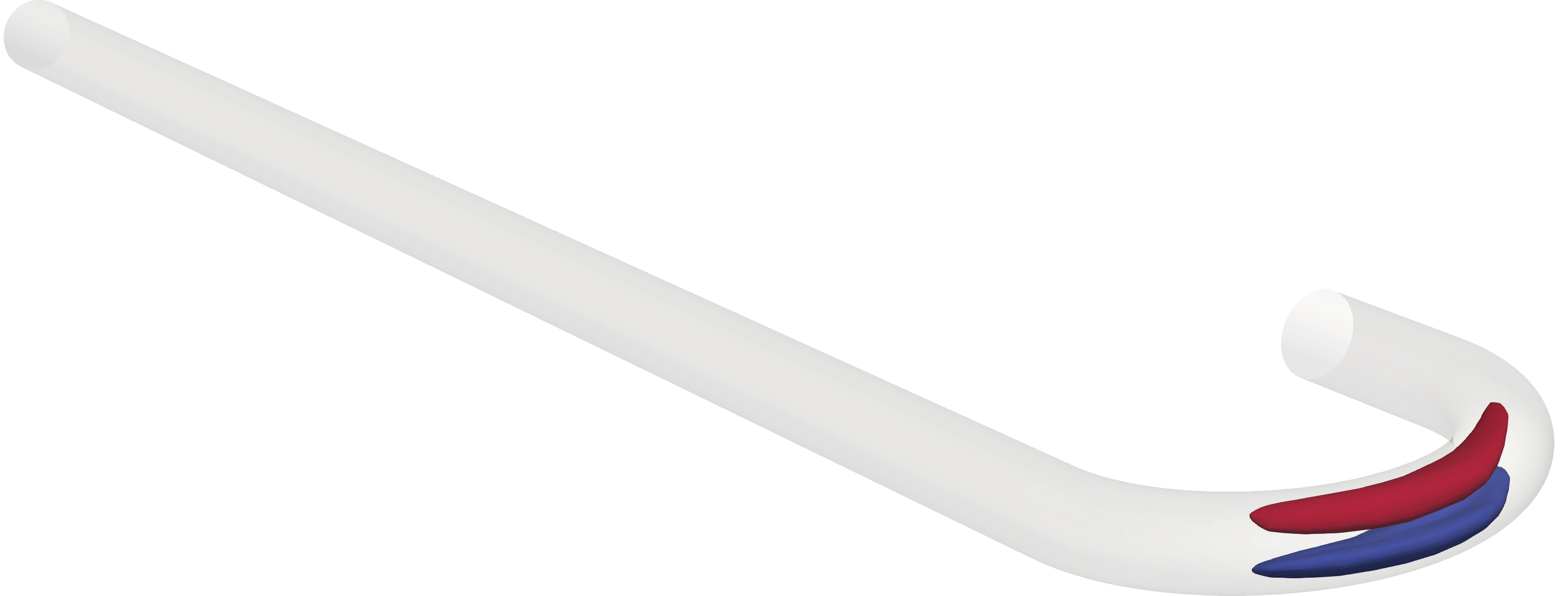}
				\put(48,-3){\small\bfseries (a)}
			\end{overpic}
		\end{subfigure}\hfill
		\begin{subfigure}[t]{0.49\textwidth}
			\begin{overpic}[width=\linewidth]{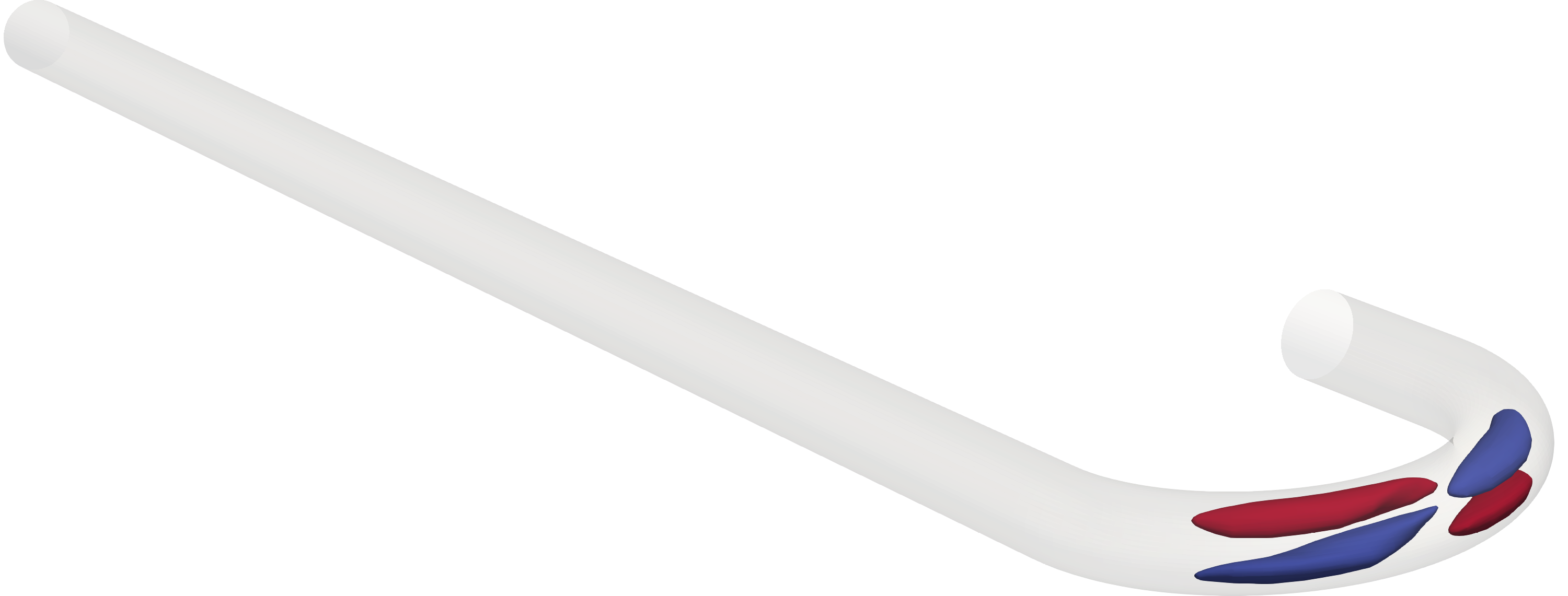}
				\put(48,-3){\small\bfseries (b)}
			\end{overpic}
		\end{subfigure}
		\caption{Phase-aligned signed streamfunction isosurfaces for the swirl-breathing modes at $Re_D=5300$ and $\gamma=0.2$: \textit{(a)} mode~\textit{SB1} ($St=0.13$) at $\Psi_j=\pm0.025\,U_bD$ and \textit{(b)} mode~\textit{SB2} ($St=0.27$) at $\Psi_j=\pm0.025\,U_bD$.}
		\label{fig:additional_cases_gamma0p2_sb_modes}
	\end{figure}
	
	\begin{figure}[!htbp]
		\centering
		\begin{subfigure}[t]{0.49\textwidth}
			\begin{overpic}[width=\linewidth]{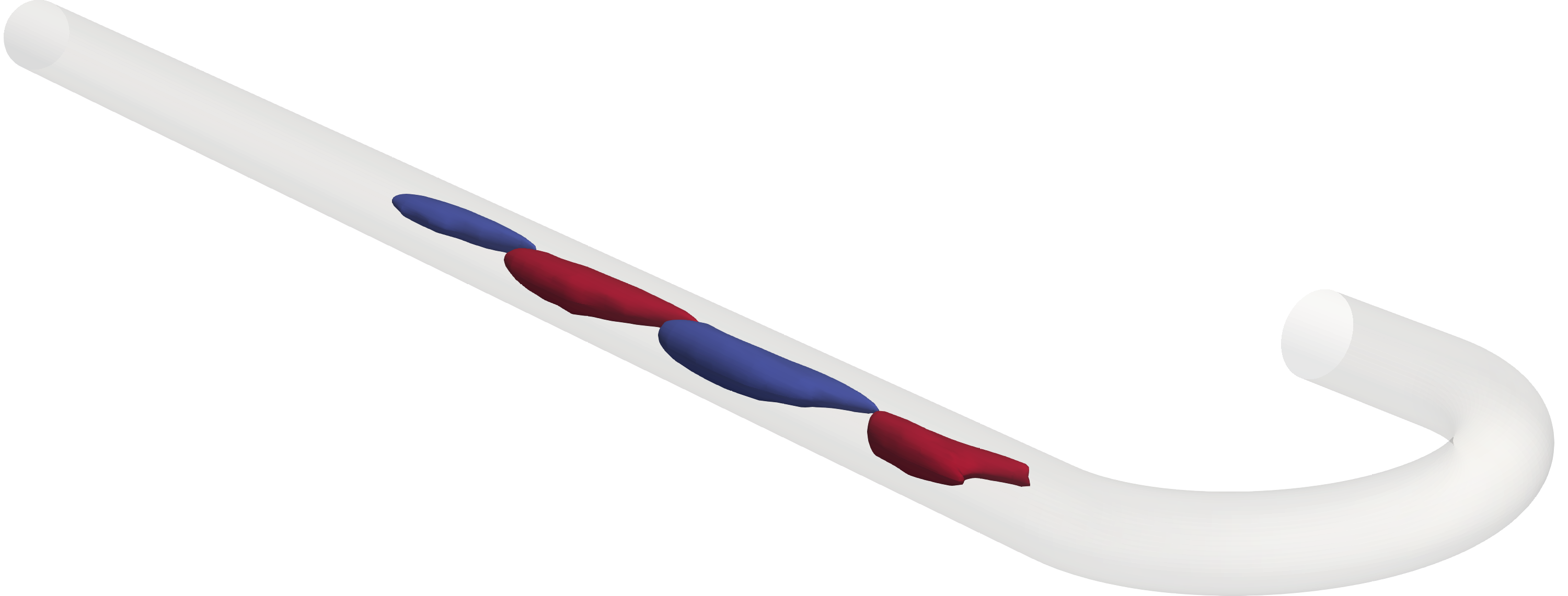}
				\put(48,-3){\small\bfseries (a)}
			\end{overpic}
		\end{subfigure}\hfill
		\begin{subfigure}[t]{0.49\textwidth}
			\begin{overpic}[width=\linewidth]{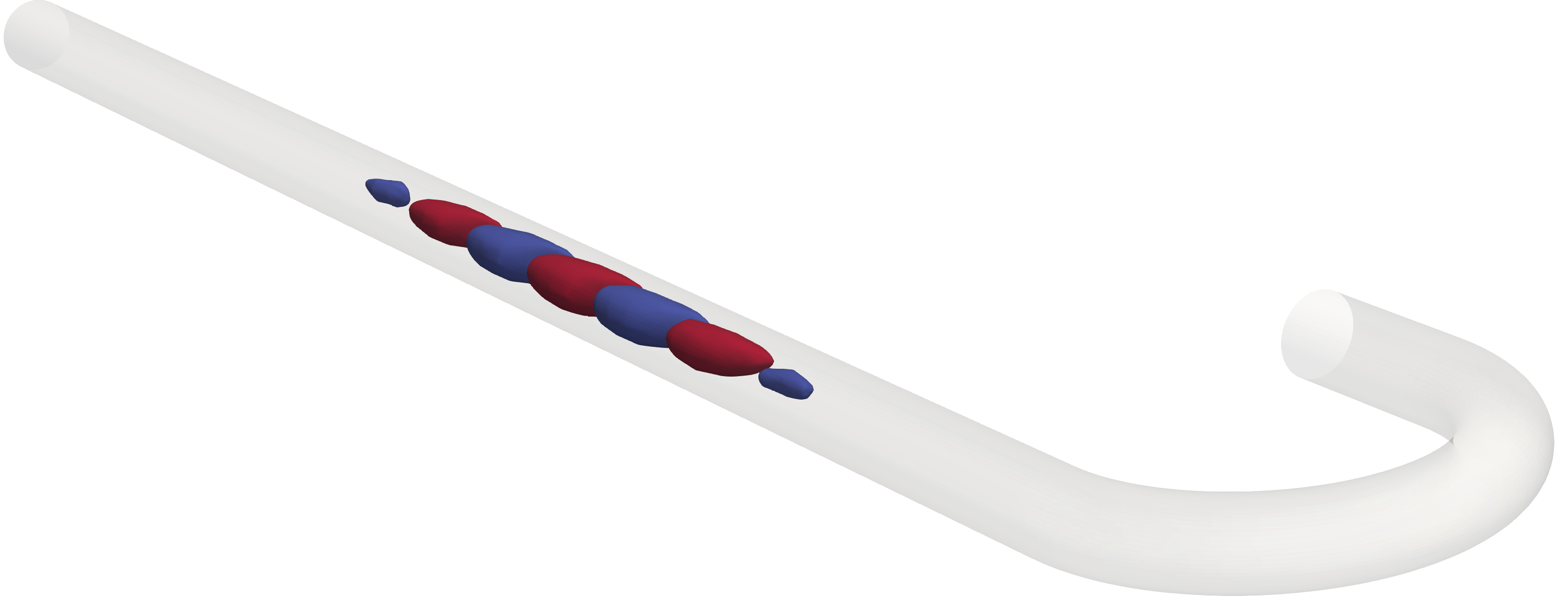}
				\put(48,-3){\small\bfseries (b)}
			\end{overpic}
		\end{subfigure}
		\caption{Phase-aligned signed streamfunction isosurfaces at $\Psi_j=\pm0.015\,U_bD$ for the downstream shear-layer modes at $Re_D=5300$ and $\gamma=0.2$: \textit{(a)} mode~\textit{DS1} ($St=0.16$) and \textit{(b)} mode~\textit{DS2} ($St=0.32$).}
		\label{fig:additional_cases_gamma0p2_downstream_modes}
	\end{figure}
	
	\begin{figure}[!htbp]
		\centering
		\begin{subfigure}[t]{0.49\textwidth}
			\begin{overpic}[width=\linewidth]{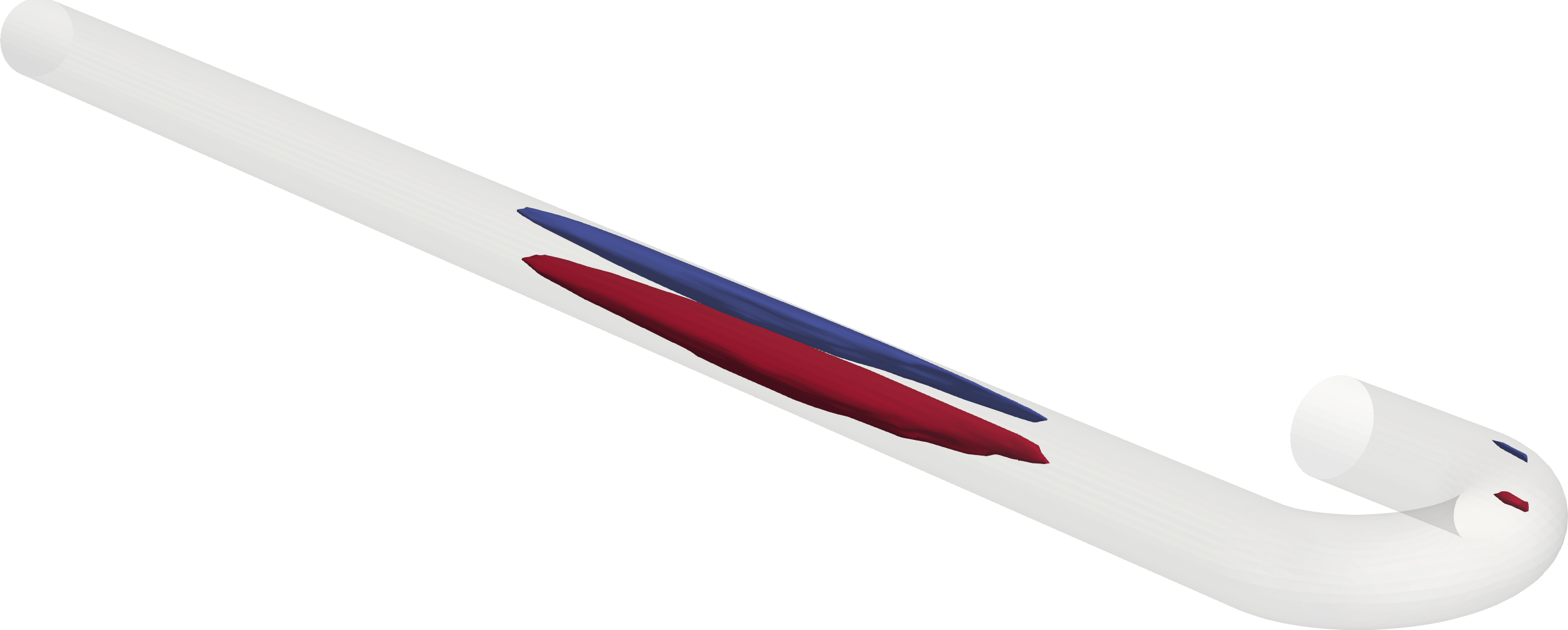}
				\inletplanearrow{62}{14}{1.25}{-.6}
				\put(48,-3){\small\bfseries (a)}
			\end{overpic}
		\end{subfigure}\hfill
		\begin{subfigure}[t]{0.49\textwidth}
			\begin{overpic}[width=\linewidth]{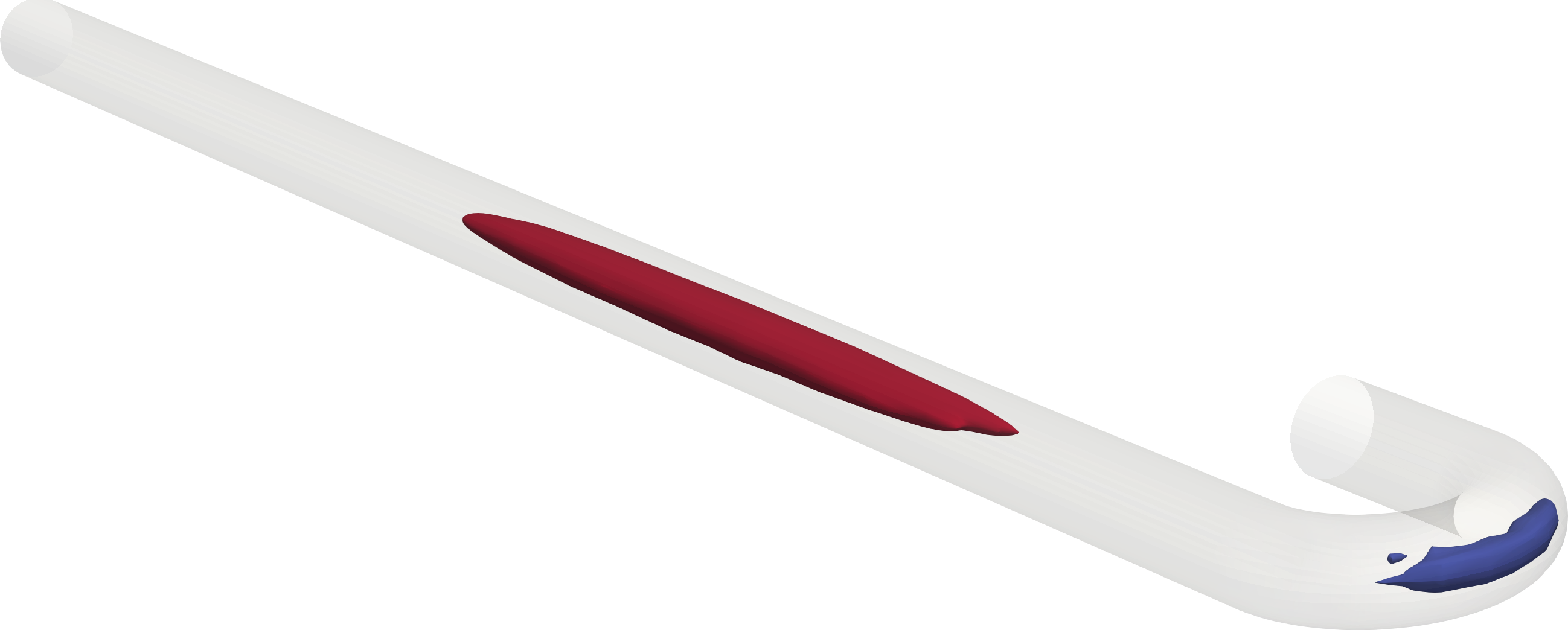}
				\put(48,-3){\small\bfseries (b)}
			\end{overpic}
		\end{subfigure}
		\caption{Phase-aligned signed isosurfaces for mode~\textit{Ax} ($St=0.03$) at $Re_D=5300$ and $\gamma=0.4$: \textit{(a)} the streamwise modal velocity $u^{\xi}_{\textit{Ax},s}=\pm0.3\,U_b$ and \textit{(b)} the modal streamfunction $\Psi_j=\pm0.02\,U_bD$.}
		\label{fig:additional_cases_gamma0p4_axial_mode}
	\end{figure}
	
	\begin{figure}[!htbp]
		\centering
		\begin{subfigure}[t]{0.49\textwidth}
			\begin{overpic}[width=\linewidth]{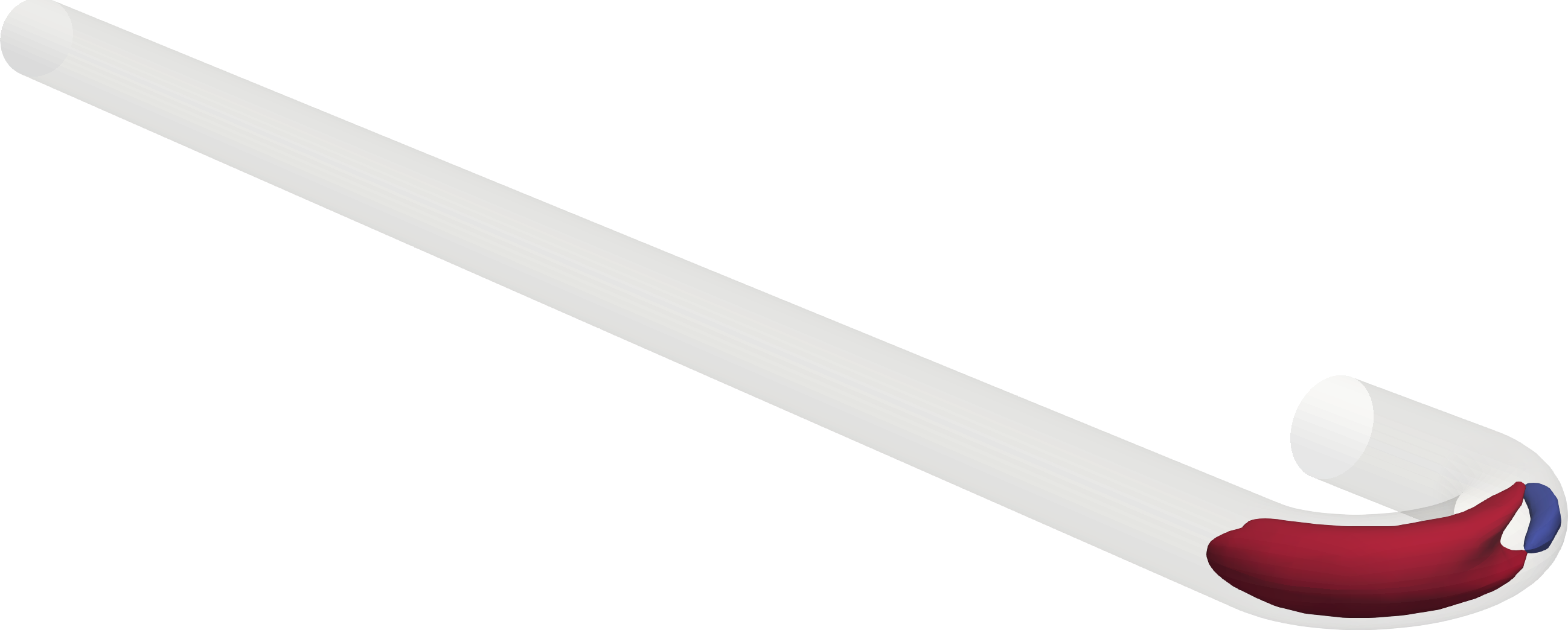}
				\put(48,-3){\small\bfseries (a)}
			\end{overpic}
		\end{subfigure}\hfill
		\begin{subfigure}[t]{0.49\textwidth}
			\begin{overpic}[width=\linewidth]{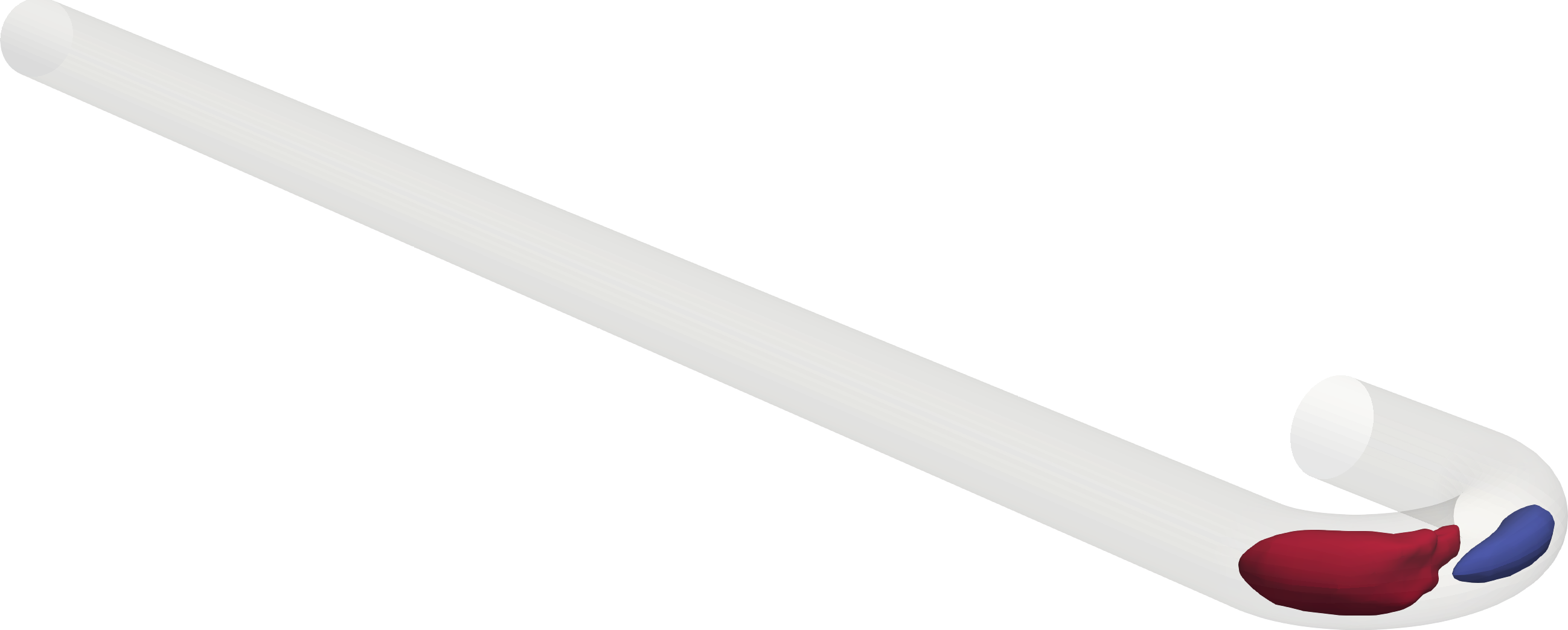}
				\put(48,-3){\small\bfseries (b)}
			\end{overpic}
		\end{subfigure}
		\caption{Phase-aligned signed streamfunction isosurfaces for the swirl-switching modes at $Re_D=5300$ and $\gamma=0.4$: \textit{(a)} mode~\textit{SS1} ($St=0.13$) at $\Psi_j=\pm0.03\,U_bD$ and \textit{(b)} mode~\textit{SS2} ($St=0.25$) at $\Psi_j=\pm0.03\,U_bD$.}
		\label{fig:additional_cases_gamma0p4_ss_modes}
	\end{figure}
	
	\begin{figure}[!htbp]
		\centering
		\begin{subfigure}[t]{0.49\textwidth}
			\begin{overpic}[width=\linewidth]{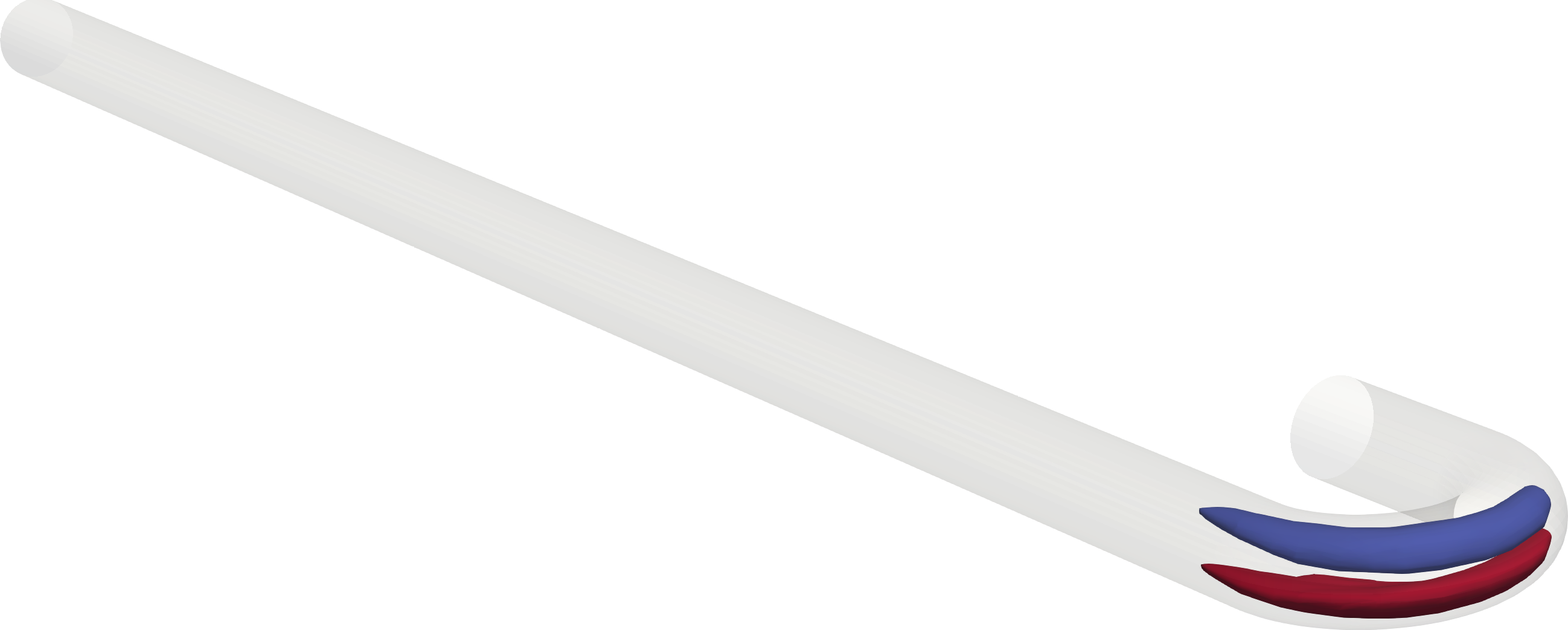}
				\put(48,-3){\small\bfseries (a)}
			\end{overpic}
		\end{subfigure}\hfill
		\begin{subfigure}[t]{0.49\textwidth}
			\begin{overpic}[width=\linewidth]{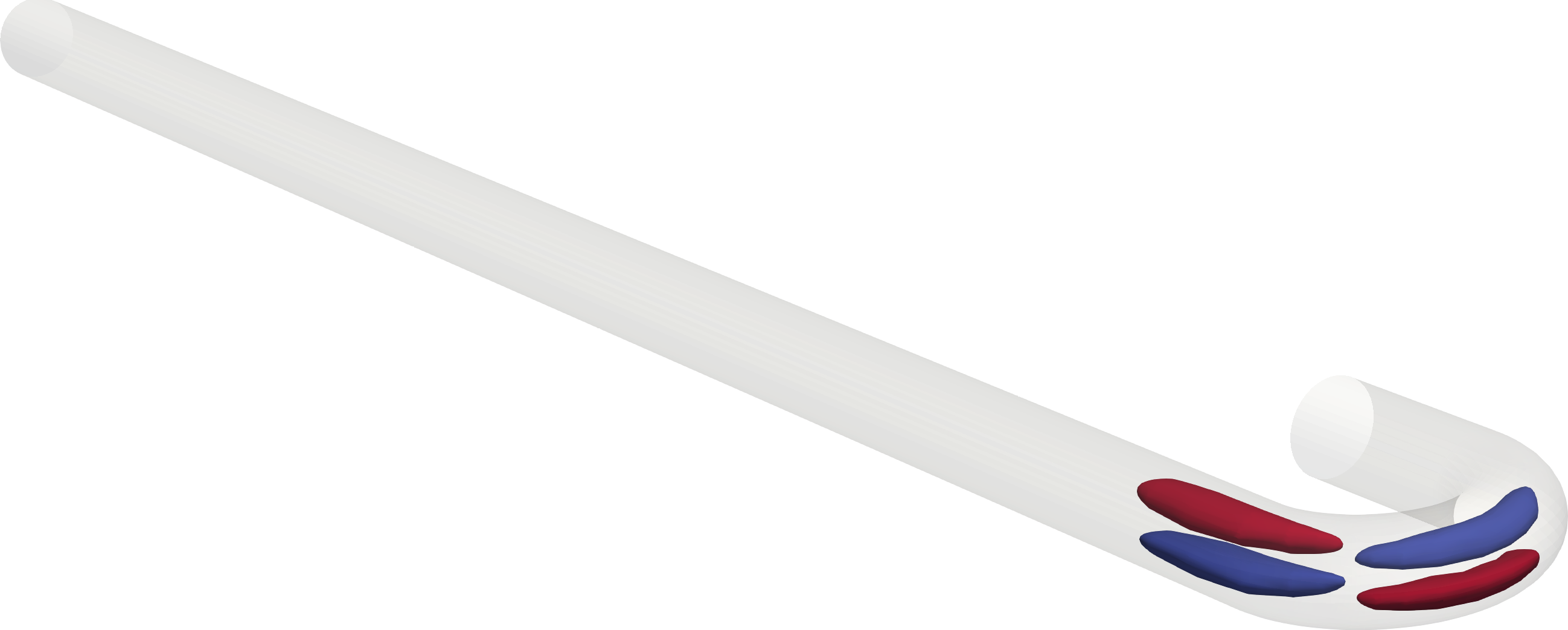}
				\put(48,-3){\small\bfseries (b)}
			\end{overpic}
		\end{subfigure}
		\caption{Phase-aligned signed streamfunction isosurfaces at $\Psi_j=\pm0.03\,U_bD$ for the swirl-breathing modes at $Re_D=5300$ and $\gamma=0.4$: \textit{(a)} mode~\textit{SB1} ($St=0.12$) and \textit{(b)} mode~\textit{SB2} ($St=0.24$).}
		\label{fig:additional_cases_gamma0p4_sb_modes}
	\end{figure}
	
	\begin{figure}[!htbp]
		\centering
		\begin{subfigure}[t]{0.49\textwidth}
			\begin{overpic}[width=\linewidth]{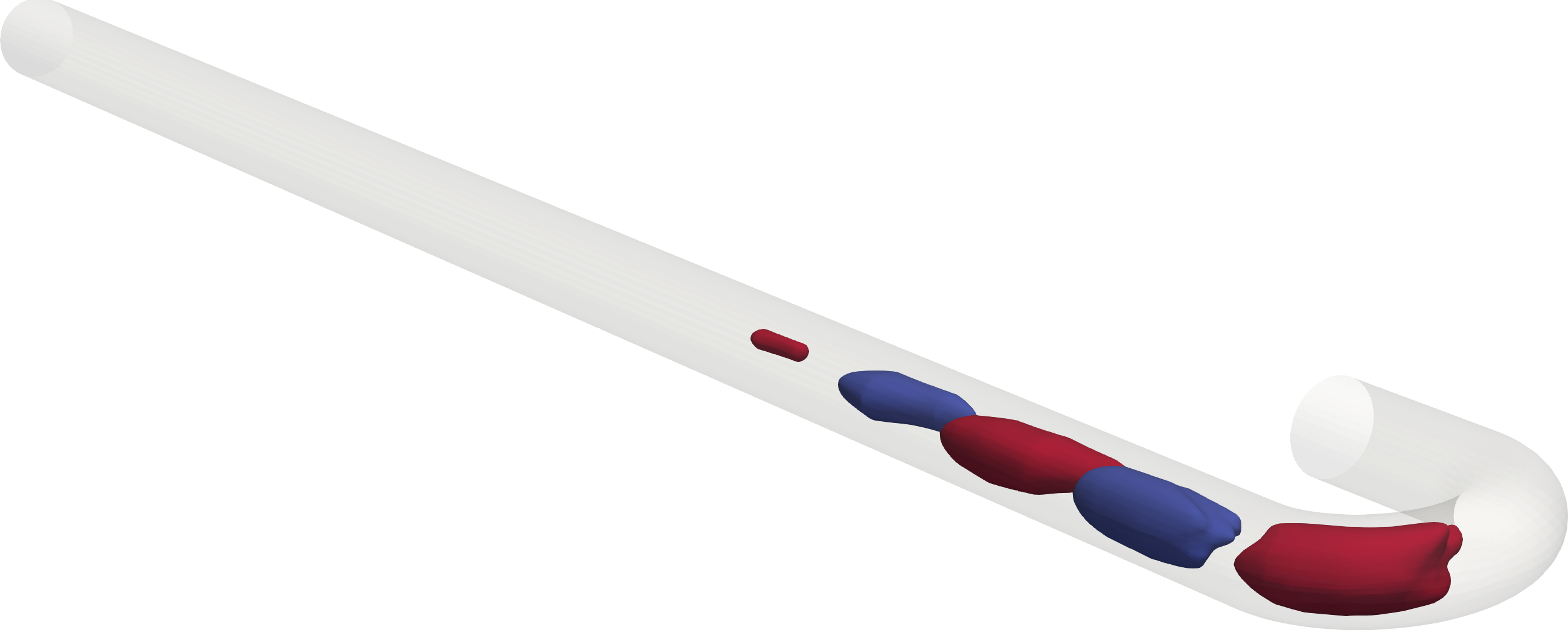}
				\inletplanearrow{62}{14}{1.25}{-.6}
				\put(48,-3){\small\bfseries (a)}
			\end{overpic}
		\end{subfigure}\hfill
		\begin{subfigure}[t]{0.49\textwidth}
			\begin{overpic}[width=\linewidth]{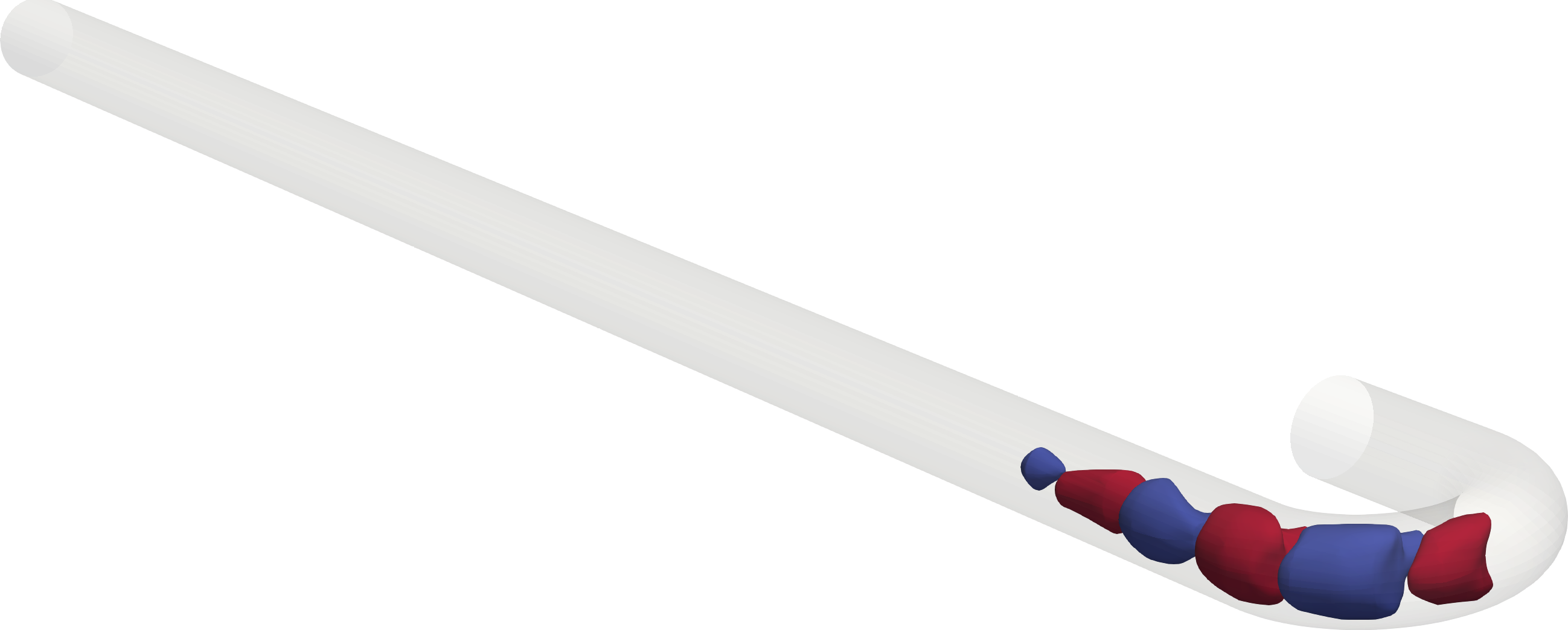}
				\put(48,-3){\small\bfseries (b)}
			\end{overpic}
		\end{subfigure}
		\caption{Phase-aligned signed streamfunction isosurfaces at $\Psi_j=\pm0.03\,U_bD$ for the downstream shear-layer modes at $Re_D=5300$ and $\gamma=0.4$: \textit{(a)} mode~\textit{DS1} ($St=0.27$) and \textit{(b)} mode~\textit{DS2} ($St=0.54$).}
		\label{fig:additional_cases_gamma0p4_downstream_modes}
	\end{figure}
	
	\FloatBarrier
	
	Figures~\ref{fig:additional_cases_gamma0p2_axial_mode}--\ref{fig:additional_cases_gamma0p4_downstream_modes} show the persistence of the seven identified modes in the additional cases. For the case with $Re_D=5300$ and $\gamma=0.2$, all seven modes remain nearly identical to their higher-Reynolds-number counterparts. For the case with stronger curvature, $\gamma=0.4$, all modes except the \textit{DS} family closely resemble their lower-curvature counterparts at both Reynolds numbers. Mode~\textit{Ax} exhibits the same low-frequency, long-wavelength structure as in figure~\ref{fig:axialMode}, while the \textit{SS} and \textit{SB} modes remain localised in the curved section of the pipe and their peak frequencies remain in the same range as those of their reference counterparts. However, the most notable changes in both mode shape and frequency occur in modes~\textit{DS1} and \textit{DS2} for the case with $\gamma=0.4$ in figure~\ref{fig:additional_cases_gamma0p4_downstream_modes}. Both instabilities shift farther upstream, closer to the bend, than the \textit{DS} modes at the lower curvature $\gamma=0.2$. Furthermore, the peak frequencies of this family increase significantly. The sensitivity of this family to curvature is consistent with the interpretation that it likely arises from the strong downstream shear layer. Increasing the curvature from $\gamma=0.2$ to $\gamma=0.4$ substantially reorganises the downstream shear layer; as a result, the location and peak Strouhal number of the \textit{DS} family are significantly modified. In contrast, the other three families remain closely related across all three cases. This contrast suggests that unlike the \textit{DS} family, the other modes are inherent instabilities of the secondary flow and are therefore only weakly affected by curvature and Reynolds number.

	\FloatBarrier

	\section{Results of the local stability analysis of the mean flow within the bend} 
	\label{sec:lsa_result}
\begin{figure}[!htbp]
		\centering
		\includegraphics[width=\textwidth]{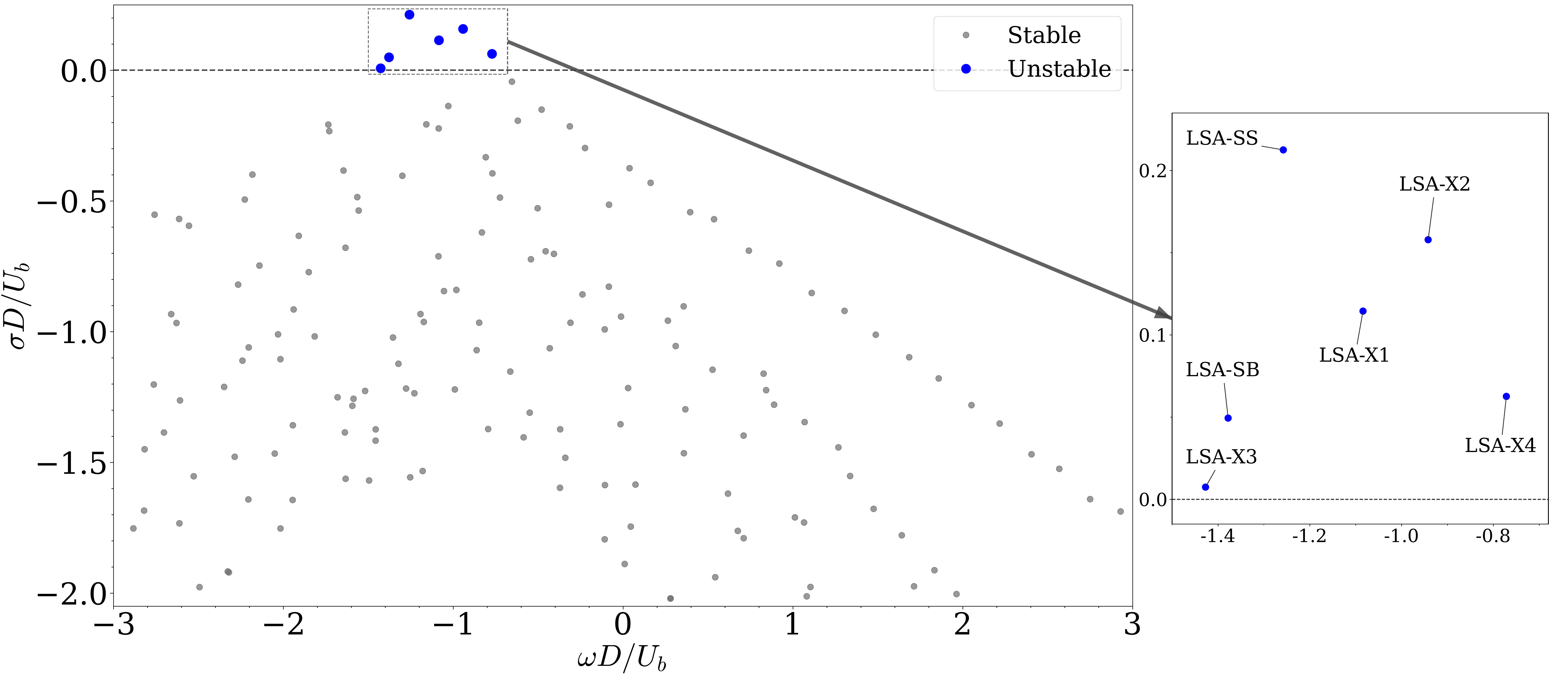}
		\caption{Local eigenspectrum at $\alpha D=1.2$ and $\phi=15^{\circ}$, shown in non-dimensional units $(\omega D/U_b,\sigma D/U_b)$. Grey and blue dots denote temporally stable and unstable eigenvalues, respectively.}
		\label{fig:lsa_eigenspectrum_phi15}
	\end{figure}

	The eigenproblem~\eqref{eq:eig} is solved at every integer bend angle $10^{\circ}\le\phi\le160^{\circ}$ for two streamwise wavenumbers estimated from the wavelengths of the FHPOD modes \textit{SS1} and \textit{SS2}, respectively:
	
	\vspace{.5em}
	\begin{itemize}
		\item $\alpha D = 0.6$, for the lower-frequency branches.
		\item $\alpha D = 1.2$, for the higher-frequency branches.
	\end{itemize}
	\vspace{.5em}
	This step reveals all unstable modes in this interval for the selected streamwise wavenumbers. The identified unstable modes are then used as seeds for the tracking algorithm described in \S\ref{sec:meth_track}. The tracking algorithm results in six unstable branches across the investigated bend-angle interval $10^{\circ}\le\phi\le160^{\circ}$. Figure~\ref{fig:lsa_eigenspectrum_phi15} shows a representative full eigenspectrum at $\phi=15^{\circ}$, where all six branches are locally unstable simultaneously. The most unstable branch, \textit{LSA-SS}, is the swirl-switching mode, while \textit{LSA-SB} denotes the unstable swirl-breathing branch. This section focuses on these two modes; the remaining unstable modes are presented in appendix~\ref{app:secondary_lsa_modes}.
	
\begin{figure}[!htbp]
		\centering
		\includegraphics[width=\textwidth]{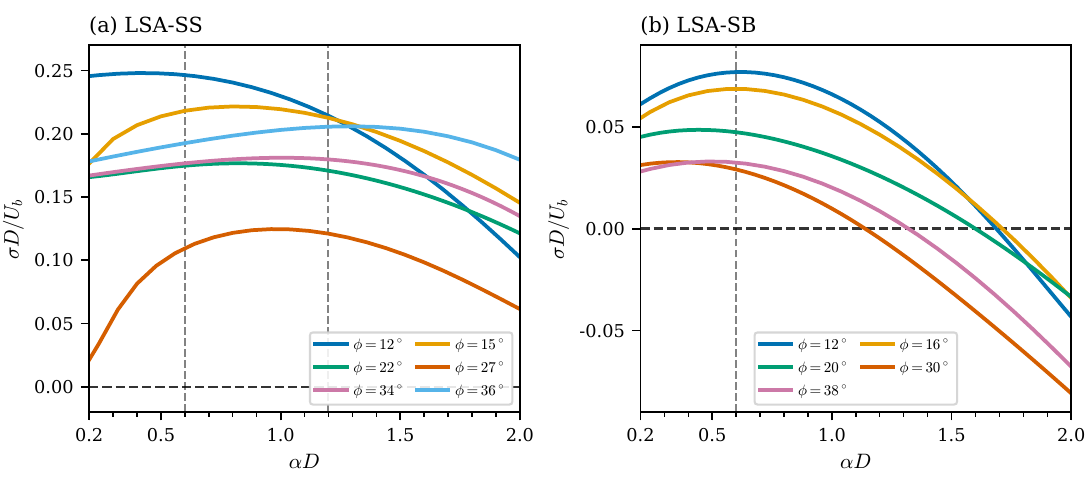}
		\caption{Dependence of the normalised temporal growth rate $\sigma D/U_b$ on the diameter-scaled streamwise wavenumber $\alpha D$ at selected bend angles for \textit{(a)} \textit{LSA-SS} (sinuous) and \textit{(b)} \textit{LSA-SB} (varicose). The curves join the computed wavenumbers without point or extremum markers. The vertical grey dashed lines identify $\alpha D=0.6$ and $1.2$ in panel~\textit{(a)}, whereas panel~\textit{(b)} retains only the $\alpha D=0.6$ reference. The horizontal black dashed lines mark neutral stability ($\sigma=0$); $\sigma<0$ is temporally stable.}
		\label{fig:alphaD_sensitivity_LSA_SS_SB}
	\end{figure}

	As the streamwise wavenumber is selected a priori by estimating the wavelength using the corresponding FHPOD modes, we investigate the dependence of the growth rate on the wavenumber for the \textit{LSA-SS} and \textit{LSA-SB} branches at various bend angles, as shown in figure~\ref{fig:alphaD_sensitivity_LSA_SS_SB}. For \textit{LSA-SS}, the maxima at the lower bend angles $\phi=12^{\circ}$ and $15^{\circ}$ lie on opposite sides of $\alpha D=0.6$, whereas the maxima at $\phi=27^{\circ}$, $34^{\circ}$ and $36^{\circ}$ lie at or close to $\alpha D=1.2$. The unstable \textit{LSA-SS} branches therefore favour two wavenumber regions close to the scales inferred from FHPOD. For the \textit{LSA-SB} branch, the maximum growth rate is close to $\alpha D=0.6$ for all investigated bend angles. As a result, figure~\ref{fig:alphaD_sensitivity_LSA_SS_SB} supports the choice of the streamwise wavenumbers $\alpha D=0.6$ and $\alpha D=1.2$.

	\FloatBarrier

\begin{figure}[!htbp]
		\centering
		\begin{subfigure}[t]{0.49\textwidth}
			\begin{overpic}[width=\linewidth]{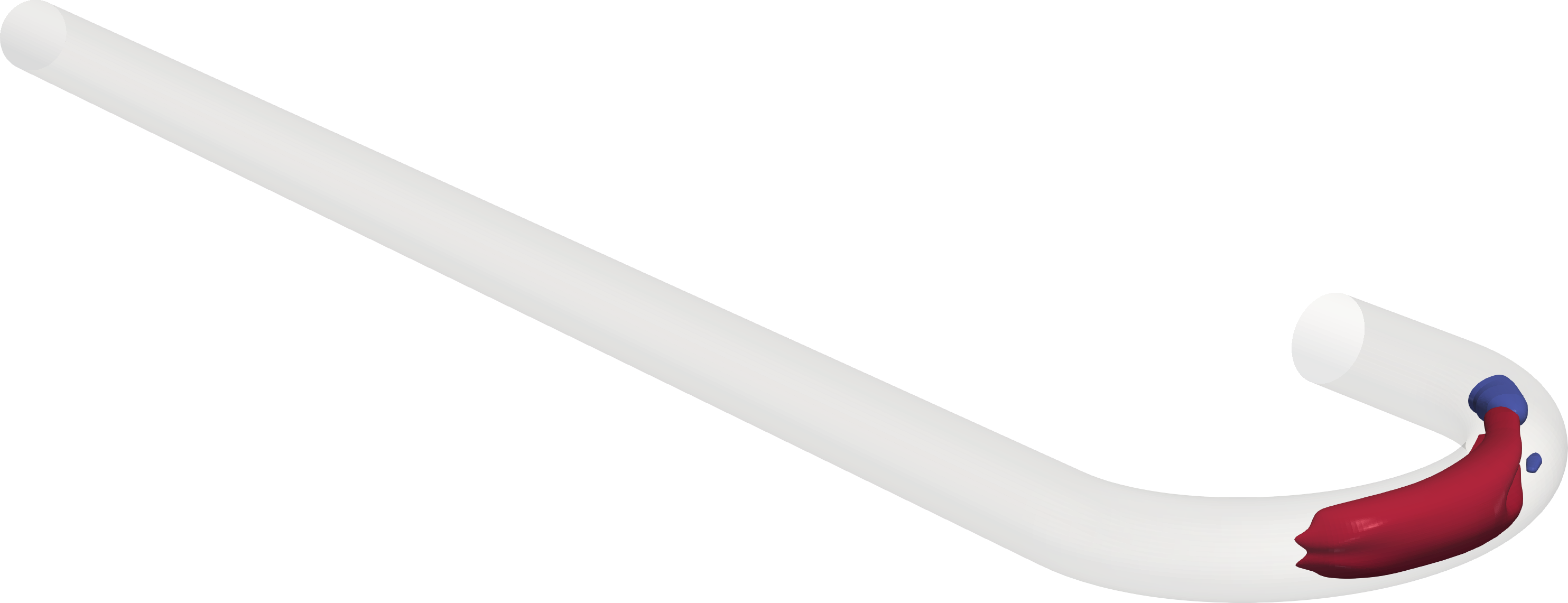}
				\inletplanearrow{62}{17.2}{1.25}{-.6}
				\put(48,0){\small\bfseries (a)}
			\end{overpic}
		\end{subfigure}\hfill
		\begin{subfigure}[t]{0.49\textwidth}
			\begin{overpic}[width=\linewidth]{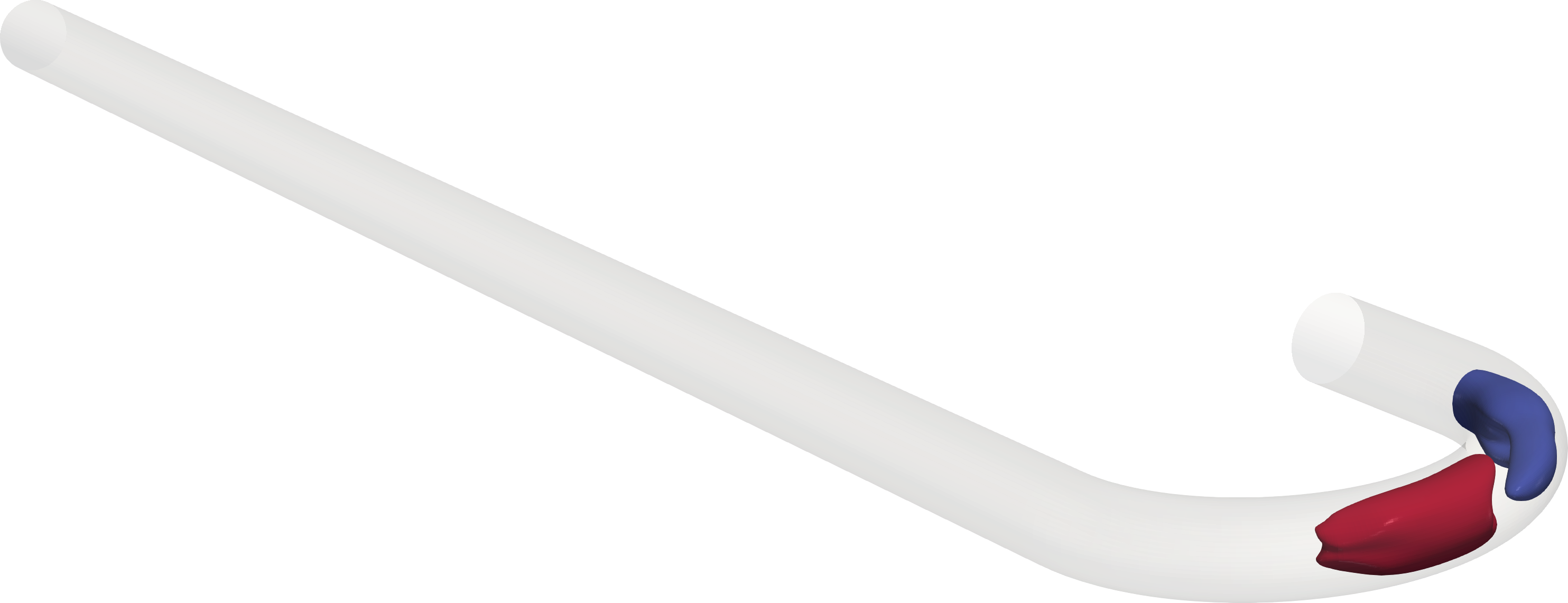}
				\put(48,0){\small\bfseries (b)}
			\end{overpic}
			\vspace{-0.5em}
		\end{subfigure}
		
		\begin{subfigure}[t]{0.49\textwidth}
			\begin{overpic}[width=\linewidth]{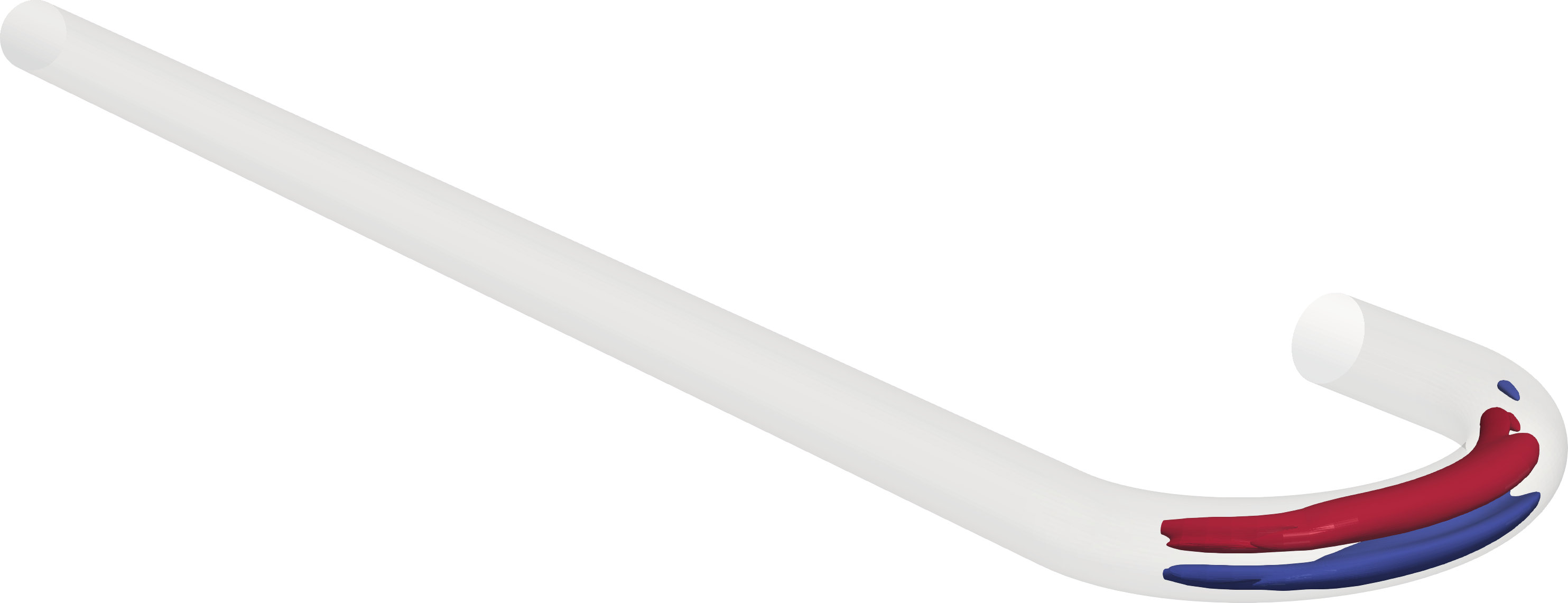}
				\put(48,0){\small\bfseries (c)}
			\end{overpic}
		\end{subfigure}\hfill
		\begin{subfigure}[t]{0.49\textwidth}
			\begin{overpic}[width=\linewidth]{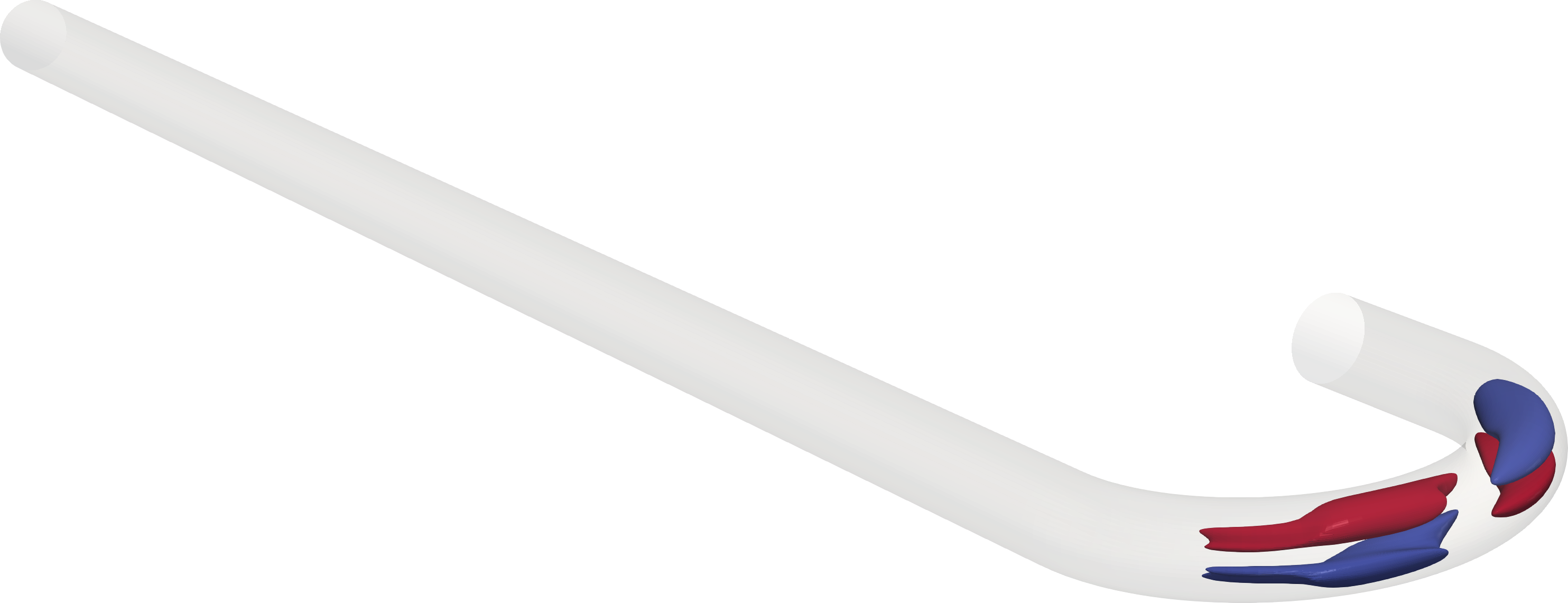}
				\put(48,0){\small\bfseries (d)}
			\end{overpic}
		\end{subfigure}
		\caption{Three-dimensional reconstructions of the four principal unstable modes obtained from the local stability analysis: \textit{(a)} \textit{LSA-SS1} at $\alpha D=0.6$, \textit{(b)} \textit{LSA-SS2} at $\alpha D=1.2$, \textit{(c)} \textit{LSA-SB1} at $\alpha D=0.6$ and \textit{(d)} \textit{LSA-SB2} at $\alpha D=1.2$. Red and blue denote equal positive and negative levels of streamfunction $\Psi=\pm 0.07 U_bD$.}
		\label{fig:stab_shapes}
	\end{figure}

\begin{figure}[!htbp]
		\centering
		\begin{subfigure}[t]{0.80\textwidth}
			\begin{overpic}[width=\linewidth]{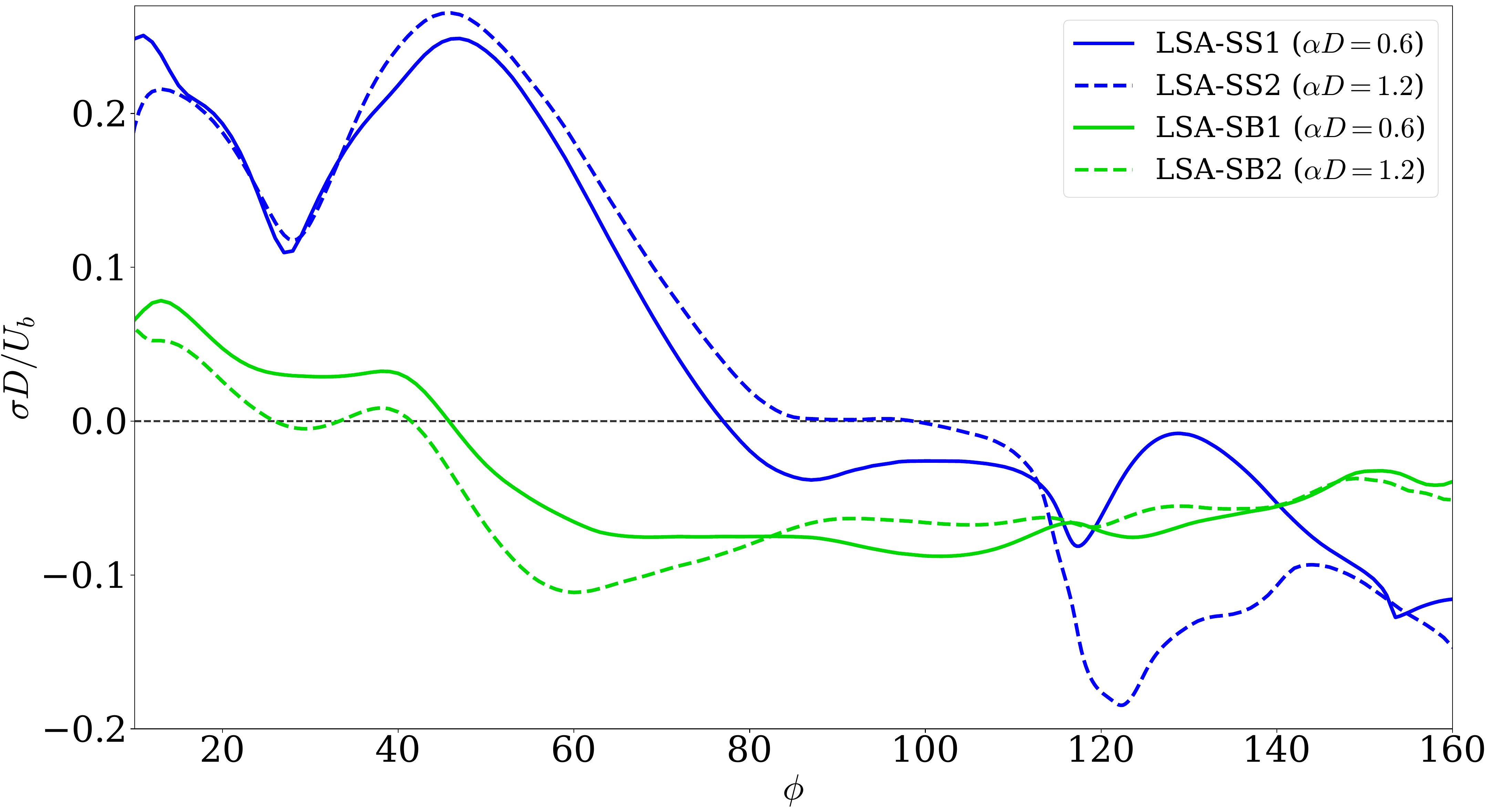}
				\put(48,-3){\small\bfseries (a)}
			\end{overpic}
		\end{subfigure}
		
		\vspace{1.5em}
		\begin{subfigure}[t]{0.80\textwidth}
			\begin{overpic}[width=\linewidth]{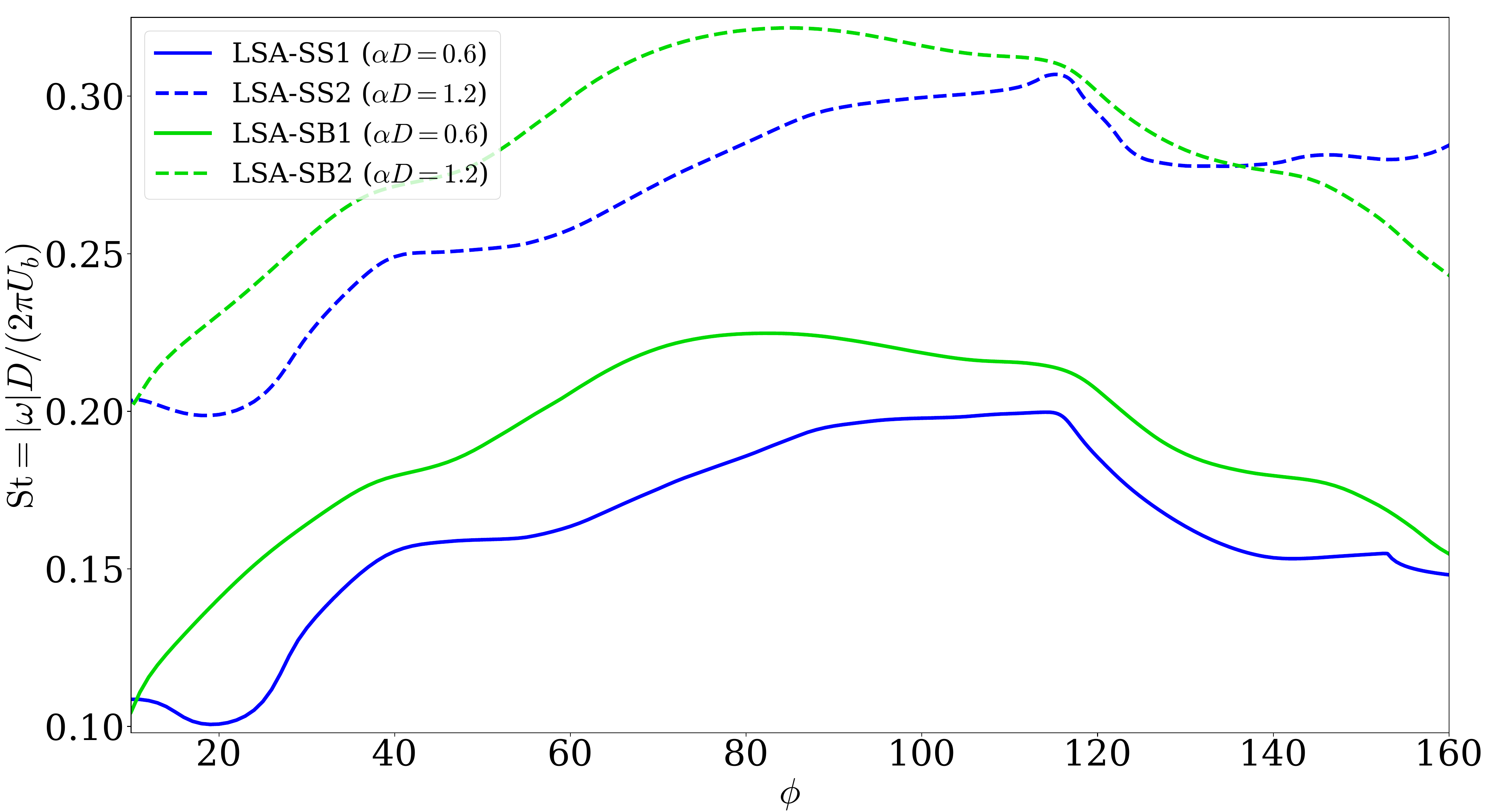}
				\put(48,-3){\small\bfseries (b)}
			\end{overpic}
		\end{subfigure}
		\caption{Evolution of the four principal unstable modes, \textit{LSA-SS1}, \textit{LSA-SS2}, \textit{LSA-SB1} and \textit{LSA-SB2}, along the bend: \textit{(a)} temporal growth rate and \textit{(b)} Strouhal number. Solid curves represent these modes evaluated at $\alpha D=0.6$, and dashed curves denote those evaluated at $\alpha D=1.2$. The horizontal black dashed line in panel~\textit{(a)} marks neutral stability ($\sigma=0$); $\sigma<0$ is temporally stable.}
		\label{fig:stab_tracks}
	\end{figure}

	Figure~\ref{fig:stab_shapes} shows the reconstructions obtained using the procedure described in \S\ref{sec:meth_wkb}. The \textit{LSA-SS1} and \textit{LSA-SS2} reconstructions constitute the sinuous instabilities of the Dean vortices, closely matching their FHPOD counterparts \textit{SS1} and \textit{SS2} in figure~\ref{fig:SS_modes}. Similarly, the \textit{LSA-SB1} and \textit{LSA-SB2} reconstructions represent the varicose instability of the Dean vortices, closely resembling their FHPOD counterparts \textit{SB1} and \textit{SB2} in figure~\ref{fig:SB_modes}. This agreement is also reflected in the Strouhal numbers of these modes. Figure~\ref{fig:stab_tracks} shows the evolution of the growth rate and Strouhal number over the bend angle $\phi$ for the \textit{LSA-SS} and \textit{LSA-SB} branches. At $\alpha D=0.6$, the tracked modes are denoted by \textit{LSA-SS1} and \textit{LSA-SB1}. The solid curves show the lower-frequency modes at $\alpha D=0.6$, while the dashed curves show the higher-frequency modes at $\alpha D=1.2$. A summary of these modes is reported in table~\ref{tab:principal_lsa_branch_statistics}.
	
	\begin{table}
		\centering
		\caption{Summary of the \textit{LSA-SS} and \textit{LSA-SB} branches over $10^\circ\leq\phi\leq160^\circ$. The Strouhal-number ranges and mean values are evaluated only over the unstable bend angles $\phi$.}
		\label{tab:principal_lsa_branch_statistics}
		\vspace{0.2em}
		\scriptsize
		\setlength{\tabcolsep}{3pt}
		\begin{tabular}{l c c c c c c}
			Mode & $\alpha D$ & Unstable angles & $St$ & Mean $St$ & $\max(\sigma D/U_b)$ & Peak-growth angle, $\phi_{\sigma,\max}$ \\
			\hline
			\textit{LSA-SS1} & 0.6 & $10.0^\circ$--$77.0^\circ$ & 0.101--0.183 & 0.145 & 0.251 & $11.0^\circ$ \\
			\textit{LSA-SS2} & 1.2 & $10.0^\circ$--$98.7^\circ$ & 0.199--0.299 & 0.253 & 0.265 & $46.0^\circ$ \\
			\textit{LSA-SB1} & 0.6 & $10.0^\circ$--$45.8^\circ$ & 0.105--0.184 & 0.155 & 0.0784 & $13.0^\circ$ \\
			\textit{LSA-SB2} & 1.2 & $10.0^\circ$--$25.9^\circ$ and $33.3^\circ$--$41.5^\circ$ & 0.201--0.272 & 0.240 & 0.0601 & $10.0^\circ$ \\
		\end{tabular}
	\end{table}

	The agreement between the FHPOD and LSA modes is especially notable, since the local stability analysis relies only on the mean cross-sectional base flow and nevertheless recovers the essential structures embedded in the DNS as unstable modes. Our analysis therefore supports the interpretation that swirl-switching is an intrinsic instability of the bent-pipe mean flow that can be amplified by turbulent fluctuations. Additionally, figure~\ref{fig:stab_tracks}\textit{(b)} indicates that the swirl-switching mode should not be interpreted as a single monochromatic oscillation, but rather as a finite-band response produced by spatially evolving local instabilities. In this sense, the present stability analysis further supports the use of the FHPOD approach to isolate coherent structures with broadened spectral support.

	\section{Summary and conclusions}
	
	In this study, we revisit the swirl-switching phenomenon in turbulent bent-pipe flows. We address the shortcomings of the classical POD by first establishing five requirements for a physically meaningful modal decomposition of turbulent flows. As these requirements are often not satisfied in POD studies, we show how distinct instabilities are distributed across multiple modes, their energy is artificially split, and their physical origin becomes ambiguous and difficult to identify. We therefore propose an alternative decomposition method based on the Hilbert transform combined with temporal band-pass filtering before the SVD step, termed filtered Hilbert POD (FHPOD). The current approach allows the decomposition of the flow into separate energetically and spectrally distinct mode families. This modal decomposition approach shows that the different spectral peaks previously reported in the literature do not necessarily correspond to a single physical phenomenon, but reflect distinct coherent structures with varying strengths along the streamwise direction. More specifically, the method separates entangled distinct structures that have often been grouped under the same ``swirl-switching'' label in the literature. 
	
	The resulting decomposition reveals seven modes in four physical families. Mode~\textit{Ax} is a low-frequency axial wave at $St\approx0.03$. The bend-localised sinuous family consists of modes~\textit{SS1} and~\textit{SS2} at $St\approx0.13$ and $St\approx0.25$, respectively, while the corresponding varicose family consists of modes~\textit{SB1} and~\textit{SB2} at $St\approx0.12$ and $St\approx0.25$. Modes~\textit{DS1} and~\textit{DS2}, at $St\approx0.15$ and $St\approx0.30$, are localised in the downstream shear layer of the four-vortex mean-flow structure and are likely induced by the strong shear layer downstream of the bend. 
	
	The separation of the \textit{SS} and \textit{SB} families from the \textit{DS} family is an essential step in understanding the behaviour of the swirl-switching and swirl-breathing modes and in establishing them as intrinsic instabilities of the mean flow in the linear stability analysis. By estimating the wavelengths of the \textit{SS} and \textit{SB} modes from FHPOD, the local stability analysis is performed at two streamwise wavenumbers, $\alpha D=0.6$ and $\alpha D=1.2$. The additional study of the growth rate with respect to the streamwise wavenumber confirms that the maximum growth rate of \textit{LSA-SS} occurs near these two wavenumbers, while \textit{LSA-SB} is amplified near $\alpha D=0.6$. The results of the FHPOD and LSA analyses agree in the frequency ranges, symmetries and spatial features of the identified modes. The continuous variation of the LSA-mode frequencies along the bend indicates that these instabilities have a finite-band response rather than a single characteristic frequency.

	The collective evidence from both the FHPOD decomposition and LSA favours the interpretation that swirl-switching and swirl-breathing are inherent instabilities of the unstable Dean vortices. Although streaks and very-large-scale motions can excite or modulate these instabilities, the linearised operator determines the amplified response. This interpretation is supported by \citet{SakakibaraMachida2012}, who found a correlation between the upstream streaks and the displacement of the inner-side stagnation point. It is also consistent with the findings of \citet{KalpakliVester2015}, in which modifying the incoming flow delayed and altered swirl-switching without necessarily eliminating it. 
	
	The present approach has the following limitations. The FHPOD procedure requires manual band selection. Although the filter is designed to be robust to modest variations in the band-edge locations, it still requires an initial inspection of POD spectra to identify the dominant peaks. As a result, the method is most effective when the underlying coherent structures have distinct spectral peaks and sufficient spatial localisation. If multiple modes strongly overlap in both space and frequency, they may not be fully separable using the FHPOD decomposition.

    \FloatBarrier
    \appendix
	\normalsize
	\renewcommand{\theHsection}{appendix.\Alph{section}}
    \section{Inflow condition and convergence of the statistics}
	\label{app:inflow_condition}
    
\begin{figure}[!htbp]
		\centering
		\begin{subfigure}[t]{0.315\textwidth}
			\begin{overpic}[width=\linewidth]{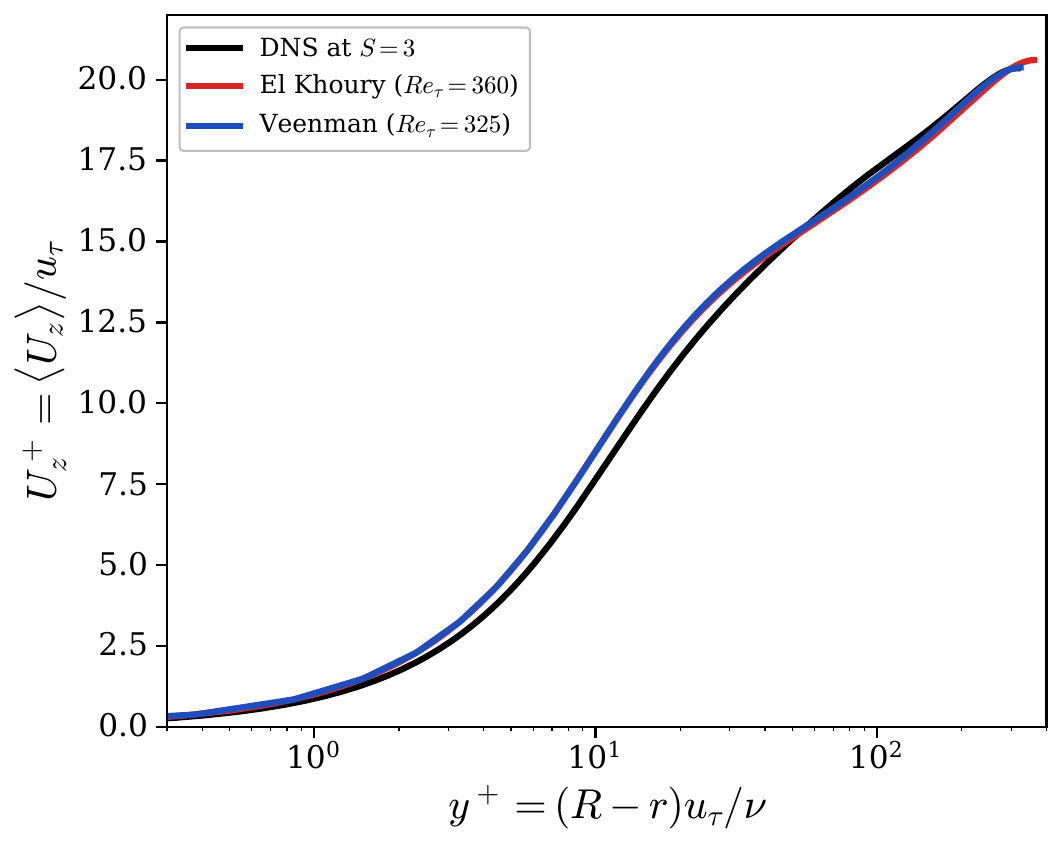}
				\put(52,-6){\small\bfseries (a)}
			\end{overpic}
			\vspace{0.6em}
		\end{subfigure}\hfill
		\begin{subfigure}[t]{0.315\textwidth}
			\begin{overpic}[width=\linewidth]{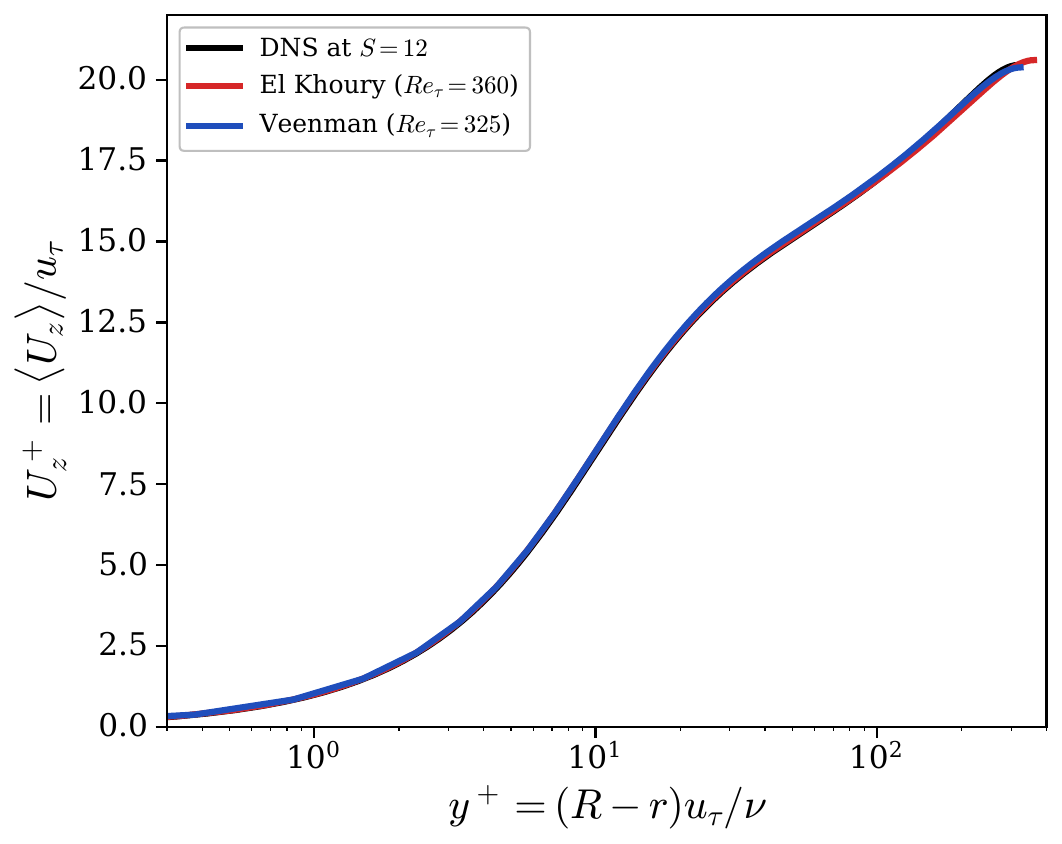}
				\put(52,-6){\small\bfseries (b)}
			\end{overpic}
		\end{subfigure}\hfill
		\begin{subfigure}[t]{0.315\textwidth}
			\begin{overpic}[width=\linewidth]{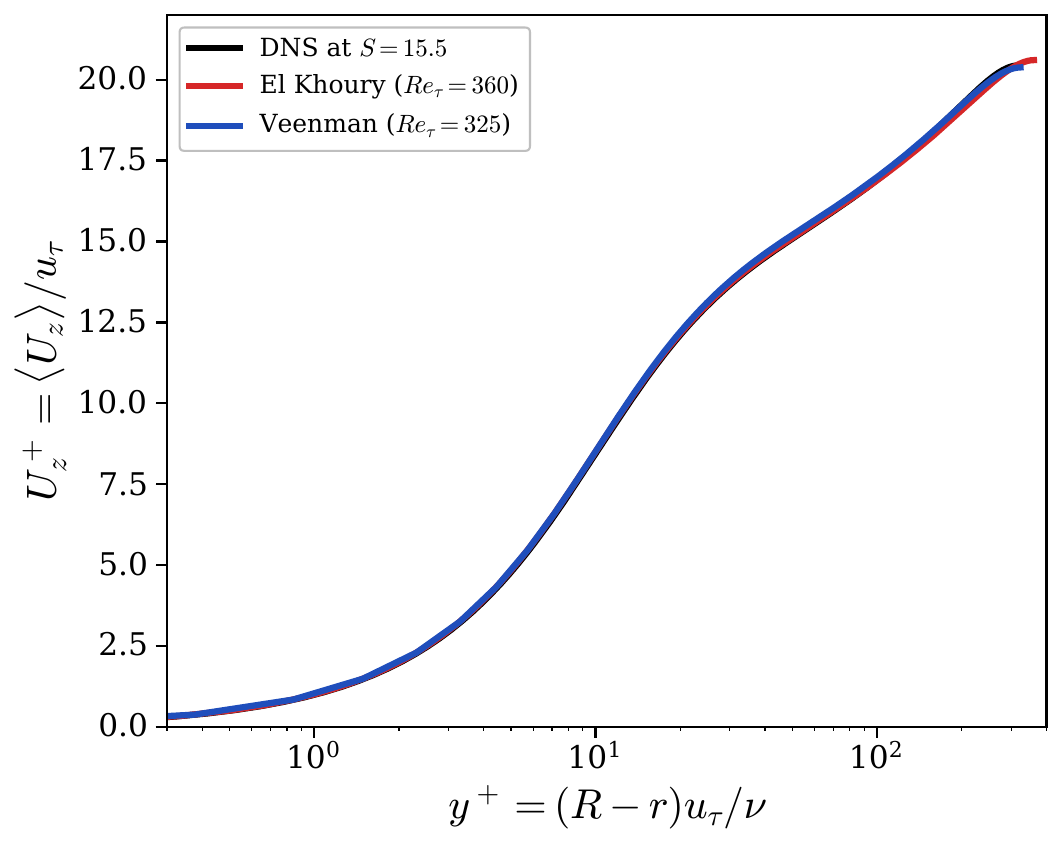}
				\put(52,-6){\small\bfseries (c)}
			\end{overpic}
		\end{subfigure}
		
		\begin{subfigure}[t]{0.315\textwidth}
			\begin{overpic}[width=\linewidth]{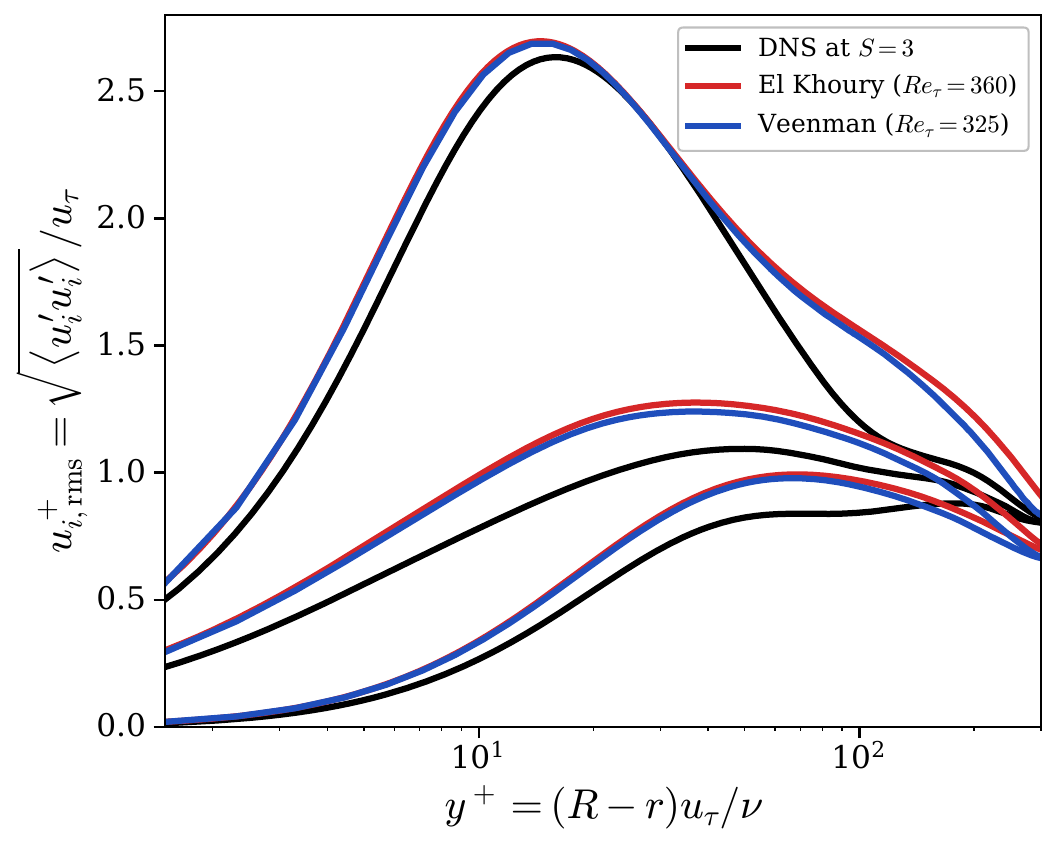}
				\put(52,-6){\small\bfseries (d)}
			\end{overpic}
		\end{subfigure}\hfill
		\begin{subfigure}[t]{0.315\textwidth}
			\begin{overpic}[width=\linewidth]{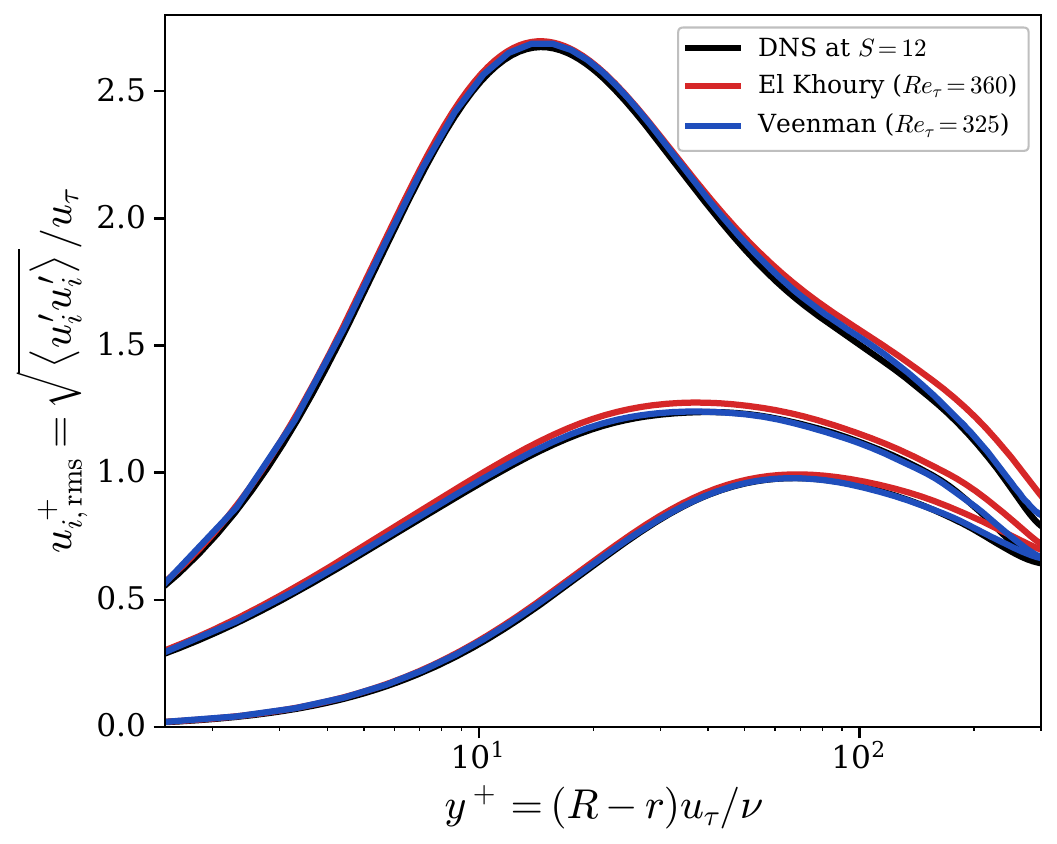}
				\put(52,-6){\small\bfseries (e)}
			\end{overpic}
		\end{subfigure}\hfill
		\begin{subfigure}[t]{0.315\textwidth}
			\begin{overpic}[width=\linewidth]{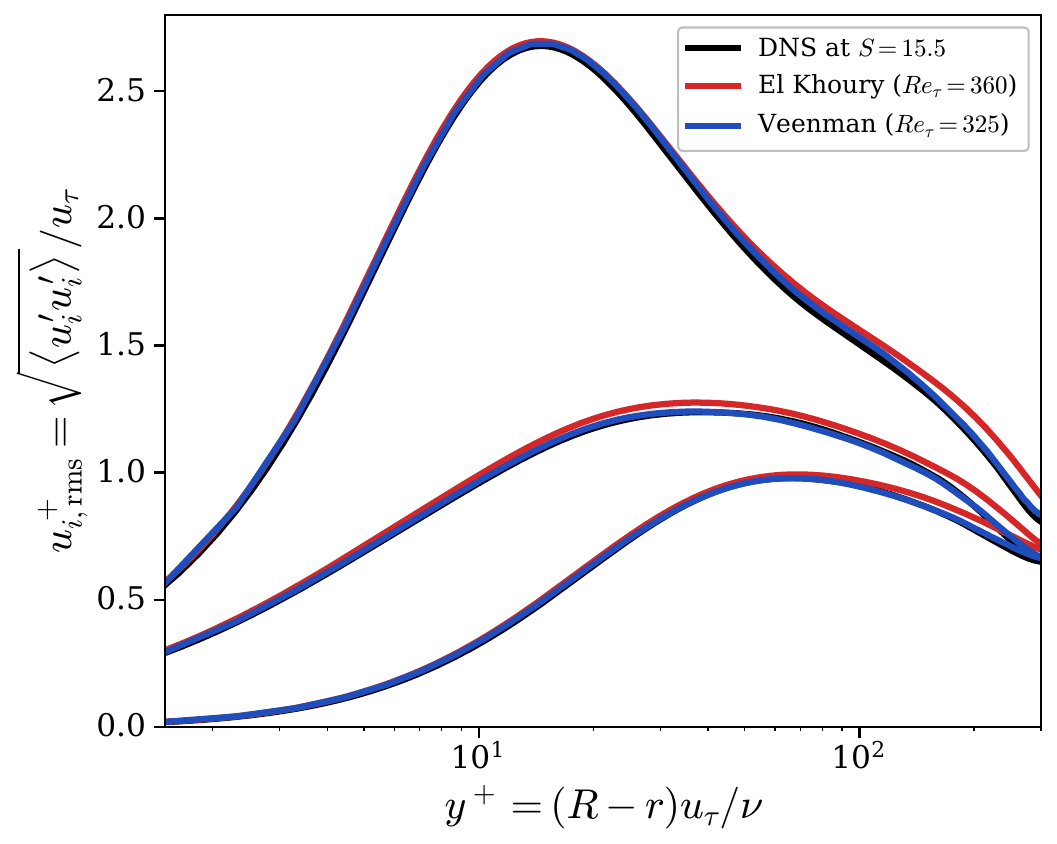}
				\put(52,-6){\small\bfseries (f)}
			\end{overpic}
		\end{subfigure}
		\caption{Streamwise development of the flow upstream of the bend. The profiles of the inner-scaled mean velocity (upper row) and r.m.s. components of the velocity (lower row) are shown against reference data from \citet{ElKhoury2013} and \citet{Veenman2004}. Panels \textit{(a)}--\textit{(c)} show the mean velocity at $S=3$, $S=12$ and $S=15.5$, respectively; panels \textit{(d)}--\textit{(f)} show the corresponding velocity r.m.s. profiles.}
		\label{fig:inflow_validation}
	\end{figure}

    Turbulence inflow generation methods usually require some finite distance for adjustment, and the near-inlet errors associated with them depend on the manner in which fluctuations are generated \citep{Stanly2026}. Considering the earlier works by \citet{Hufnagel2018,Lupi2025} in which the same synthetic eddy inflow generation method was used, the length of the straight section upstream of the bend was conservatively chosen so that the turbulence can fully develop upstream of the curved section of the pipe while avoiding spurious excitations that can arise from recycling or precursor methods. As figure~\ref{fig:inflow_validation} shows, $12D$ downstream of the inflow plane, both the mean flow and Reynolds stresses converge to the reference data at similar Reynolds numbers. 
	
	\FloatBarrier
	\section{Numerical treatment of the eddy-viscosity}
    \label{app:eddy_visc}
    
\begin{figure}[!htbp]
		\centering
		\begin{overpic}[width=0.5\linewidth]{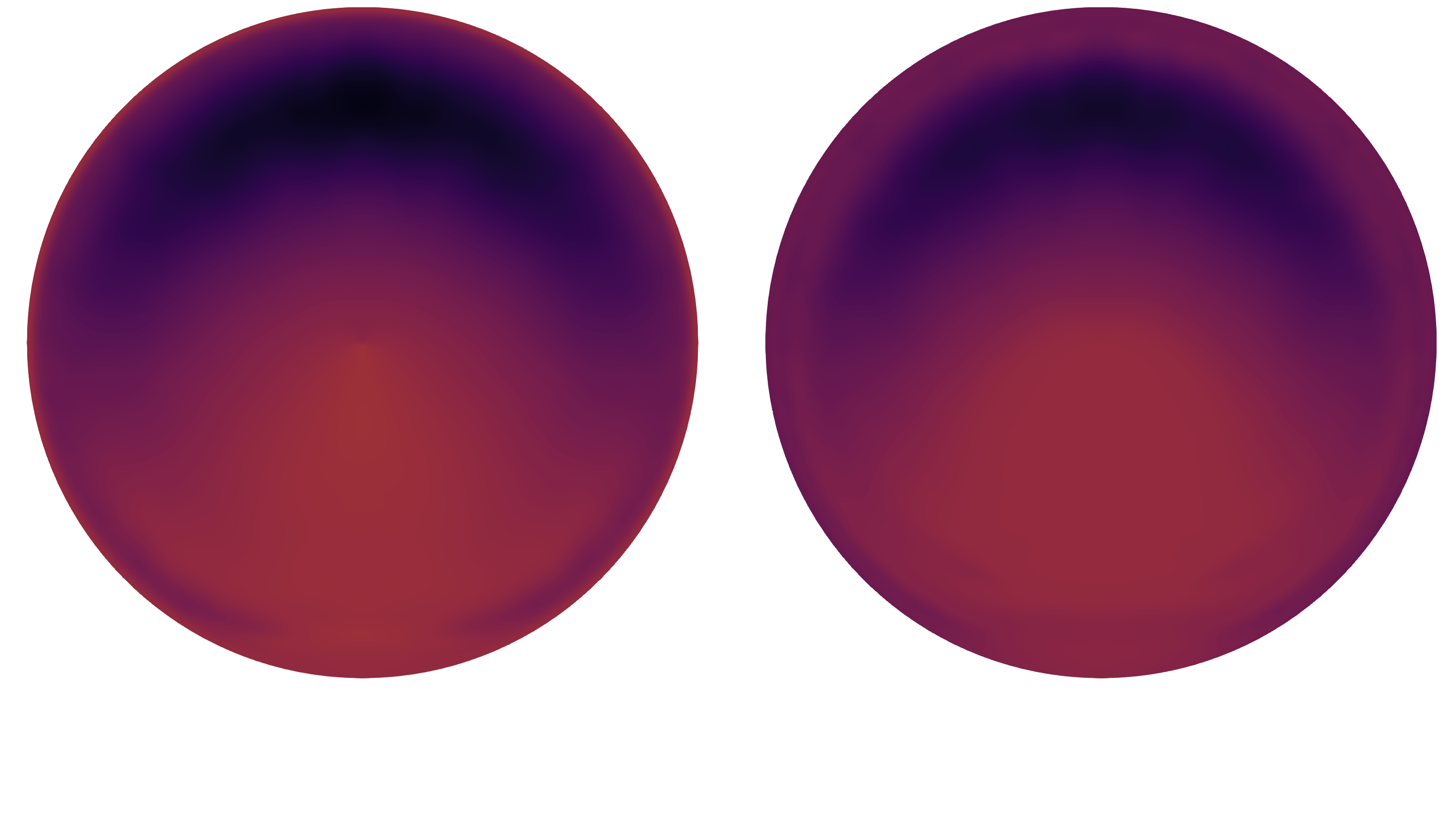}
			\put(23,5){\small\bfseries (a)}
			\put(73,5){\small\bfseries (b)}
		\end{overpic}
		\vspace{2mm}
		\begin{overpic}[width=0.5\linewidth]{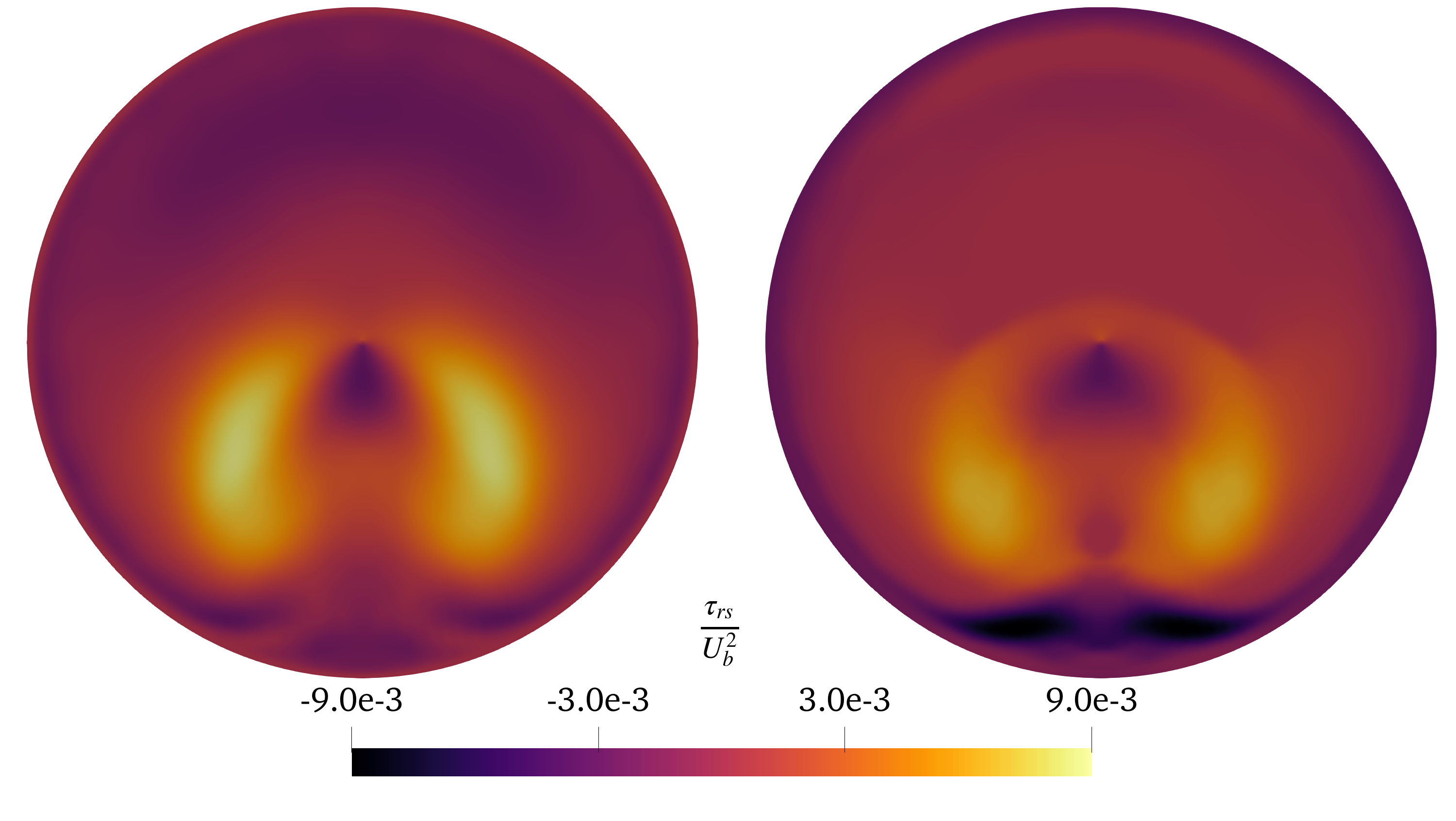}
			\put(23,-3){\small\bfseries (c)}
			\put(73,-3){\small\bfseries (d)}
		\end{overpic}
		
		\caption{Comparison of the radial--streamwise shear stress calculated in DNS $\tau_{rs}^{\mathrm{DNS}}
			=-\overline{u'_r u'_s}$ with the Boussinesq shear stress model $\tau_{rs}^{\mathrm{model}}
			=2\nut[\Smat(\Uvec)]_{rs}$ reconstructed from the fitted eddy-viscosity for two different bend angles: \textit{(a)} DNS at $\phi=30^{\circ}$, \textit{(b)} eddy-viscosity model at $\phi=30^{\circ}$, \textit{(c)} DNS at $\phi=90^{\circ}$ and \textit{(d)} eddy-viscosity model at $\phi=90^{\circ}$.}
		\label{fig:eddy_visc_dns_comparison}
	\end{figure}

    As $\nut$ enters the operator in the linear stability equation (see \eqref{eq:NS_lin}), the resulting regularised and clipped closure should be differentiable. Similar to the positive-clipping methods used for data-fitted eddy-viscosity fields \citep{Rukes2016,vonSaldern2024}, we implement a differentiable variant by combining the Chen--Harker--Kanzow--Smale smooth approximation of the positive part of $\nut^{\mathrm{raw}}$  \citep{ChenMangasarian1996} with a cubic Hermite ramp. The smooth and clipped eddy-viscosity field $\nu_t$ is therefore obtained from
	
	\begin{equation}
		\label{eq:nut_smooth}
		\begin{aligned}
			\nut
			&= \frac{g(\eta)}{2} \left(\nut^{\mathrm{raw}}+\sqrt{(\nut^{\mathrm{raw}})^{2}+\epsilon^{2}}\right),
			&\quad g(\eta)&=\eta^{2}(3-2\eta),\\
			\eta
			&=\min\!\left(
			\max\!\left(
			\frac{|\Smat_{\mathrm{dev}}|-c_0\,\Sscale(\phi)}
			{(c_1-c_0)\Sscale(\phi)},0
			\right),1
			\right).
		\end{aligned}
	\end{equation}
	The strain rate scale $\Sscale(\phi)$ in \eqref{eq:nut_smooth} is defined by
	\begin{equation}
		\Sscale(\phi)
		:=\left[
		\frac{\displaystyle\int_{\mathcal A(\phi)}
			|\Smat_{\mathrm{dev}}|^2 h_s\,\diff A}
		{\displaystyle\int_{\mathcal A(\phi)}h_s\,\diff A}
		\right]^{1/2}, \qquad |\Smat_{\mathrm{dev}}|=(2\Smat_{\mathrm{dev}}:\Smat_{\mathrm{dev}})^{1/2}
		\label{eq:strain_scale}
	\end{equation}
	where $\diff A$ is the cross-sectional area element and $h_s$ is the streamwise metric scale factor. Here, $|\Smat_{\mathrm{dev}}|$ is the magnitude of the deviatoric mean strain-rate tensor, and $g(\eta)$ is the ramp function. The constants $c_0=0.05$, $c_1=0.10$ and $\epsilon/(U_bD)=1\times10^{-5}$ were selected empirically for the present DNS fields so that the ramp function suppresses the poorly conditioned regions below 5\% of $\Sscale(\phi)$, smoothly transitions from zero to one between 5\% and 10\% of $\Sscale(\phi)$, and approaches the fitted eddy-viscosity $\nut^{\mathrm{raw}}$ above 10\% of $\Sscale(\phi)$, while its negative values remain smoothly clipped. Figure~\ref{fig:eddy_visc_dns_comparison} shows the comparison between the true Reynolds shear stress from DNS and the modelled shear stress using the Boussinesq approximation from the fitted regularised eddy-viscosity at two representative bend angles. 
	
    \FloatBarrier
    \section{Rayleigh quotient iteration}
    \label{app:RQI} 

    After solving \eqref{eq:eig} for every integer angle $10^{\circ}\le\phi\le160^{\circ}$, unstable modes are found at different bend angles $\phi_{\mathrm{ref}}$. Using each identified unstable mode as an initial seed, we continue the search in both directions, $10^{\circ}\le\phi<\phi_{\mathrm{ref}}$ and $\phi_{\mathrm{ref}}<\phi\le160^{\circ}$. The eigenvalue at the bend angular position $\phi_k$  is denoted by $\mu_k$. The eigenproblem is rebuilt at every $\phi_k$
	\begin{equation}
		\label{eq:eig_phi}
		-\mu_k\mathcal{M}\qhvec_\alpha(\phi_k)=\mathcal{L}(\phi_k)\qhvec_\alpha(\phi_k).
	\end{equation}
	The eigenvector from the preceding angular position provides the initial guess. For the first continuation step, the seed eigenvalue itself is used as the initial shift; for the second step, the shift is linearly extrapolated from the two available eigenvalues. Thereafter, a three-point Lagrange extrapolation is used to determine the initial shift: 
	\begin{equation}
		\label{eq:eig_predictor}
		\begin{aligned}
			\qhvec_{\alpha,k}^{\mathrm{pred}}
			&=\qhvec_\alpha(\phi_{k-1}),\\
			\mu_k^{\mathrm{pred}}
			&=\sum_{\ell=1}^{3}\mu_{k-\ell}
			\prod_{\substack{m=1\\m\ne \ell}}^{3}
			\frac{\phi_k-\phi_{k-m}}{\phi_{k-\ell}-\phi_{k-m}}.
		\end{aligned}
	\end{equation}
	These predicted quantities serve as the initial guess in the corrector step using Rayleigh quotient iteration (RQI) for the eigenproblem:
	
	\begin{equation}
		\label{eq:rqi}
		\begin{aligned}
			\left[\mathcal L(\phi_k)+\mu_k^{(\iota)}\mathcal M\right]
			\evec_k^{(\iota+1)}
			&=\mathcal M\widehat{\evec}_k^{(\iota)},\\
			\widehat{\evec}_k^{(\iota+1)}
			=\frac{\evec_k^{(\iota+1)}}
			{\left(\left(\evec_k^{(\iota+1)}\right)^\dagger
				\mathcal M\evec_k^{(\iota+1)}\right)^{1/2}},
			\quad \mu_k^{(\iota+1)}
			&=-\frac{
				\left(\widehat{\evec}_k^{(\iota+1)}\right)^\dagger\mathcal L(\phi_k)
				\widehat{\evec}_k^{(\iota+1)}}{
				\left(\widehat{\evec}_k^{(\iota+1)}\right)^\dagger\mathcal M
				\widehat{\evec}_k^{(\iota+1)}} 
			,\\
			\widehat{\evec}_k^{(0)}
			=\frac{\qhvec_{\alpha,k}^{\mathrm{pred}}}
			{\left(\left(\qhvec_{\alpha,k}^{\mathrm{pred}}\right)^\dagger
				\mathcal M\qhvec_{\alpha,k}^{\mathrm{pred}}\right)^{1/2}},
			\qquad\mu_k^{(0)}&=\mu_k^{\mathrm{pred}}
			.
		\end{aligned}
	\end{equation}
	Here, $(\cdot)^\dagger$ denotes the Hermitian transpose, $\evec_k^{(\iota+1)}$ is the unnormalised RQI eigenvector iterate at angular position $\phi_k$ and iteration $\iota+1$, while $\widehat{\evec}_k^{(\iota+1)}$ is its normalised counterpart. The continuity of a tracked branch between two consecutive angular positions is assessed using the correlation
	\begin{equation}
		\mathcal C_{k,m}=
		\frac{
			\left|
			\displaystyle\int_{\mathcal A}
			\left(\widehat{\boldsymbol u}_{k-1}^{\mathrm{acc}}\right)^{*}\!\cdot
			\widehat{\boldsymbol u}_{k,m}\,\mathrm d\mathcal A
			\right|
		}{
			\left(
			\displaystyle\int_{\mathcal A}
			\left|\widehat{\boldsymbol u}_{k-1}^{\mathrm{acc}}\right|^2
			\,\mathrm d\mathcal A
			\right)^{1/2}
			\left(
			\displaystyle\int_{\mathcal A}
			\left|\widehat{\boldsymbol u}_{k,m}\right|^2
			\,\mathrm d\mathcal A
			\right)^{1/2}
		},
		\qquad m=1,\ldots,N_k,
		\label{eq:adjacent_mode_correlation}
	\end{equation}
	where $(\cdot)^*$ denotes complex conjugation, $\widehat{\boldsymbol u}_{k-1}^{\mathrm{acc}}$ is the velocity field of the previously accepted eigenvector and $\widehat{\boldsymbol u}_{k,m}$ is the velocity field of candidate eigenvector $m$ among the $N_k$ eigenvectors in the computed local eigenspectrum at $\phi_k$. The dot product contains all three velocity components, and the integration is over the pipe cross-section. At every integer bend angle, all eigenvectors in the local eigenspectrum are ranked according to $\mathcal C_{k,m}$. A continuation step is accepted only when the eigenvector to which RQI converges is the candidate with the highest correlation. If RQI converges to another candidate, the interval from the preceding accepted angular position is recomputed after halving $\Delta\phi$. The mean-flow and eddy-viscosity fields required at refined bend angles are interpolated from the DNS as before. This procedure is repeated until RQI returns an eigenvector that has the highest correlation with the previously accepted eigenvector.

	\FloatBarrier
    \section{The four additional unstable local modes}
	\label{app:secondary_lsa_modes}
	
\begin{figure}[!htbp]
		\centering
		\begin{subfigure}[t]{0.7\textwidth}
			\begin{overpic}[width=\linewidth]{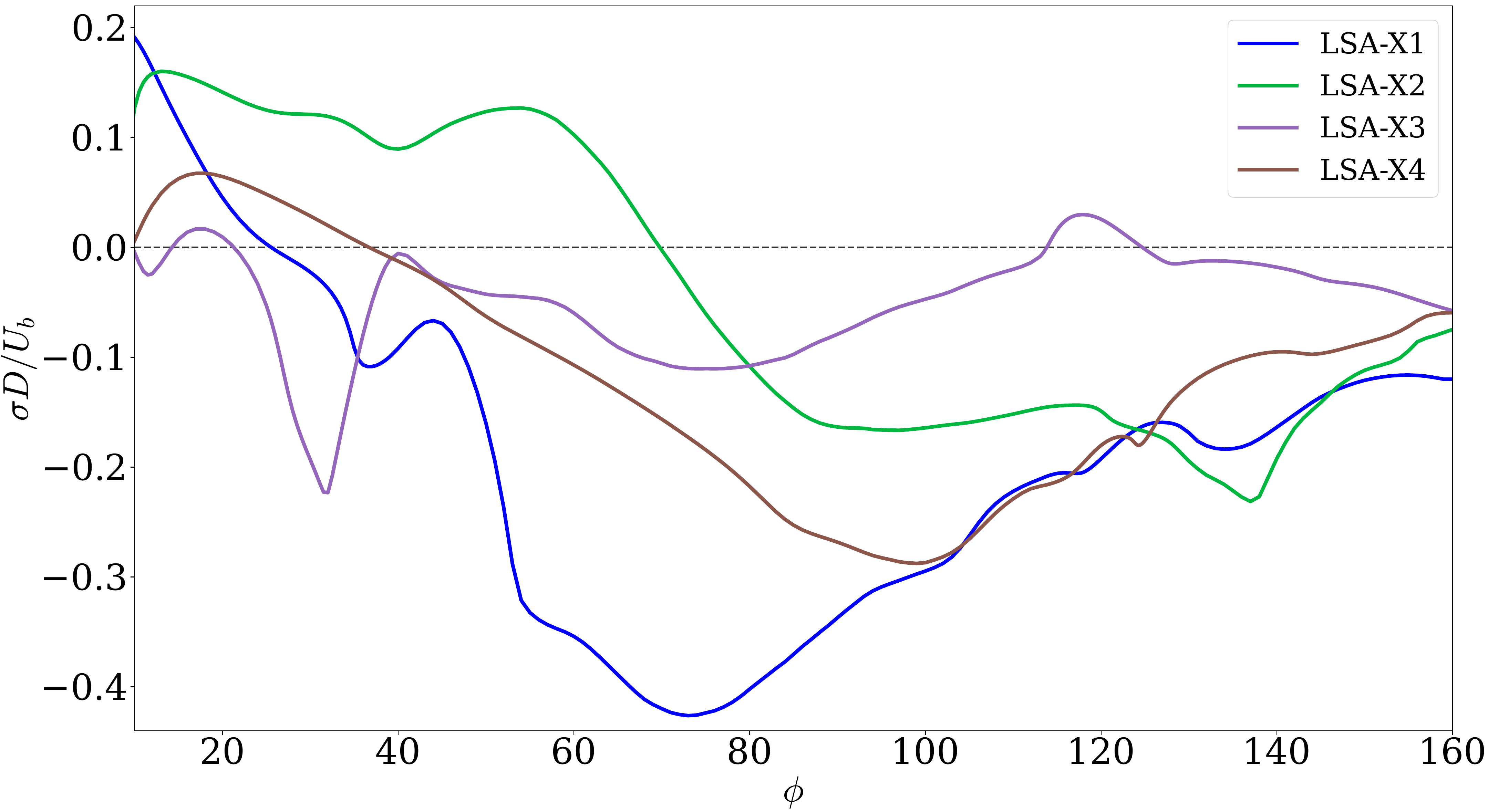}
				\put(48,-3){\small\bfseries (a)}
			\end{overpic}
		\end{subfigure}
		
		\vspace{1.5 em}
		\begin{subfigure}[t]{0.7\textwidth}
			\begin{overpic}[width=\linewidth]{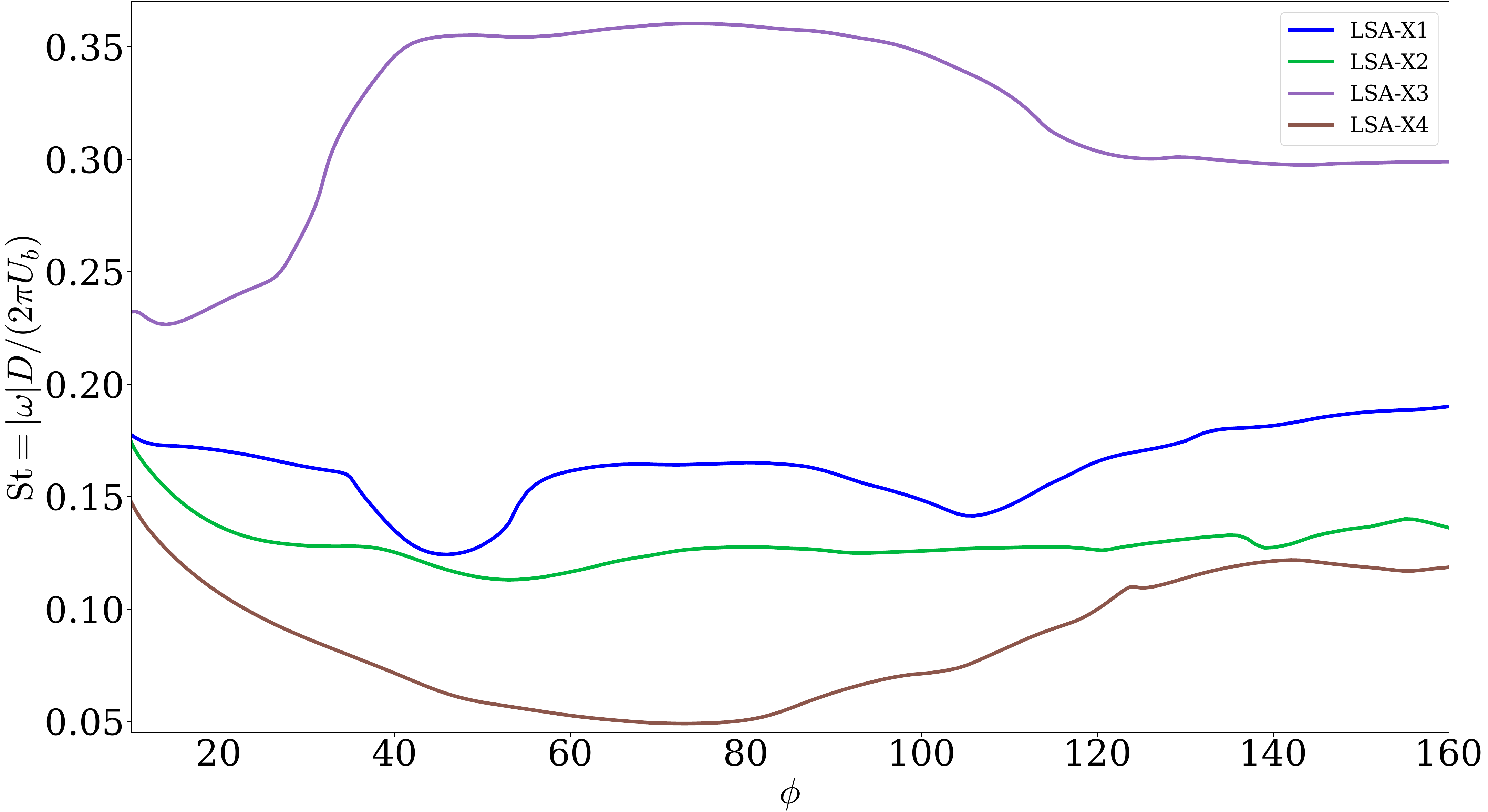}
				\put(48,-3){\small\bfseries (b)}
			\end{overpic}
		\end{subfigure}
		\caption{Evolution of the four additional unstable modes \textit{LSA-X1}--\textit{LSA-X4}: \textit{(a)} temporal growth rate and \textit{(b)} Strouhal number.}
		\label{fig:appendix_x_tracks}
	\end{figure}

	The local stability analysis reveals six unstable branches. The \textit{LSA-SS} and \textit{LSA-SB} branches are discussed in detail in \S\ref{sec:lsa_result}; the remaining four branches, \textit{LSA-X1}--\textit{LSA-X4}, are presented here. Figure~\ref{fig:appendix_x_tracks} shows the evolution of the growth rate and Strouhal number with bend angle $\phi$ for these modes. The additional modes are summarised in table~\ref{tab:appendix_lsa_branch_statistics}.

	\FloatBarrier
	\FloatBarrier

	\begin{table}
		\centering
		\caption{Summary of the additional unstable modes over $10^\circ\leq\phi\leq160^\circ$. The Strouhal-number ranges and mean values are evaluated only over the unstable bend angles $\phi$.}
		\label{tab:appendix_lsa_branch_statistics}
		\vspace{0.2em}
		\scriptsize
		\setlength{\tabcolsep}{3pt}
		\begin{tabular}{l c c c c c c}
			Mode & $\alpha D$ & Unstable angles & $St$ & Mean $St$ & $\max(\sigma D/U_b)$ & Peak-growth angle, $\phi_{\sigma,\max}$ \\
			\hline
			\textit{LSA-X1} & 1.2 & $10.0^\circ$--$25.6^\circ$ & 0.167--0.178 & 0.171 & 0.191 & $10.0^\circ$ \\
			\textit{LSA-X2} & 1.2 & $10.0^\circ$--$69.8^\circ$ & 0.113--0.174 & 0.127 & 0.160 & $13.0^\circ$ \\
            \textit{LSA-X2} & 0.6 & $10.0^\circ$--$68.2^\circ$ & 0.024--0.076 & 0.038 & 0.209 & $11.0^\circ$ \\
			\textit{LSA-X3} & 1.2 & $14.2^\circ$--$21.3^\circ$ and $113.8^\circ$--$124.7^\circ$ & 0.227--0.315 & 0.277 & 0.0299 & $118.0^\circ$ \\
			\textit{LSA-X4} & 1.2 & $10.0^\circ$--$36.7^\circ$ & 0.077--0.148 & 0.103 & 0.0675 & $18.0^\circ$ \\
		\end{tabular}
	\end{table}

\begin{figure}[!htbp]
		\centering
		\begin{subfigure}[t]{0.49\textwidth}
			\begin{overpic}[width=\linewidth]{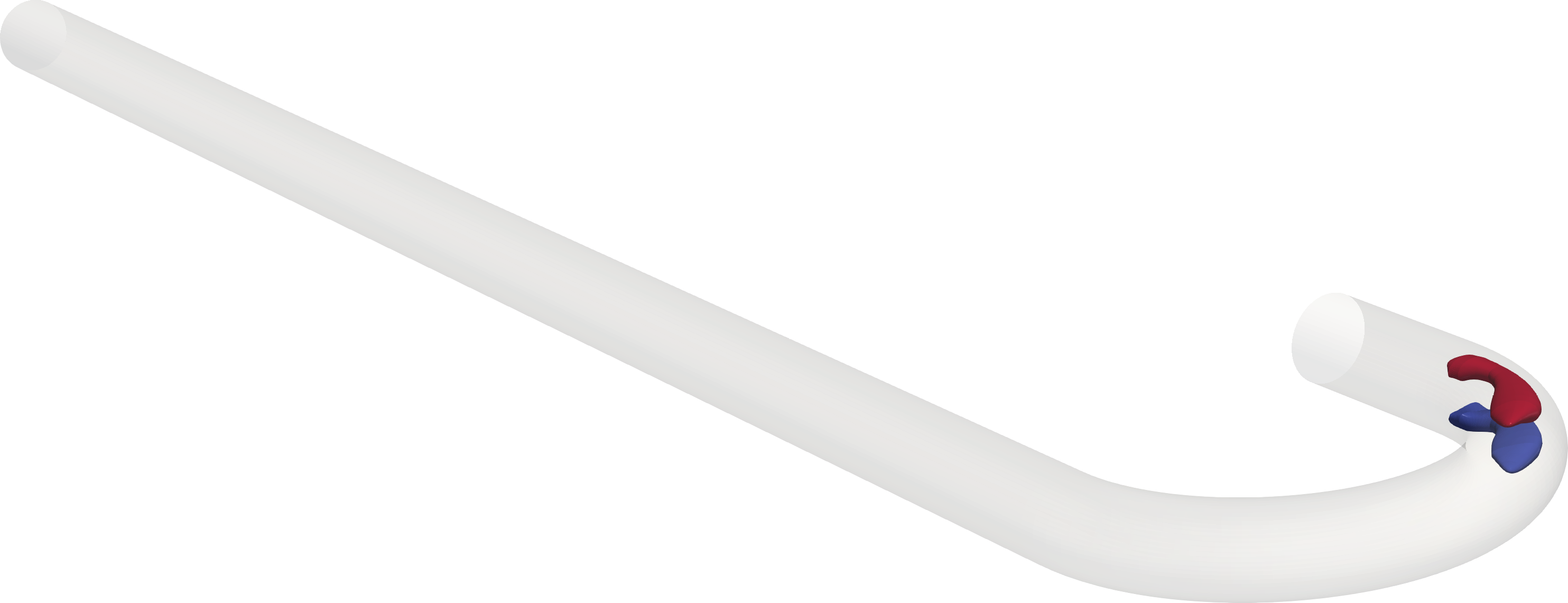}
				\inletplanearrow{62}{17.2}{1.25}{-.6}
				\put(48,0){\small\bfseries (a)}
			\end{overpic}
		\end{subfigure}\hfill
		\begin{subfigure}[t]{0.49\textwidth}
			\begin{overpic}[width=\linewidth]{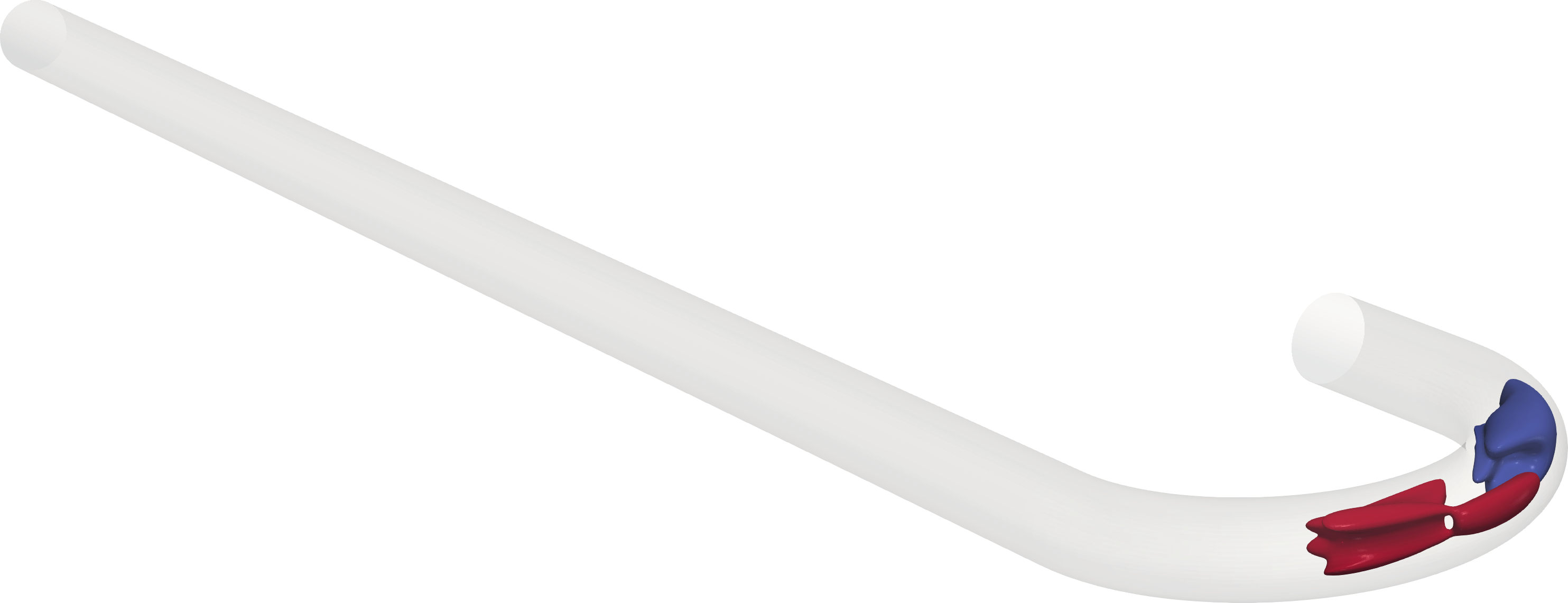}
				\put(48,0){\small\bfseries (b)}
			\end{overpic}
			\vspace{-0.5em}
		\end{subfigure}
		\begin{subfigure}[t]{0.49\textwidth}
			\begin{overpic}[width=\linewidth]{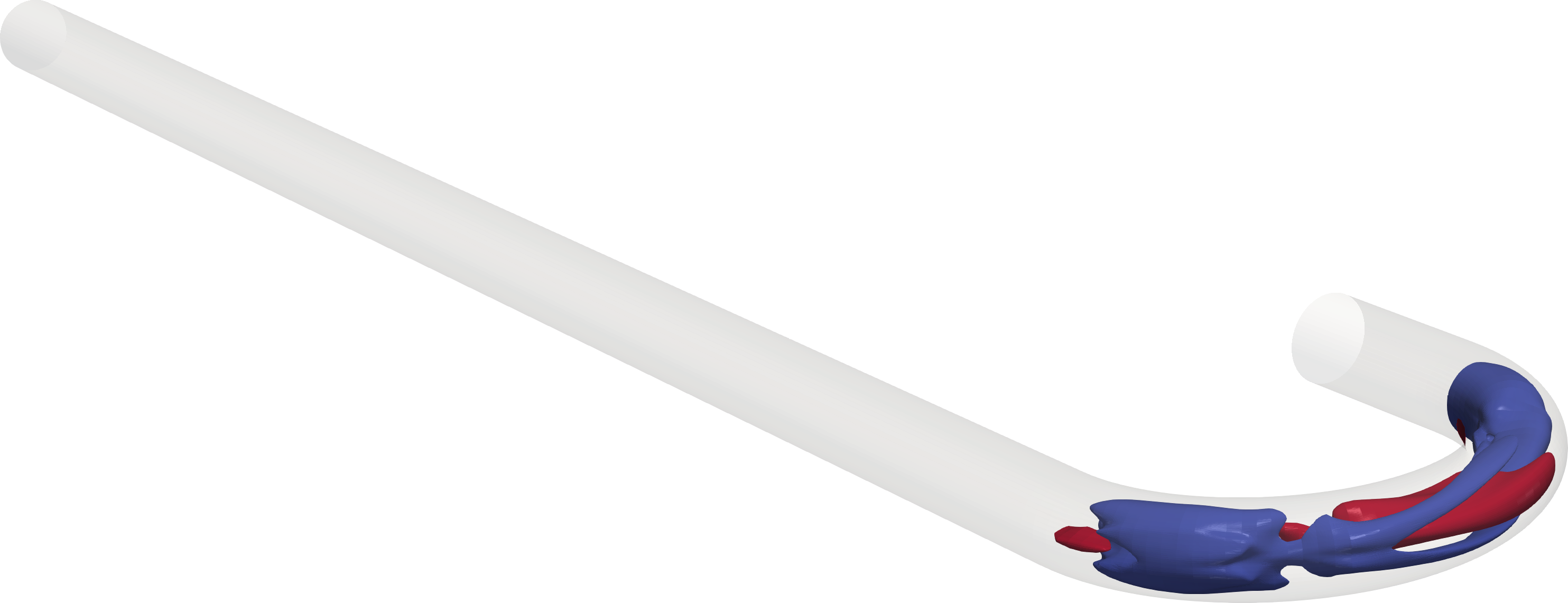}
				\put(48,0){\small\bfseries (c)}
			\end{overpic}
		\end{subfigure}\hfill
		\begin{subfigure}[t]{0.49\textwidth}
			\begin{overpic}[width=\linewidth]{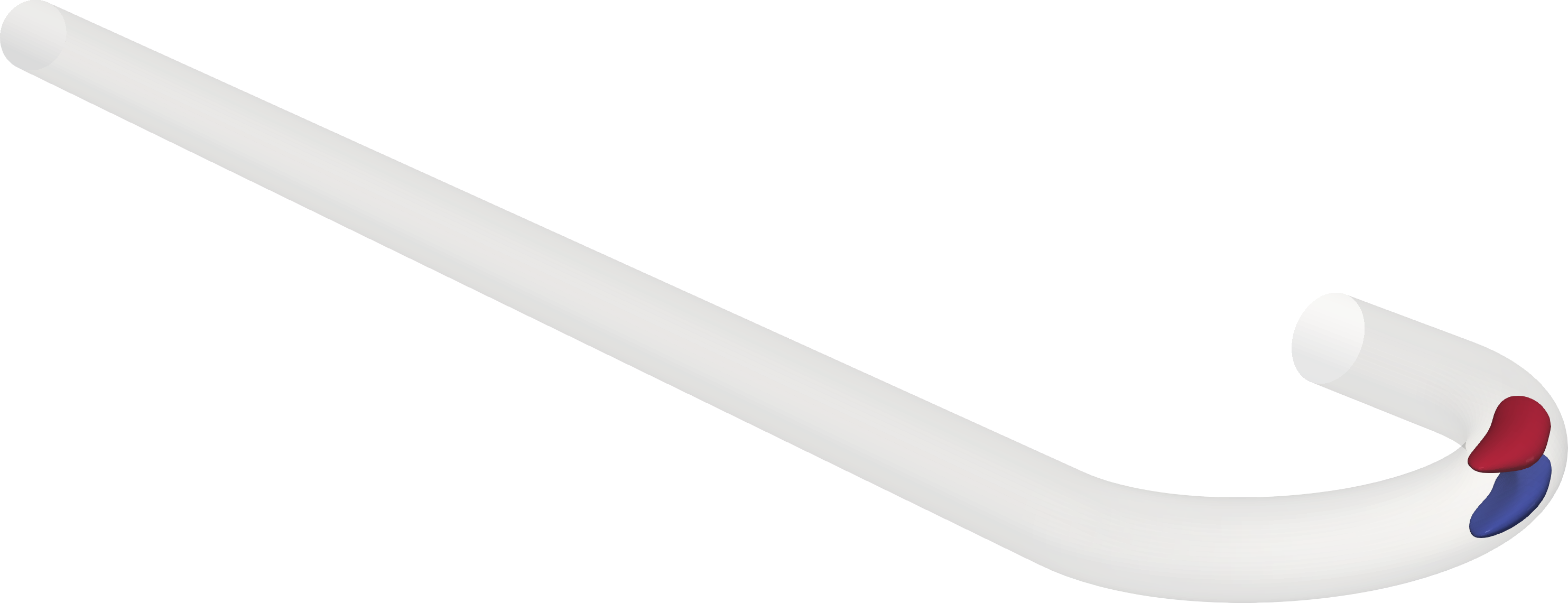}
				\put(48,0){\small\bfseries (d)}
			\end{overpic}
		\end{subfigure}
		\caption{Three-dimensional reconstructions of the four additional unstable modes at $\alpha D=1.2$: \textit{(a)} \textit{LSA-X1}, \textit{(b)} \textit{LSA-X2}, \textit{(c)} \textit{LSA-X3} and \textit{(d)} \textit{LSA-X4}. Red and blue denote equal positive and negative levels of streamfunction $\Psi=\pm 0.07 U_bD$.}
		\label{fig:appendix_x_shapes}
	\end{figure}

	The reconstructions of these modes are shown in figure~\ref{fig:appendix_x_shapes}. Modes~\textit{LSA-X1} and~\textit{LSA-X4} are localised at the entrance to the bend, where the secondary flow starts to form. Both modes are short-lived and become stable at $\phi=25.6^\circ$ and $\phi=36.7^\circ$, respectively. 
	
\begin{figure}[!htbp]
		\centering
		
		\begin{subfigure}[t]{0.49\textwidth}
			\centering
			\begin{overpic}[width=\linewidth]{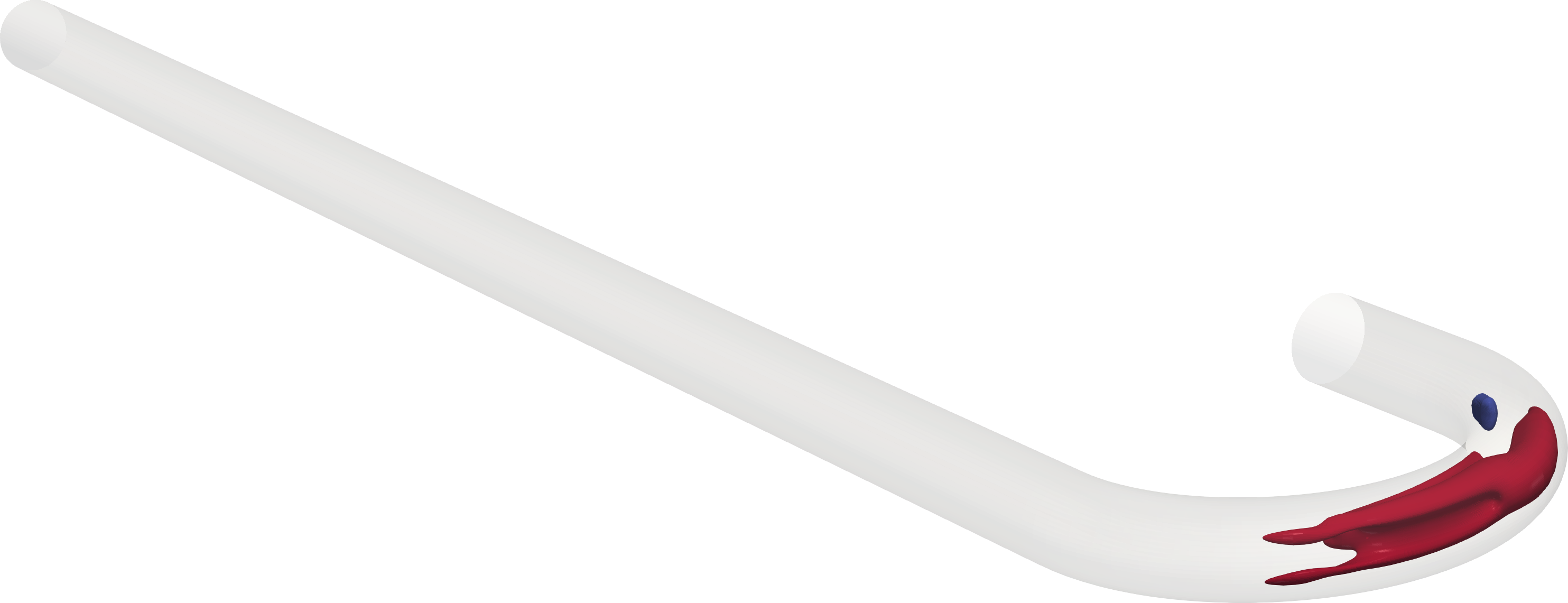}
				\inletplanearrow{62}{17.2}{1.25}{-.6}
				\put(48,-3){\small\bfseries (a)}
			\end{overpic}
		\end{subfigure}
		
		\vspace{1.5em}
		
		\begin{subfigure}[t]{0.6\textwidth}
			\centering
			\begin{overpic}[width=\linewidth]{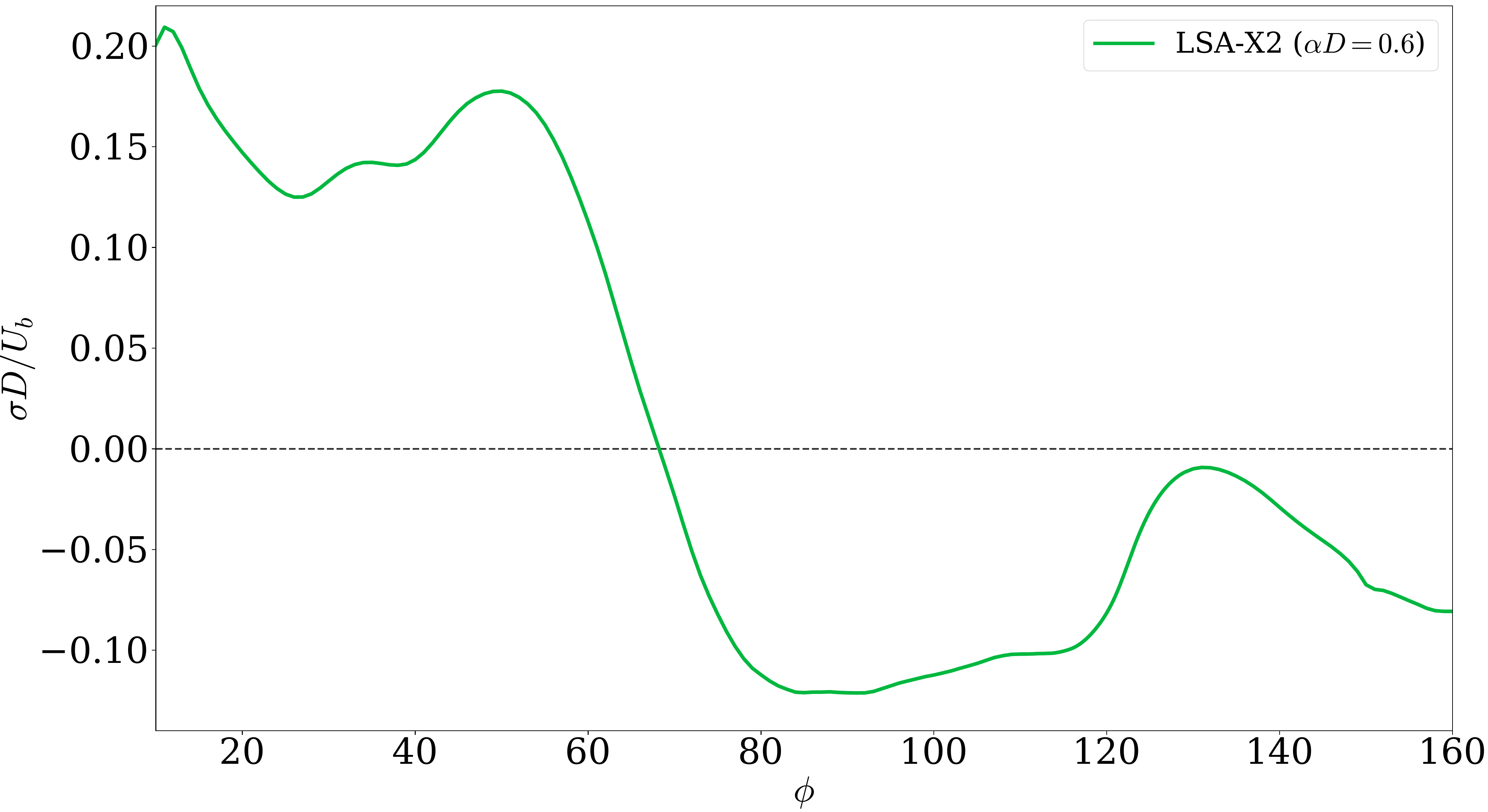}
				\put(48,-3){\small\bfseries (b)}
			\end{overpic}
		\end{subfigure}
		
		\vspace{1.5em}
		
		\begin{subfigure}[t]{0.6\textwidth}
			\centering
			\begin{overpic}[width=\linewidth]{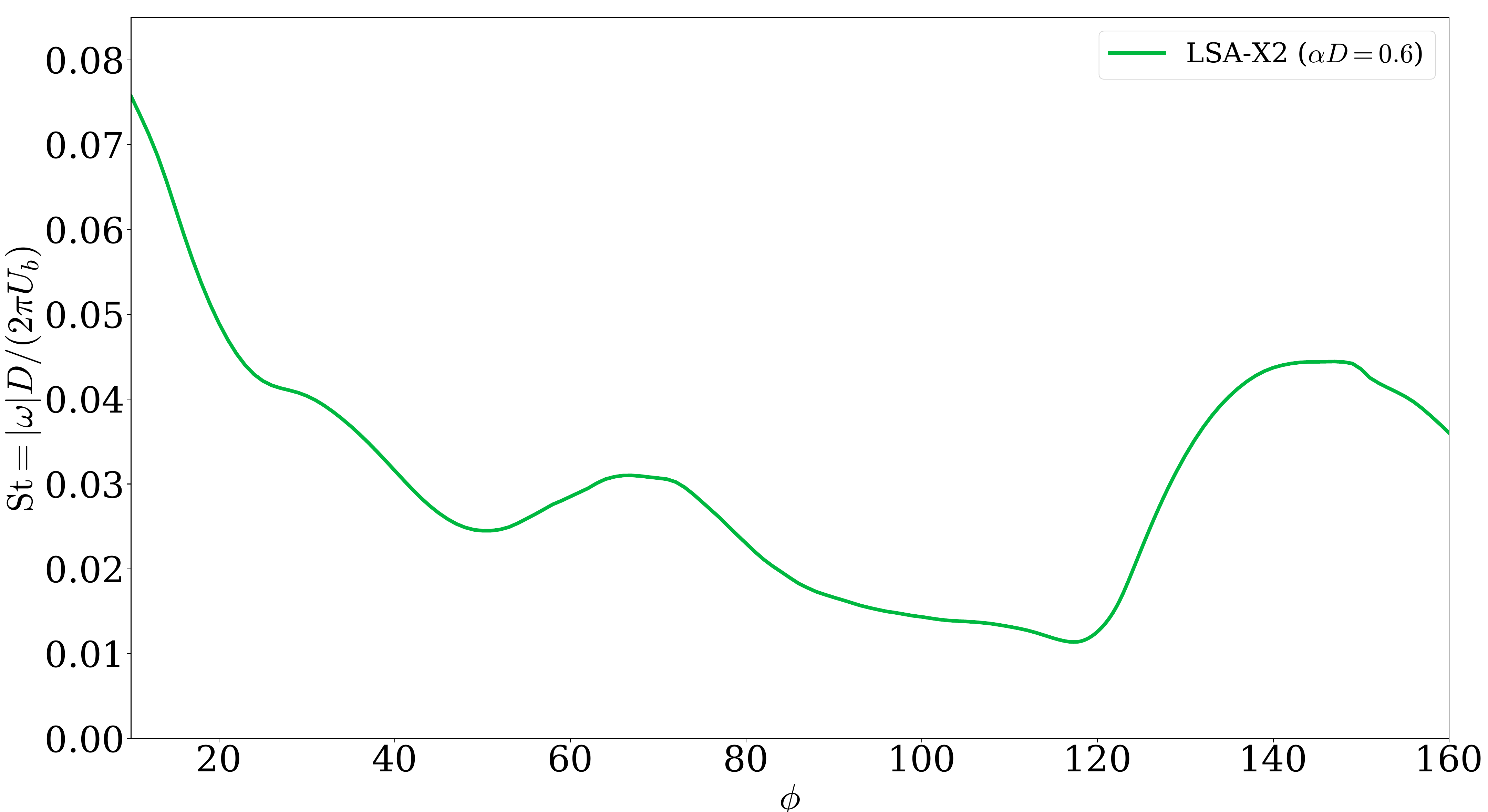}
				\put(48,-3){\small\bfseries (c)}
			\end{overpic}
		\end{subfigure}
		
		\caption{\textit{(a)} Three-dimensional reconstruction of \textit{LSA-X2} at $\alpha D=0.6$, where red and blue denote positive and negative streamfunction levels. The temporal growth rate and Strouhal number are shown in \textit{(b)} and \textit{(c)}, respectively.}
		\label{fig:appendix_lsa_x2_lowalpha}
	\end{figure}

	The \textit{LSA-X2} branch is likely related to the FHPOD mode~\textit{Ax}. Figure~\ref{fig:appendix_lsa_x2_lowalpha}\textit{(a)} shows the spatial reconstruction of this mode for $\alpha D=0.6$, with its temporal growth rate and Strouhal number shown in panels~\textit{(b)} and~\textit{(c)}. Its mean Strouhal number over the unstable bend angles is $St\approx0.038$, which is close to the peak frequency of the FHPOD mode~\textit{Ax} at $St=0.03$. This association is, however, speculative, and further analysis is required to investigate the possible relation between \textit{LSA-X2} and the FHPOD mode~\textit{Ax}.

\begin{figure}[!htbp]
		\centering
		\begin{subfigure}[t]{0.49\textwidth}
			\begin{overpic}[width=\linewidth]{figure_60.png}
				\inletplanearrow{62}{17.2}{1.25}{-.6}
				\put(48,0){\small\bfseries (a)}
			\end{overpic}
		\end{subfigure}\hfill
		\begin{subfigure}[t]{0.49\textwidth}
			\begin{overpic}[width=\linewidth]{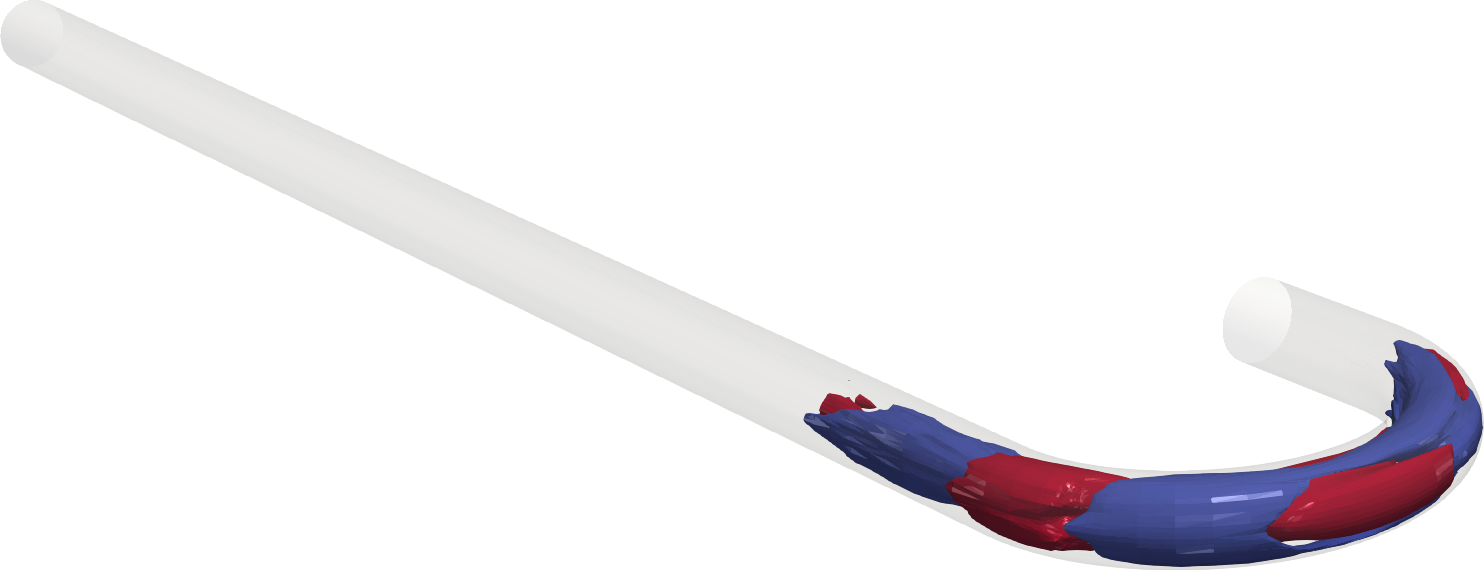}
				\put(48,0){\small\bfseries (b)}
			\end{overpic}
		\end{subfigure}
		
		\vspace{1.2em}
		\begin{subfigure}[t]{0.64\textwidth}
			\begin{overpic}[width=\linewidth]{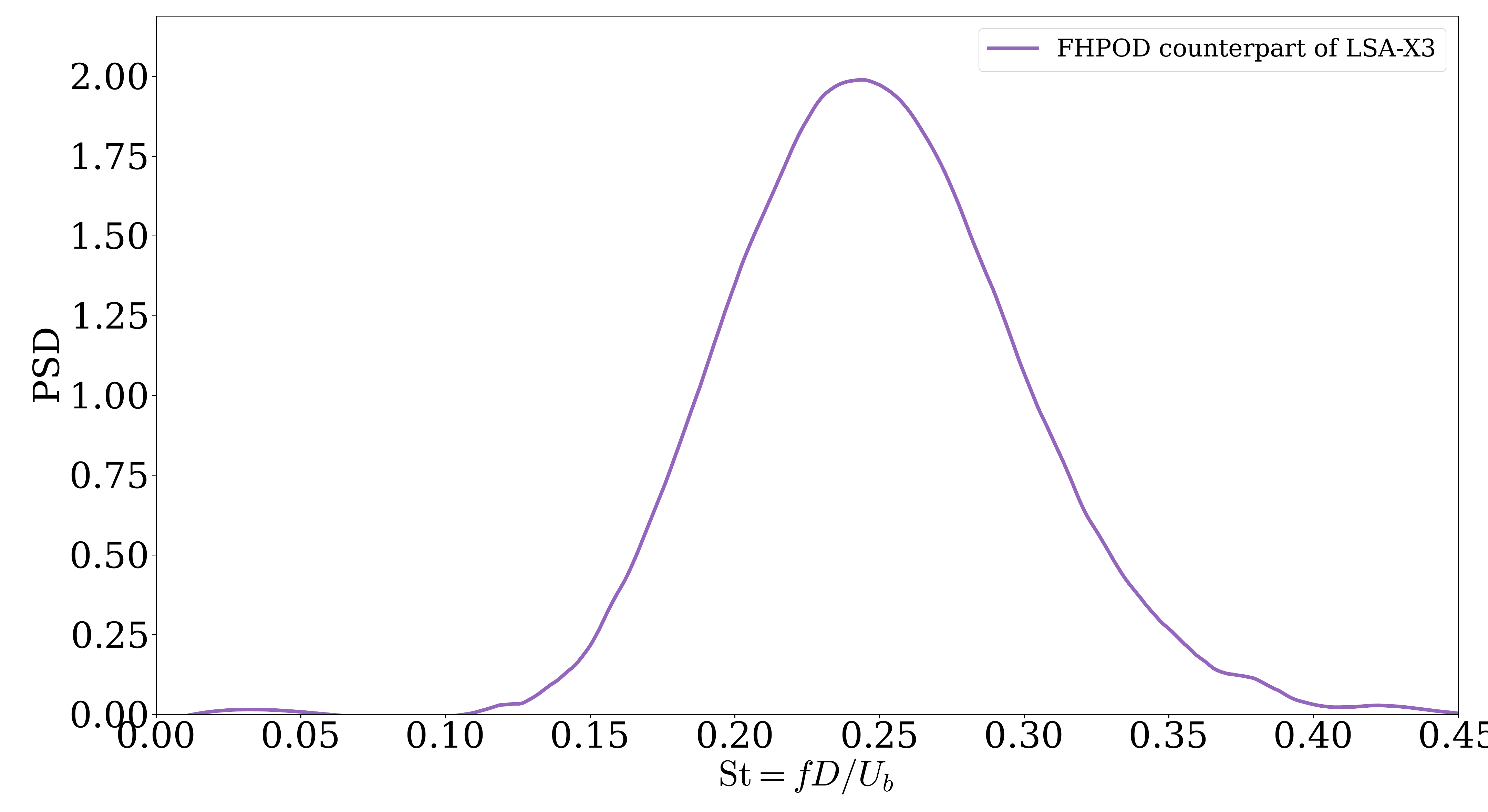}
				\put(48,-3){\small\bfseries (c)}
			\end{overpic}
		\end{subfigure}
		\caption{Comparison of \textit{(a)} the three-dimensional reconstruction of \textit{LSA-X3}, \textit{(b)} its FHPOD counterpart and \textit{(c)} the spectra of the FHPOD mode.}
		\label{fig:appendix_lsa_x3}
	\end{figure}

	Mode~\textit{LSA-X3} represents the radial instability of the Dean vortices in addition to the sinuous and varicose instabilities discussed earlier. Its frequency range is close to those of \textit{LSA-SS2} and~\textit{LSA-SB2}. Above $\phi=100^\circ$, all other modes are stable, whereas \textit{LSA-X3} becomes unstable over $113.8^\circ<\phi<124.7^\circ$. A low-energy FHPOD mode is also found, which is likely the counterpart of \textit{LSA-X3}. It shares similar spatial features and frequency with \textit{LSA-X3} and its mode shape and its spectrum are shown in figure~\ref{fig:appendix_lsa_x3}\textit{(b,c)}.
	
	\FloatBarrier
	
	\begin{bmhead}[Funding.]
		This project was funded by the European High Performance Computing Joint Undertaking (JU), together with Sweden, Germany, Spain, Greece, and Denmark, under Grant Agreement No. 101093393. The computing hardware was funded by the German Research Foundation (DFG). 
	\end{bmhead}
	
	\begin{bmhead}[Acknowledgements.]
		The authors gratefully acknowledge the scientific support and high-performance computing resources provided by the Erlangen National High Performance Computing Center (NHR@FAU) of the Friedrich--Alexander--Universität Erlangen--Nürnberg (FAU).
	\end{bmhead}
	
	\begin{bmhead}[Declaration of Interests.] The authors report no conflict of interest.
	\end{bmhead}
	
	\bibliography{bibl}

\end{document}